\def\arxivversion{}
\ifdefined\arxivversion
\documentclass[acmsmall,screen,nonacm]{acmart}
\date{}
\else
\documentclass[acmsmall,screen,review,anonymous,nonacm]{acmart}
\fi

\usepackage[T1]{fontenc}

\usepackage{algorithm}
\usepackage{placeins}
\usepackage{algpseudocodex}

\algrenewcommand{\algorithmicrequire}{\textbf{Input:}}
\algrenewcommand{\algorithmicensure}{\textbf{Output:}}
\algrenewcommand{\algorithmicindent}{1.2em}
\usepackage{graphicx}
\usepackage{array}
\usepackage{tabularx}
\usepackage{xcolor}
\usepackage{colortbl}
\usepackage{multirow}
\usepackage{xspace}
\usepackage[most]{tcolorbox}

\usepackage{pgfplots}
\usepgfplotslibrary{groupplots}
\pgfplotsset{compat=1.18}
\usetikzlibrary{decorations.pathmorphing, fit, positioning}

\theoremstyle{acmdefinition}

\newcommand{\opn}{\mathord{\lhd}}
\newcommand{\cls}{\mathord{\rhd}}
\newcommand{\ACCEPT}{\textcolor{green!50!black}{\checkmark}}
\newcommand{\REJECT}{\textcolor{red}{\text{\sffamily X}}}

\newcommand{\test}{\mathit{\Psi}}

\def\|#1|{\textit{#1}} \def\<#1>{{\setlength{\fboxsep}{0.8pt}\colorbox{black!6}{\texttt{#1}}}}
\newcommand{\kw}[1]{\textbf{#1}}

\newcommand{\good}[1]{\mbox{\textcolor{green!50!black}{#1}}}
\newcommand{\bad}[1]{\mbox{\textcolor{red}{#1}}}

\newcommand{\ddmin}{\texttt{ddmin}\xspace}
\newcommand{\drdd}{\texttt{drdd}\xspace}
\newcommand{\autodidact}{\textsc{AutoDiDact}\xspace}

\newcommand{\vs}{\textvisiblespace}

\newcommand{\sq}{\textquotesingle}

\newlength{\chunkcell}

\makeatletter
\newcommand{\chunkchars}[1]{\settowidth{\chunkcell}{\texttt{M}}\addtolength{\chunkcell}{1pt}\@tfor\chunk@c:=#1\do{\makebox[\chunkcell][c]{\texttt{\chunk@c}}}}
\makeatother

\makeatletter
\newcommand{\chunk@unit}[1]{\fcolorbox{black!45}{white}{\strut\chunkchars{#1}}\kern-\fboxrule}
\def\chunk@units#1|{\ifx\chunk@end#1\else\chunk@unit{#1}\expandafter\chunk@units\fi}

\newcommand{\chunk}[2][black]{\kern1pt\tikz[baseline=(chunk.base)]{\node[inner sep=0pt, outer sep=0pt](chunk){\setlength{\fboxsep}{0.5pt}\setlength{\fboxrule}{0.3pt}\chunk@units#2|\chunk@end|\kern\fboxrule};
		\draw[line width=1pt, #1]([xshift=0.6pt, yshift=-2pt]chunk.south west) -- ([xshift=-0.6pt, yshift=-2pt]chunk.south east);}\kern1pt}
\makeatother

\newcommand{\chunkunit}[1]{{\setlength{\fboxsep}{0.5pt}\setlength{\fboxrule}{0.3pt}\fcolorbox{black!45}{white}{\strut\chunkchars{#1}}}}

\definecolor{spanone}{rgb}{0.00, 0.45, 0.70}
\definecolor{spantwo}{rgb}{0.80, 0.47, 0.65}

\newcommand{\badchunk}[1]{\chunk[red, decorate, decoration={snake, amplitude=0.8pt, segment length=3pt}]{#1}}

\newcommand{\tablestyle}{\centering\footnotesize \setlength{\tabcolsep}{2pt}\renewcommand{\arraystretch}{0.9}}

\newcommand{\tablemedium}{\fontsize{6.8pt}{8.3pt}\selectfont\setlength{\tabcolsep}{1.6pt}}

\definecolor{lightyellow}{rgb}{1.0, 1.0, 0.75}

\definecolor{pastelblue}{rgb}{0.70, 0.85, 1.0}

\newcolumntype{N}{r}
\newcolumntype{L}{>{\raggedright\arraybackslash}X}

\newcommand{\case}[1]{{\renewcommand{\ldots}{{\rmfamily\textellipsis}}\texttt{#1}}}

\newlength{\subtablesep}
\newcommand{\subtablegap}{\noalign{\vskip\subtablesep}}

\newcommand{\halftable}{\dimexpr(\linewidth-\subtablesep)/2\relax}

\newcommand{\tablenote}[1]{\par\medskip\parbox{\linewidth}{#1}}

\newcommand{\continuedcaption}[2]{\addtocounter{table}{-1}\renewcommand{\theHtable}{\arabic{table}.#1.continued}\caption[]{#2 (continued)}}

\newtcolorbox{result}{
	colback   = gray!10,
	colframe  = blue!40!gray,
	boxrule   = 0.5pt,
	arc       = 2pt,
	left      = 8pt,
	right     = 8pt,
	top       = 6pt,
	bottom    = 6pt,
	center,
	width     = 1.0\linewidth,
	fontupper = \small,
	shadow    = {1pt}{-1pt}{0pt}{gray!20},
}

\newcommand{\todo}[1]{\ifinner\else\marginpar{$|$}\fi \stepcounter{footnote}\textcolor{red}{\footnote[\value{footnote}]{\textcolor{red}{\textbf{TODO}} \textit{#1}}}}

\IfFileExists{SUBMIT}{\renewcommand{\todo}[1]{}}{}

\hypersetup{
	colorlinks = true,
	urlcolor   = blue,
	linkcolor  = blue,
	citecolor  = blue
}
\usepackage{cleveref}

\makeatletter
\def\theHALG@line{\thealgorithm.\arabic{ALG@line}}
\makeatother

\newcommand{\inputs}{125}
\newcommand{\formats}{18}
\newcommand{\textFormats}{ten}

\newcommand{\textInputs}{74}
\newcommand{\binaryInputs}{51}
\newcommand{\reproducers}{113}
\newcommand{\controls}{12}
\newcommand{\smallestInput}{16}
\newcommand{\largestInput}{224,151}

\newcommand{\worded}{75}
\newcommand{\abstained}{50}

\newcommand{\discoveryTime}{11.0}
\newcommand{\textDiscoveryCalls}{278}
\newcommand{\binaryDiscoveryCalls}{284}
\newcommand{\abstentionLeast}{284}
\newcommand{\abstentionMost}{289}
\newcommand{\abstentionPooled}{2.9}
\newcommand{\abstentionMean}{18.2}
\newcommand{\abstentionCostlier}{12}
\newcommand{\abstentionWorst}{6.9}

\newcommand{\wordedGroups}{50}
\newcommand{\budgetInputs}{37}
\newcommand{\budgetTextInputs}{20}

\newcommand{\budgetSufficient}{200}

\newcommand{\budgetStable}{300}
\newcommand{\probedGaps}{52,943}
\newcommand{\internalGaps}{2,102,563}
\newcommand{\probedShare}{2.5}

\newcommand{\smallShare}{38.9}
\newcommand{\largeInputs}{26}
\newcommand{\largeShare}{2.1}

\newcommand{\controlInputs}{20}
\newcommand{\controlFamilies}{11}
\newcommand{\controlGaps}{251,728}

\newcommand{\controlCalls}{7,579}
\newcommand{\controlExhaustive}{223,630}
\newcommand{\controlCallShare}{3.4}
\newcommand{\controlDisagreements}{304}
\newcommand{\controlMispredicted}{298}
\newcommand{\controlPooled}{six}

\newcommand{\controlAgreement}{99.88}

\newcommand{\controlLowest}{94.3}
\newcommand{\controlHighest}{100}
\newcommand{\controlErrorless}{SMT, SQL, and YAML}
\newcommand{\fixpointInputs}{75}
\newcommand{\fixpointSmaller}{60}
\newcommand{\fixpointCheaper}{36}

\newcommand{\fixpointSizeSaving}{83}
\newcommand{\fixpointCallSaving}{37}
\newcommand{\fixpointSmallInputs}{34}

\newcommand{\fixpointSmallCheaper}{5}

\newcommand{\fixpointSmallCallRatio}{1.79}
\newcommand{\fixpointSmallSizeSaving}{30}

\newcommand{\fixpointMidSizeSaving}{69}
\newcommand{\fixpointMidCallSaving}{7}
\newcommand{\fixpointLargeInputs}{26}
\newcommand{\fixpointLargeSmaller}{23}
\newcommand{\fixpointLargeCheaper}{22}

\newcommand{\fixpointLargeSizeSaving}{98}
\newcommand{\fixpointLargeCallSaving}{87}

\newcommand{\shrinkInputs}{75}
\newcommand{\shrinkImproved}{55}
\newcommand{\shrinkAdditionalSaving}{16}
\newcommand{\shrinkAdditionalCost}{11}
\newcommand{\fixpointEqual}{14}
\newcommand{\fixpointLarger}{one}
\newcommand{\fixpointTimeRatio}{0.89}

\newcommand{\fixpointLargeTimeRatio}{0.25}
\newcommand{\minimalInputs}{75}
\newcommand{\minimalSizeRatio}{0.17}

\newcommand{\minimalCallSaving}{30}
\newcommand{\minimalShrunk}{28}
\newcommand{\minimalCost}{11}
\newcommand{\wordShare}{3.1}
\newcommand{\tokensShare}{7.7}
\newcommand{\pairsShare}{0.8}

\newcommand{\inferenceSmallShare}{50.1}

\newcommand{\inferenceLargeShare}{8.9}

\newcommand{\inferenceShare}{11.6}
\newcommand{\pairCalls}{5,841}
\newcommand{\pairMedian}{71}
\newcommand{\pairMost}{255}
\newcommand{\pairTime}{7.7}
\newcommand{\smallAtomCalls}{414}

\newcommand{\largestTokenizeCalls}{454}
\newcommand{\largestTokenizeShare}{0.2}

\newcommand{\pairInputs}{75}
\newcommand{\pairLearned}{59}
\newcommand{\pairAdmitted}{4,478}
\newcommand{\pairTags}{4,369}
\newcommand{\pairUntagged}{109}
\newcommand{\pairDelimiters}{96}
\newcommand{\pairConditional}{six}
\newcommand{\pairOther}{seven}

\newcommand{\creduceInputs}{75}
\newcommand{\creduceSmaller}{26}
\newcommand{\creduceCheaper}{59}

\newcommand{\creduceSizeSaving}{30}
\newcommand{\creduceCallSaving}{54}
\newcommand{\creduceReduceCallRatio}{0.19}
\newcommand{\creduceRawSizeRatio}{0.56}
\newcommand{\creduceMedianSizeRatio}{1.03}
\newcommand{\creduceEqual}{11}
\newcommand{\creduceLarger}{38}

\newcommand{\creduceNoXmlSizeRatio}{1.23}
\newcommand{\creduceNoXmlCallRatio}{0.63}

\newcommand{\creduceXmlSizeRatio}{0.02}

\newcommand{\hddFamilies}{eight}
\newcommand{\hddInputs}{50}
\newcommand{\hddSmaller}{28}
\newcommand{\hddCheaper}{6}
\newcommand{\hddSizeRatio}{0.96}
\newcommand{\hddCallRatio}{3.85}

\newcommand{\hddReduceCallRatio}{1.53}
\newcommand{\hddRawSizeRatio}{0.85}
\newcommand{\hddMedianSizeRatio}{0.92}

\newcommand{\hddSyntaxInputs}{2}

\newcommand{\hddAtomSizeRatio}{9.2}

\newcommand{\hddNoXmlSizeRatio}{0.89}
\newcommand{\hddNoXmlCallRatio}{2.51}

\newcommand{\hddXmlSizeRatio}{1.29}
\newcommand{\hddXmlCallRatio}{21.18}

\newcommand{\hddCleanSizeRatio}{0.97}
\newcommand{\hddCleanCallRatio}{4.06}

\newcommand{\hddCohortAtomSizeSaving}{90}
\newcommand{\hddCohortAtomCallSaving}{63}

\newcommand{\hddCohortCreduceCallSaving}{65}
\newcommand{\lexerInputs}{69}
\newcommand{\lexerFamilies}{nine}

\newcommand{\lexerExact}{45.9}

\newcommand{\lexerWithin}{99.2}
\newcommand{\lexerRecall}{97.7}
\newcommand{\lexerInsideSpace}{16.3}
\newcommand{\lexerInsideQuoted}{42.9}
\newcommand{\lexerCrossings}{12,492}
\newcommand{\lexerMergedShare}{99.6}
\newcommand{\lexerCrossFamilies}{XML, Python, and YAML}
\newcommand{\lexerCrossShare}{99.7}

\newcommand{\largeHddInputs}{18}
\newcommand{\showcaseKiB}{95.4}
\newcommand{\showcaseAtomSize}{45,061}
\newcommand{\showcaseAtomCalls}{232,803}
\newcommand{\showcaseSize}{151}
\newcommand{\showcaseCalls}{3,023}
\newcommand{\structureBytes}{99}
\newcommand{\grammarBytes}{18,234}

\begin{document}

\title{AutoDiDact: Input Reduction over Inferred Structure}

\ifdefined\arxivversion
\author{Aarush Kumbhakern}
\affiliation{\institution{Ashoka University}
    \city{Sonipat}
    \country{India}}
\email{kern.academic@mailbox.org}

\author{Rahul Gopinath}
\affiliation{\department{School of Computer Science}
    \institution{University of Sydney}
    \city{Sydney}
    \country{Australia}}
\email{rahul.gopinath@sydney.edu.au}

\else
\author{Anonymous Author(s)}
\fi

\begin{abstract}
Delta debugging reduces a failing input by deleting chunks of it, keeping a deletion only if an oracle confirms the failure persists. It chunks the input based only on its length, so its deletions cut through units such as identifiers and delimited groups, wasting calls and leaving outputs large. Current state-of-the-art structural reducers avoid this by supplying grammars or built-in format knowledge. We show that structured reduction needs neither: a neutral word and delimiter pairs suffice, and both can be inferred from the failing input by asking only what delta debugging already asks its oracle: does the failure persist?

A \emph{neutral word}, such as a space, preserves the failure when inserted between such units but not inside them, so it splits the input into \emph{tokens}. A \emph{delimiter pair} is two tokens that preserve the failure only when added or removed together, so the span they bound can be reduced as a unit. We formalize both, state when the oracle can reveal them, and learn each with a bounded number of calls. \autodidact{} reduces the resulting tree of spans by hierarchical delta debugging (HDD) and shrinks each token across its occurrences.

\autodidact{} finds a neutral word in \worded{} of \inputs{} inputs from \formats{} format families. On these, it returns outputs \fixpointSizeSaving\% smaller than byte-level delta debugging with \fixpointCallSaving\% fewer calls on geometric mean, inference included. The savings grow with input size, to \fixpointLargeSizeSaving\% and \fixpointLargeCallSaving\% on inputs of at least 10\,KiB. Without a grammar, \autodidact{} matches grammar-based HDD in output size at \hddCallRatio{}$\times$ its calls, and uses \creduceCallSaving\% fewer calls than C-Reduce's language-agnostic mode. Because it needs only the oracle, \autodidact{} can minimize whatever a fuzzer finds, malformed inputs included.
\end{abstract}


\maketitle

\section{Introduction}\label{sec:introduction}

Delta debugging~\cite{zeller2002simplifying} reduces a failing input by deleting chunks of it, keeping a deletion only when an \emph{oracle} confirms the failure persists. Byte-level delta debugging chunks the input based only on its length, blind to its structure. A chunk that cuts through a unit such as an identifier or splits a pair of parentheses is rejected, wasting a call. A whole unit is deleted only when a chunk happens to align with it, and edits that must touch separate places at once, such as removing both parentheses of a pair, are never tried. On a \showcaseKiB\,KiB C program from our benchmark that crashes GCC~4.3.0 (\Cref{tab:results}), byte-level delta debugging spends \showcaseAtomCalls{} calls and stops at \showcaseAtomSize{} bytes, not counting whitespace: no single byte can be removed, yet the output is far from small.

Structural reducers edit whole units, using format knowledge supplied in advance. Hierarchical delta debugging (HDD)~\cite{misherghi2006hdd} and Perses~\cite{sun2018perses} obtain units from a grammar. Such a grammar may be unavailable, even for a widely used language~\cite{perlfaq7} or a specified format~\cite{gopinath2020mining}, and one that exists may not cover the dialects and extensions an input uses~\cite{park2021jest,momot2016seven}. Other reducers build format knowledge into the tool. C-Reduce~\cite{regehr2012test} transforms C and C++ with the Clang front end, and for other formats, its \texttt{-{}-not-c} mode keeps generic passes, such as deleting lines, C tokens, or bracket pairs from a fixed list, with or without their contents. This fixed list cannot pair an XML start tag \<{<name>}> with its end tag \<{</name>}>, since the names vary with the input.

We show that structured reduction needs neither a grammar nor built-in format knowledge: a neutral word and delimiter pairs suffice. Both can be inferred from the failing input by asking only what delta debugging already asks its oracle, without a lexer, example corpus, or syntax oracle. A \emph{neutral word}, such as a space, preserves the failure when inserted between units but not inside them, so the offsets where it is accepted split the input into \emph{tokens}. Each \emph{delimiter pair}, such as matching parentheses or an XML start and end tag, encloses a \emph{span}. Its delimiters are \emph{coupled}: they preserve the failure when added or removed together but not either one alone. We implement this approach in \autodidact{}, which applies HDD to the inferred tokens and spans, deleting whole spans or only their delimiters, and then shortens each distinct token at all its occurrences. The reducer keeps an edit only if the oracle confirms that the failure persists. Incorrect structure may leave a larger output or require more calls. On the GCC input, \autodidact{} reaches \showcaseSize{} bytes in \showcaseCalls{} calls, inference included.

When no neutral word is found, reduction works on bytes, and when no pairs are found, on flat tokens. Because the inferred structure comes from the oracle, it reflects what the failure depends on. In a JavaScript test that crashes an instrumented library, the imported module path \<"../instrumented/lodash/hasPath.js"> stays one token, while the printed test title \<\sq{}SynTest Test Suite\sq{}>, where a space is accepted at every offset, splits into single characters.

We evaluate \autodidact{} on \inputs{} inputs from \formats{} format families. Inference finds a space as the neutral word on every text input and abstains on all but one binary input. Given a neutral word, \autodidact{} returns outputs \textbf{\fixpointSizeSaving\% smaller} than byte-level delta debugging with \textbf{\fixpointCallSaving\% fewer calls} on geometric mean, inference included. The savings grow with input size: on inputs of at least 10\,KiB they reach \fixpointLargeSizeSaving\% and \fixpointLargeCallSaving\%, respectively, while below 1\,KiB outputs are still \fixpointSmallSizeSaving\% smaller for \fixpointSmallCallRatio{}$\times$ the calls. Even without a grammar, \autodidact{} matches grammar-based HDD in output size, at \hddSizeRatio{}$\times$ on \hddInputs{} inputs, for \hddCallRatio{}$\times$ the calls, or \hddReduceCallRatio{}$\times$ excluding inference. Against C-Reduce's language-agnostic mode, it uses \creduceCallSaving\% fewer calls. Its outputs are \creduceSizeSaving\% smaller overall, driven by XML, and \creduceNoXmlSizeRatio{}$\times$ C-Reduce's on the other families. Inference finds \lexerRecall\% of the token boundaries that a standard lexer finds. It makes \inferenceShare\% of \autodidact{}'s calls overall and \inferenceLargeShare\% on inputs of at least 10\,KiB.

\autodidact{} parses nothing, so it runs on malformed and binary inputs alike. Fuzzers such as AFL~\cite{zalewski2014afl} and libFuzzer~\cite{libfuzzer} minimize the crashing inputs they find at the byte level, since they cannot assume an input format. With \autodidact{}, a fuzzer can carry a general minimizer that works on any input.

\noindent\textbf{Contributions:}
\begin{itemize}
	\item The observation that a neutral word and delimiter pairs, both learnable from the failure oracle, suffice for structured reduction without a grammar or built-in format knowledge.
	\item Formal definitions of neutral words and delimiter pairs as oracle behavior, the conditions under which the oracle reveals them, and bounded, sampling-based learners that infer them or abstain, together with a tokenizer that predicts boundary verdicts from earlier probes and tests likely boundaries together.
	\item Laminar reduction, HDD over inferred tokens and spans, with a tree construction in which $k$ stray delimiters change the number of spans by at most $k$.
	\item An empirical evaluation on \inputs{} inputs from \formats{} format families against byte-level delta debugging, HDD, and C-Reduce, with a released benchmark of oracles, pinned target builds, our implementation \autodidact{}, and experiment scripts.
\end{itemize}

\section{Input Reduction}\label{sec:minimization}

Given a failing input $S$, an \emph{input reducer} deletes \emph{atoms}, here bytes, while preserving the failure. An oracle $\test$ returns $\test(s)=\ACCEPT$ if candidate $s$ triggers the failure and $\test(s)=\REJECT$ otherwise, and a reducer keeps a deletion only if the result is accepted. We measure cost in oracle calls and in wall time, since calls differ in cost: one may run several target processes.

Ideally, a reducer would return a \emph{global minimum}: an accepted subsequence of smallest size. Without assumptions about $\test$, finding one can require exponentially many oracle calls~\cite{zeller2002simplifying}. Reducers therefore settle for local minimality. An input is \emph{1-minimal} over a given sequence of units, such as atoms or tokens, if no single unit can be deleted while preserving the failure.

\noindent\textbf{Running example.} Consider a C compiler that crashes on any call to a function named \<trace>, but only after the call parses and type-checks. Suppose a reported function body contains the statements \<{\kw{int} count = 0; \kw{if} (count > 100) \{ trace(count); \}}>. To keep the example small, we reduce only these 49 bytes and keep the rest of the function fixed, with \<\kw{void} trace(\kw{int});> declared. The oracle compiles the function with the candidate in place and accepts it if the compiler crashes. It rejects a candidate that does not parse or type-check, just as it rejects one that compiles without crashing, so the reducer cannot tell the two apart. For the measurements below, we simulate the crash: the oracle accepts a candidate that type-checks and contains a call to \<trace>.

\noindent\textbf{Delta debugging.} Delta debugging~\cite{zeller2002simplifying} deletes chunks of the input, keeping each deletion the oracle accepts. We use \drdd{}~\cite{kumbhakern2026drdd} (\Cref{alg:drdd}), a variant of the original \ddmin{} algorithm that defers its restarts. It sweeps the input with chunks of halving size $s$, from half the input down to two units. Each sweep cuts the input into chunks of $s$ units, the last possibly shorter, and tries deleting each in turn, from left to right, keeping every accepted deletion. A closing scan then tries deleting each unit alone, repeating until a scan deletes nothing. The result is 1-minimal over the chosen units, so the choice of units determines both where the cuts fall and what the result guarantees.

\begin{figure}[tp]
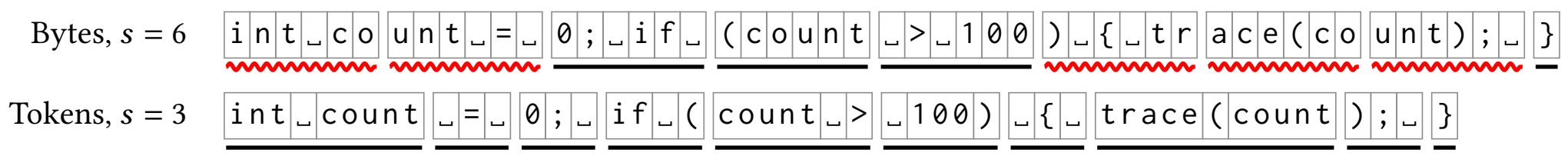

	\centering
	\footnotesize
	\begin{tabular}{@{}r@{\hspace{8pt}}l@{}}
		Bytes, $s=6$ & \badchunk{i|n|t|\vs|c|o}\badchunk{u|n|t|\vs|=|\vs}\chunk{0|;|\vs|i|f|\vs}\chunk{(|c|o|u|n|t}\chunk{\vs|>|\vs|1|0|0}\badchunk{)|\vs|\{|\vs|t|r}\badchunk{a|c|e|(|c|o}\badchunk{u|n|t|)|;|\vs}\chunk{\}} \\[9pt]
		Tokens, $s=3$ & \chunk{int|\vs|count}\chunk{\vs|=|\vs}\chunk{0|;|\vs}\chunk{if|\vs|(}\chunk{count|\vs|>}\chunk{\vs|100|)}\chunk{\vs|\{|\vs}\chunk{trace|(|count}\chunk{)|;|\vs}\chunk{\}}
	\end{tabular}
	\caption{The third \drdd{} sweep over the running example,
		in 6-byte chunks of its bytes and 3-token chunks of its tokens.
		Boxes are units, underlines are chunks,
		red squiggles split an identifier, and \vs{} is a space.}
	\label{fig:chunks}
\end{figure}

In the byte-level run, the sweeps with 24- and 12-byte chunks delete nothing, since no candidate compiles. With 6-byte chunks (\Cref{fig:chunks}), the sweep successfully deletes \<{\vs> 100}>, leaving \<(count)>.

\noindent\textbf{Avoiding bad divisions.} Byte-counted cuts fall inside identifiers. Five of the nine chunks at $s=6$ split \<count> or \<trace>. Deleting one leaves a broken or undeclared name, such as \bad{\<unt>} from deleting \bad{\<\kw{int} co>}, which the compiler rejects before reaching the crash, each at the cost of a compilation. Byte-level \drdd{} issues 111 queries, of which only 8 are accepted and 103 do not compile, and ends at the 36-byte \good{\<{\kw{int} count=0;\kw{if}(count)\{trace(count);\}}>}. However, if given the token boundaries of a C lexer, \drdd{} can instead count tokens when forming chunks. Every candidate then keeps or deletes whole tokens, never splitting an identifier, and the result is 1-minimal over tokens. On the example, this takes 58 queries and ends at the 34-byte \good{\<{\kw{int} count=0;\kw{if}(100)\{trace(count);\}}>}.

Tokens alone still waste queries. With 3-token chunks, the chunk \bad{\<{\kw{if} (}>} separates \<(> from its \<)>, and deleting it leaves an unbalanced condition that does not compile. If the reducer also knew that the parentheses and braces enclose \emph{spans}, it could delete a whole span, or \emph{unwrap} it by deleting both its delimiters in one query. Given these spans as well as the tokens, laminar \drdd{} (\Cref{sec:reduction}) deletes the \<\kw{if}> and its condition and unwraps the braces, reaching the 25-byte \good{\<{\kw{int} count=0;trace(count);}>} in 47 queries. One relation remains: \<count> must be spelled alike at its declaration and remaining use, so no deletion inside one occurrence alone is accepted. Shortening both at once, as shrinking does (\Cref{sec:reduction}), reaches the 17-byte \good{\<{\kw{int} t=0;trace(t);}>} in 15 more queries.

These counts assume the structure is given. \Cref{sec:reduction} defines reduction over tokens and spans, and \Cref{sec:learning} infers both from the failing input and its oracle.

\section{Reduction over Laminar Structure}\label{sec:reduction}

Reduction takes the failing input $S$, its oracle, and two kinds of structure. The \emph{tokens} partition $S$ into nonempty, contiguous pieces, and each delimiter \emph{pair} is an opening and a closing token. For a symmetric delimiter, such as a quote, both are the same token. Reduction runs in two stages: \emph{laminar reduction} deletes tokens and the spans that pairs enclose, and \emph{shrinking} shortens tokens.

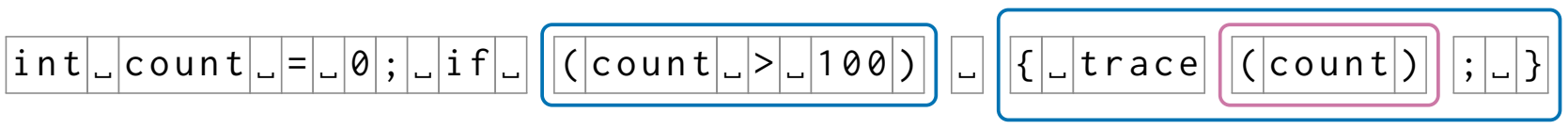
\begin{figure}[tp]
	\centering
	\footnotesize
	\begin{tikzpicture}[
		tok/.style   = {inner sep=0pt, outer sep=0pt},
		span1/.style = {draw=spanone, line width=0.7pt, rounded corners=2pt, inner xsep=2.25pt},
		span2/.style = {draw=spantwo, line width=0.7pt, rounded corners=2pt, inner xsep=2.25pt},
	]
		\node[tok] (t0) {\chunkunit{int}};
		\node[tok, base right=-0.30pt of t0] (t1) {\chunkunit{\vs}};
		\node[tok, base right=-0.30pt of t1] (t2) {\chunkunit{count}};
		\node[tok, base right=-0.30pt of t2] (t3) {\chunkunit{\vs}};
		\node[tok, base right=-0.30pt of t3] (t4) {\chunkunit{=}};
		\node[tok, base right=-0.30pt of t4] (t5) {\chunkunit{\vs}};
		\node[tok, base right=-0.30pt of t5] (t6) {\chunkunit{0}};
		\node[tok, base right=-0.30pt of t6] (t7) {\chunkunit{;}};
		\node[tok, base right=-0.30pt of t7] (t8) {\chunkunit{\vs}};
		\node[tok, base right=-0.30pt of t8] (t9) {\chunkunit{if}};
		\node[tok, base right=-0.30pt of t9] (t10) {\chunkunit{\vs}};
		\node[tok, base right=5.20pt of t10] (t11) {\chunkunit{(}};
		\node[tok, base right=-0.30pt of t11] (t12) {\chunkunit{count}};
		\node[tok, base right=-0.30pt of t12] (t13) {\chunkunit{\vs}};
		\node[tok, base right=-0.30pt of t13] (t14) {\chunkunit{>}};
		\node[tok, base right=-0.30pt of t14] (t15) {\chunkunit{\vs}};
		\node[tok, base right=-0.30pt of t15] (t16) {\chunkunit{100}};
		\node[tok, base right=-0.30pt of t16] (t17) {\chunkunit{)}};
		\node[tok, base right=5.20pt of t17] (t18) {\chunkunit{\vs}};
		\node[tok, base right=5.20pt of t18] (t19) {\chunkunit{\{}};
		\node[tok, base right=-0.30pt of t19] (t20) {\chunkunit{\vs}};
		\node[tok, base right=-0.30pt of t20] (t21) {\chunkunit{trace}};
		\node[tok, base right=5.20pt of t21] (t22) {\chunkunit{(}};
		\node[tok, base right=-0.30pt of t22] (t23) {\chunkunit{count}};
		\node[tok, base right=-0.30pt of t23] (t24) {\chunkunit{)}};
		\node[tok, base right=5.20pt of t24] (t25) {\chunkunit{;}};
		\node[tok, base right=-0.30pt of t25] (t26) {\chunkunit{\vs}};
		\node[tok, base right=-0.30pt of t26] (t27) {\chunkunit{\}}};
		\node[span1, inner ysep=2.25pt, fit=(t11) (t17)] {};
		\node[span1, inner ysep=5.20pt, fit=(t19) (t27)] {};
		\node[span2, inner ysep=2.25pt, fit=(t22) (t24)] {};
	\end{tikzpicture}
	\caption{The running example's tokens as boxes,
		and the spans of its parentheses and braces
		as frames colored by depth.
		The root holds the unframed tokens and the two outer spans.}
	\label{fig:tree}
\end{figure}

\noindent\textbf{Building the tree.} The tree (\Cref{fig:tree}) matches opening delimiters with closing ones. Some delimiters may have no match: failing inputs are often malformed, comments and strings contain delimiters, and pair learning may admit a wrong pair. We want each such \emph{stray} delimiter to change the number of spans by at most one. A stack that matches each close with the latest unmatched open lacks this property: in \<\{ /* (1 */ trace(count); \}>, the parenthesis in the comment is still open when the closing brace arrives, so the brace span is lost, and with it every span enclosing it. We therefore choose as spans a largest laminar set of matched pairs~\cite{korte2018combinatorial}, leaving unmatched delimiters, like that parenthesis, as ordinary tokens. Finding it is Dyck edit distance restricted to deletions~\cite{aho1972minimum,backurs2016parsing}. Inserting or deleting $k$ delimiters changes the number of spans by at most $k$, and admitting a further pair never reduces it. A stack first matches the unambiguous pairs, and a cubic interval dynamic program then matches the rest exactly (\Cref{app:matching}, \Cref{alg:tree}). With the parentheses and braces of the running example as pairs (\Cref{fig:tree}), the root holds the declaration's tokens, \<\kw{if}>, spaces, and two spans, \<{(count > 100)}> and \<\{ trace(count); \}>, which holds the span \<(count)>.

\noindent\textbf{Reducing a node.} At each node, a delta debugging reducer deletes tokens and child spans, keeping the node's own delimiters and everything outside it fixed. \emph{Laminar \drdd{}} uses \drdd{}. For example, deleting \<{\kw{if} (count > 100)\vs}> at the root leaves \good{\<{\kw{int} count = 0; \{ trace(count); \}}>}, which still crashes. Once a non-root node is reduced, two edits try to remove its delimiters while keeping some of its contents. The node is first \emph{hoisted}: its nearest descendant spans of the same pair are tried in its place, one query each, and the first one accepted replaces it, following Picireny's hoisting~\cite{vince2021hoisting}. A node that is not hoisted is then \emph{unwrapped} by deleting both its delimiters in one query. Unwrapping the braces of \good{\<\{ trace(count); \}>} results in \< trace(count); >.

\noindent\textbf{Traversal and guarantees.} Laminar reduction (\Cref{alg:laminar}) is HDD*~\cite{misherghi2006hdd} over the constructed tree: it reduces nodes level by level from the root, descends only into surviving spans, and repeats until a traversal changes nothing. It reduces each node separately, where HDD reduces a whole level at once.

For any tokens and pairs, laminar reduction guarantees that it (1)~preserves the failure, (2)~terminates after at most $|S|$ traversals that change the input, since each deletes a nonempty token, and (3)~ends with every node 1-minimal over its items and every remaining span's hoists and unwrapping rejected.

Each traversal runs \drdd{}'s closing scan once per node, so it makes at most three queries per item for the sweeps and the scan and two per span to hoist or unwrap it, and by (2) laminar reduction makes $O(|S|^2)$ queries, as \drdd{} does in its own worst case.

\noindent\textbf{Shrinking tokens.} After laminar reduction, shrinking (\Cref{alg:shrink}) runs one pass of \drdd{} that shortens each distinct multi-atom token by deleting from all of its occurrences at once, which turns \<{\kw{int} count=0;trace(count);}> into \good{\<{\kw{int} t=0;trace(t);}>}.

\noindent\textbf{1-minimality.} Like HDD*~\cite{misherghi2006hdd}, laminar reduction ends 1-minimal over nodes by (3). It is not 1-minimal over atoms, since it never deletes a lone delimiter or part of a token. Shrinking may also re-enable a deletion that laminar reduction rejected. Repeating \drdd{}'s closing scan over atoms until it deletes nothing makes any output 1-minimal over atoms, at one query or more per remaining atom. \autodidact{} does not run this scan, and \Cref{sec:rqValue} reports what it adds.

\section{Inferring the Structure}\label{sec:learning}

The oracle answers only yes or no for a whole input, so we read structure from how its verdict changes under edits of the failing input: insertions reveal tokens, and paired edits reveal delimiters.

\subsection{Oracle-Relative Structure}\label{sec:core}

Tokens and delimiter pairs follow one pattern of oracle verdicts: an edit that respects the structure preserves the failure, while \emph{controls}, minimal variants that break the structure, lose it. Only this contrast shows that a word separates or a pair matches: accepting both edits, as inside a comment, or rejecting both reveals nothing. Neither definition mentions a format or needs $S$ to be well formed.

\noindent\textbf{Tokens.} We seek a word that the failure treats as a separator: inserted between units, it preserves the failure, and inserted inside one, it does not. With such a word, one query per offset locates every boundary. Lexers often discard such a separator, like whitespace, between tokens. In the running example, inserting a space between \<trace> and \<(> gives \good{\<trace (count);>}, which still crashes. Inserting a byte absent from the input, such as \<@>, gives \bad{\<trace@(count);>}, which is rejected, as is \bad{\<tr ace(count);>}, with a space inside the identifier. Spaces at any number of token boundaries, in any amount, compile and crash. We take this contrast as the definition of neutral words.

Let $N=|S|$ for the failing input $S$, and number each internal offset $g$ by the count of atoms before it, so that $0<g<N$. For a word $u$, we write $S\oplus_g u = S[0:g]\cdot u\cdot S[g:N]$ for $S$ with $u$ inserted at $g$. Fix a control atom $\kappa$ that is as rare in $S$ as possible, so it is unlikely to be neutral. An offset $g$ is \emph{selective} for a word $w$ if the oracle accepts $S\oplus_g w$ but rejects both controls $S\oplus_g\kappa$ and $S\oplus_g\kappa^{|w|}$, the latter with $|w|$ copies of $\kappa$. Selectivity rules out regions the oracle ignores and regions where any text of the right length suffices. A nonempty word $w$ is \emph{neutral} for $S$ if, for some nonempty set $G$ of selective offsets, the oracle accepts $S$ with $c_g\ge0$ copies of $w$ inserted at each $g\in G$, for every choice of the counts $c_g$. Neutrality asks more than one accepted insertion: copies must be accepted in any amount and at all offsets of $G$ at once, as a separator's are. Given a neutral word $w$, the \emph{token boundaries} of $S$ are the offsets accepting a single insertion of it,
\begin{equation}\label{eq:boundaries}
	B_w(S)=\{\,g : 0<g<N,\ \test(S\oplus_g w)=\ACCEPT\,\},
\end{equation}
and these boundaries cut $S$ into its \emph{tokens}. Every selective offset is a token boundary, and offsets that the failure ignores, such as those inside a comment, are token boundaries without being selective. In the running example, the space is neutral, and the tokens are the C tokens and the spaces.

An input may have several neutral words, such as a space and a newline in C, and their boundaries can differ: inside a string literal, a space may be accepted where a newline is not. If the oracle sees its input only through a lexer that discards $w$ between tokens and rejects $\kappa$ there, every lexical boundary is a token boundary, and a lexical token is split exactly where the failure does not need it whole (\Cref{app:lexer}), as with the JavaScript test title of \Cref{sec:introduction}. Any separator such a lexer discards thus recovers the lexical boundaries, and finding one is a matter of inference (\Cref{sec:discovery}).

\noindent\textbf{Spans.} Parsers need brackets to match: adding or removing an open and close together can keep them matched, and adding or removing only one does not. In the running example, wrapping \<(count)> gives the accepted \good{\<trace((count));>}, and adding only one parenthesis, as in \bad{\<trace((count);>} or \bad{\<trace(count));>}, is rejected. Again we take this contrast as the definition. Because delimiters are tokens, a pair can be any two tokens that the failure needs matched, not only bracket characters: \<begin> and \<end>, a quotation mark with itself, or the \<{<name}> and \<{</name}> that open an XML start and end tag, whose names vary with the input and so escape any fixed list of brackets. The candidates thus narrow from every substring of $S$ to its distinct tokens.

Write $S=t_1\cdots t_n$ for its tokens. For tokens $\opn$ and $\cls$, a \emph{matching} of a token sequence $x$ pairs positions $i<j$ with $x_i=\opn$ and $x_j=\cls$, using each position at most once, so that any two pairs are nested or disjoint. Let $\nu(x)$ be the size of a largest matching and $\delta(x)$ the number of positions holding $\opn$ or $\cls$. Then $x$ is \emph{matched} if a largest matching leaves no delimiter out, that is,
\begin{equation}\label{eq:matched}
	x\text{ matched}
	\iff 2\,\nu(x)=\delta(x)
	\iff
	\begin{cases}
		|x|_{\opn}\text{ is even} & \text{if }\opn=\cls,\\
		|x|_{\opn}=|x|_{\cls}\ \wedge\ \forall y\sqsubseteq x:\ |y|_{\opn}\ge|y|_{\cls} & \text{otherwise,}
	\end{cases}
\end{equation}
where $|y|_t$ counts the occurrences of a token $t$ in $y$, and $y\sqsubseteq x$ ranges over the prefixes of $x$. The difference $\delta(x)-2\nu(x)$ counts stray delimiters, which the tree of \Cref{sec:reduction} minimizes over all pairs at once. The \emph{spans} of $(\opn,\cls)$ are the intervals $t_a\cdots t_b$ with $a<b$, $t_a=\opn$, $t_b=\cls$, and a matched interior $t_{a+1}\cdots t_{b-1}$, which for a symmetric delimiter holds none of it, since quotes do not nest. We write a span as $\opn m\cls$, with opening delimiter $\opn$, interior $m$, and closing delimiter $\cls$. Each row below replaces a span, leaving the rest of $S$ fixed, by a \emph{paired} edit, which adds or removes $\opn$ and $\cls$ together, and by two \emph{controls}, each off by one delimiter, an $\opn$ in one and a $\cls$ in the other:
\begin{equation}\label{eq:coupling}
	\begin{aligned}
		&\text{wrap:} &&\opn\,\opn m\cls\,\cls &&\text{against}\quad \opn\,\opn m\cls \;\text{ and }\; \opn m\cls\,\cls, \\
		&\text{empty or delete:} &&\opn\cls \text{ or the empty string} &&\text{against}\quad \cls \;\text{ and }\; \opn, \\
		&\text{repeat:} &&\opn m\cls\,\opn m\cls &&\text{against}\quad \opn m\cls\,\opn m \;\text{ and }\; \opn m\cls\,m\cls.
	\end{aligned}
\end{equation}
The span \emph{couples} if, in some row, the oracle accepts the paired edit and rejects both controls, telling balanced delimiters from unbalanced ones. The pair $(\opn,\cls)$ is a \emph{delimiter pair} of $S$ if at least two of its spans couple, since one coupled span could be a coincidence. Parentheses that the oracle ignores, as in a comment, are not a delimiter pair. Suppose the oracle, like a parser, rejects every unmatched input, and that $S$ is matched. Every paired edit then keeps $S$ matched, and every control does not, since one delimiter more or fewer makes $\delta$ odd, so every control is rejected. A parser's pair is thus a delimiter pair once two of its spans survive a paired edit (\Cref{app:parser}). For the same reason, deleting or unwrapping a span keeps the delimiters matched, and spans spare reduction the doomed deletion of one delimiter alone (\Cref{sec:minimization}). Nor must $S$ be matched: coupling is judged span by span, and the tree of \Cref{sec:reduction} leaves stray delimiters as ordinary tokens.

\noindent\textbf{What the learners admit.} The learners of \Cref{sec:discovery,sec:pairs} admit a word or pair on finite evidence for these definitions. Neutral-word discovery admits a word that is selective at observed offsets and passes a few \emph{reuse tests}, each inserting copies of it at some of those offsets. A neutral word passes these tests within its set $G$. Pair learning admits a pair once it observes coupling at two candidate intervals, pooling the spans of a \emph{family} of pairs that differ only in a name, such as XML tags. Two further checks guard admission: an entropy guard vetoes a word confirmed on weak evidence in a repetitive input, and a pair is tested only if duplicating either delimiter of a representative pair is rejected.

\noindent\textbf{What queries cannot show.} Both learners generalize from finitely many queries, which cannot rule out a different verdict on an untested edit or name. The definitions also leave four cases open. A neutral word need not separate anything: a digit is neutral when lengthening a number keeps the failure. Genuine delimiters yield no witness when every paired edit destroys the failure. Tokens that must appear together without enclosing anything, such as \<?> and \<:> in C, can also couple. If the control $\kappa$ is accepted at every sampled offset, none is selective, and discovery abstains.

\subsection{Neutral-Word Discovery}\label{sec:discovery}

Neutral-word discovery (\Cref{alg:discover}) proposes repeated substrings, observes where each is selective, and tests the most promising for reuse. It judges neutrality from a sample of offsets, and the evidence a sample provides depends only on the sample size. If a word rejects reuse at a fraction $p$ of the offsets where it is selective, $k$ such offsets drawn uniformly at random all accept it with probability at most $(1-p)^k$, however long $S$ is. The implementation therefore uses a fixed budget of 300 calls, with 16 reserved for confirmation. The bound assumes uniform sampling, and \textsc{Schedule} skews it toward varied contexts, so the bound only motivates the budget.

\begin{algorithm}[h]
	\caption{Discover a neutral word}
	\label{alg:discover}
	\small
	\begin{algorithmic}[1]
		\Require Failing input $S$, oracle $\test$
			\Ensure An admitted word, or $\bot$ to abstain
		\Statex \textit{Before any query would exceed 300 calls in total, return $\bot$.}
		\State $\mathit{sources}\gets$ first 40 words of $\Call{Candidates}{S}$, and $W\gets$ the first 24
		\State \textbf{if} $W=\varnothing$ \textbf{then} \textbf{return} $\bot$
		\State $\kappa\gets$ the lowest byte from 1 to 255 absent from $S$, or else the rarest byte of $S$, lowest on ties
		\State $D\gets\Call{Observe}{S,\test,W,\kappa,\Call{Schedule}{S,\mathit{sources}},284}$
		\State \textbf{if} $D=\varnothing$ \textbf{then} \textbf{return} $\bot$
		\State $\mathit{ranked}\gets$ words $u_i\in W$ with $s_i>0$ and $e_i\ge0$ \eqref{eq:discovery-score}, stably by decreasing $e_i$
		\State $\mathit{refuted}\gets\varnothing$
		\While{$\mathit{ranked}\ne\varnothing$}
			\State Remove from $\mathit{ranked}$ its first word $w$ whose root $\rho(w)\notin\mathit{refuted}$, or else its first word
			\State $A\gets\Call{Sites}{w,D}$, and $T\gets\Call{RefineSites}{w,A}$
			\State Test $\Call{ReuseTests}{w,T}$ in order, stopping at the first rejection
			\State \textbf{if} all reuse tests accepted \textbf{then return} $\bot$ if $|A|<3$ and $H(S)<2/3$, else $w$
			\State \textbf{if} a reuse test rejected \textbf{then} $\mathit{refuted}\gets\mathit{refuted}\cup\{\rho(w)\}$
		\EndWhile
		\State \Return $\bot$
	\end{algorithmic}
\end{algorithm}

\noindent\textbf{Candidates.} A separator tends to recur beside varied tokens. In the running example, the space occurs 10 times, after 7 distinct atoms and before 10. \textsc{Candidates} therefore proposes \emph{maximal repeats}: substrings that occur at least twice, with at least two distinct atoms before them and two distinct atoms after them. Non-separators such as \<(count> also qualify, but \<ount> does not, since it always follows \<c>. A candidate with $f$ occurrences, $d_\ell$ distinct atoms before, and $d_r$ after ranks by $\min(d_\ell,d_r)\sqrt{f}$, with shorter words first on ties. The space thus ranks first at $7\sqrt{10}\approx22$, well ahead of \<0> at $3\sqrt{3}\approx5.2$. The top 40 candidates steer where \textsc{Schedule} samples offsets, and the top 24 of them form the tested words $W$. An offset where $\kappa$ is rejected costs one query per tested word, so with 24 words the observation budget covers about ten such offsets. The other 16 only steer sampling, at no cost. Both limits are hand-picked. The \emph{root} $\rho(u)$ is the shortest word whose repetitions form $u$, and words with the same root form a \emph{family}, such as one and two spaces.

\noindent\textbf{Assumptions.} Discovery finds a neutral word only if it occurs in $S$ as a maximal repeat among the top 24 candidates. It cannot learn a word that $S$ does not repeat, such as a space in minified code, and then abstains, so reduction proceeds over bytes. Formats without a recurring separator, such as most binary formats, account for all \abstained{} abstentions in our benchmark (\Cref{sec:rqRecovery}).

\noindent\textbf{Observation and scoring.} At each offset that \textsc{Schedule} picks, \textsc{Observe} first inserts the control $\kappa$. If $\kappa$ is accepted, the offset reveals nothing. Otherwise, it inserts each word of $W$ and credits the words that are selective there. A word's score $s_i$ averages its credit, weighting offsets that credit few families more, since an offset that credits many says little about any one. Its adjusted score $e_i$ subtracts a rank penalty, so later candidates need more evidence. Scores filter and order candidates for confirmation.

\noindent\textbf{Confirmation.} Selectivity shows only that one copy is accepted at one offset, and a neutral word must also preserve the failure with any number of copies at the offsets of $G$ together. In the running example, adding three more spaces after \<\kw{int}> gives \good{\<\kw{int}\vs\vs\vs\vs{}count = 0;>}, which still crashes. Such a query is a reuse test, inserting copies of the word at one site or at several together. In score order, discovery confirms the first word that passes its reuse tests. Each word gets at most five tests, at up to four accepted offsets, favoring distinct contexts (\Cref{app:discovery}). A rejection refutes reuse at those sites, and discovery tries other families first. Passing cannot rule out non-neutral words or interactions at untested sites. On the running example, the space ranks first and is confirmed in 5 calls after 284 observation calls.

\noindent\textbf{Abstention.} Neutral-word discovery abstains when no word passes confirmation within its 300 calls. It also abstains on weak evidence in repetitive inputs, where a word may pass only by lengthening padding: if fewer than three selective sites support the confirmed word and $H(S)<2/3$. Here $H(S)$ is the entropy of the atoms of $S$, normalized by its maximum for their number of distinct values, and zero when fewer than two distinct atoms occur.

\subsection{Tokenization}\label{sec:prediction}

Given the neutral word $w$, tokenization finds the token boundaries $B_w(S)$ \eqref{eq:boundaries}. Probing each internal offset $g$ alone, by querying $S\oplus_g w$, finds them exactly, at one query per offset. The default tokenizer (\Cref{alg:tokenize}) saves queries in two ways: it predicts verdicts from those it has already settled on the same input, and it tests several likely boundaries with one query.

\begin{algorithm}[htbp]
	\caption{Tokenize with boundary prediction and pooling}
	\label{alg:tokenize}
	\small
	\begin{algorithmic}[1]
		\Require Failing input $S$, neutral word $w$, oracle $\test$
		\Ensure The tokens of $S$
		\Statex \textit{Every verdict that the oracle answers or implies trains \textsc{Estimate}.}
		\State $V\gets$ empty verdict map, and $R\gets$ empty pool
		\For{$g=1,\ldots,N-1$}
			\State $\hat p_g\gets\Call{Estimate}{g,V}$
			\If{$\hat p_g\ge0.98$ \textbf{or} $\hat p_g\le0.02$}
				\State $V[g]\gets\ACCEPT$ if $\hat p_g\ge0.98$, else $\REJECT$ \Comment{predict}
			\ElsIf{$\hat p_g<1/2$}
				\State $\Call{TestPool}{R}$, then $V[g]\gets\test(S\oplus_g w)$ \Comment{probe alone}
			\Else
					\State Add $g$ to $R$, and $V[g]\gets\ACCEPT$ \Comment{tentative boundary}
				\State \textbf{if} $\prod_{h\in R}\hat p_h\le1/2$ \textbf{then} $\Call{TestPool}{R}$
			\EndIf
		\EndFor
		\State $\Call{TestPool}{R}$
		\State \Return the substrings of $S$ between consecutive offsets of $\{0,N\}\cup\{g:V[g]=\ACCEPT\}$
	\end{algorithmic}
\end{algorithm}

\noindent\textbf{Prediction.} Inputs repeat local patterns, and so do verdicts. In the running example, the offsets inside \<count> have the same atoms around them at all three occurrences, as does the offset between \<(> and \<count> at two. \textsc{Estimate} predicts an offset's verdict from the settled verdicts of offsets that resemble it in the atoms around them, the verdicts next to them, or the share of boundaries nearby. Each kind of resemblance is a \emph{context}. \textsc{Estimate} combines the boundary rates of an offset's 14 contexts into $\hat p$, the estimated chance that the offset is a boundary. It trusts a context more when more settled offsets share it, and learns on each input which contexts predict well (\Cref{app:tokenization}). The tokenizer predicts the verdict at fixed cutoffs $\hat p\ge0.98$ or $\hat p\le0.02$. The contexts and cutoffs are hand-picked and shared across formats.

\noindent\textbf{Pooling.} A neutral word can appear at many offsets at once, so the tokenizer tests likely boundaries together: one query inserts $w$ at every offset of a \emph{pool}. In the running example, \good{\<trace\vs(\vs{}count\vs)\vs;>} still crashes, settling the four offsets between the tokens of \<trace(count);> in one query. A rejected pool holds at least one non-boundary, which binary splitting finds (\Cref{app:tokenization}), as in adaptive group testing~\cite{dorfman1943detection}. Pooling assumes, beyond what neutrality guarantees, that a pool is accepted exactly when each of its insertions alone would be. Where this fails, a rejected pool of boundaries yields a false non-boundary, merging two tokens.

\noindent\textbf{What to query.} A query tells the most when its outcome is a coin flip and the least when it is nearly certain (\Cref{app:tokenization}). The tokenizer therefore predicts nearly certain verdicts, and probes a likely non-boundary, with $\hat p<1/2$, alone, since a pool containing it would probably be rejected. It adds each likely boundary to the pool instead, and tests the pool once the estimated chance that all its offsets are boundaries, $\prod\hat p$, falls to one half.

\noindent\textbf{What prediction affects.} The thresholds give no error guarantee. A wrong prediction can only merge tokens, hiding a delimiter, or split a token that exhaustive probing keeps whole, changing the inferred pairs, output size, and oracle cost. The resulting tokens are then frozen for pair learning and reduction. \Cref{sec:rqRecovery} measures them against exhaustive probing and against real lexers.

\subsection{Pair Learning}\label{sec:pairs}

\Cref{sec:core} showed that \<(count)> couples. In the running example, so does \<{(count > 100)}>, the other span of the parentheses. Pair learning (\Cref{alg:pairs}) finds such pairs on its own. It proposes \emph{hypotheses}, each a single pair or a family of related pairs, such as XML tags, together with \emph{candidate intervals} where they might delimit spans. It ranks hypotheses without queries, skips unpromising ones, and admits a hypothesis once two of its candidate intervals couple, as the parentheses do.

\begin{algorithm}[htbp]
	\caption{Learn delimiter pairs}
	\label{alg:pairs}
	\small
	\begin{algorithmic}[1]
		\Require Input $S=t_1\cdots t_n$ with frozen tokens, word $w$, oracle $\test$
		\Ensure Inferred delimiter pairs $P$
		\State $P\gets\varnothing$, and $Z\gets\varnothing$ \Comment{admitted intervals}
		\For{$\mathit{kind}\in(\text{stacked},\text{symmetric})$}
			\State $Q\gets\Call{Hypotheses}{S,w,\mathit{kind},Z}$, and $\mathit{streak}\gets0$
			\For{each $(H,U,p)\in Q$}
				\State \textbf{if} $\mathit{streak}=4$ \textbf{then break}
				\State $A\gets\{(a,b)\in U:a,b\text{ lie in the same region of }Z\}$, and $\pi\gets2+\lfloor p\rfloor$
				\State \textbf{if} $H$ reuses a token of $P$, $|U|-|A|>\pi$, no interval in $U$ encloses a non-blank token, or not $\Call{Sensitive}{H}$ \textbf{then continue} \Comment{screens}
				\State \textbf{if} $\Call{Test}{A,\pi}$ \textbf{then} $P\gets P\cup H$, $Z\gets Z\cup A$, and $\mathit{streak}\gets0$ \textbf{else} $\mathit{streak}\gets\mathit{streak}+1$
			\EndFor
		\EndFor
		\State \Return $P$
	\end{algorithmic}
\end{algorithm}

\noindent\textbf{Hypotheses.} Delimiters are \emph{non-blank} tokens, those not made only of copies of $w$. A hypothesis is \emph{stacked} if its opening and closing delimiters differ, so that a stack can match them like brackets, and \emph{symmetric} if a token pairs with itself, like a quotation mark. Every two distinct tokens that each occur at least twice form a stacked single pair, in whichever orientation balances better, and every token occurring at least four times forms a symmetric one. An \emph{affix family} is a set of pairs whose delimiters share a name $\nu$ and differ only in a fixed prefix or suffix, such as \<<>$\nu$ and \<</>$\nu$ for XML tags. It pools evidence from names too rare to propose alone. Formally, it groups the pairs $(a\nu,b\nu)$ or $(\nu a,\nu b)$ over at least two names $\nu$, in either orientation, for any distinct nonempty affixes $a,b$. \textsc{Hypotheses} ranks each kind and keeps its $K=256$ best, a hand-picked cap that binds on most inputs. The stacked list is tested first, and the symmetric list is built from the intervals admitted so far.

\noindent\textbf{Assumptions.} Pair learning can admit a pair only if tokenization kept its delimiters as tokens, the pair is proposed as above, and its hypothesis ranks among the top of its kind. A pair occurring only once can be learned only through an affix family. In the running example, the braces occur once each, so pair learning never proposes them.

\noindent\textbf{Candidate intervals.} Each hypothesis proposes intervals $U$ for coupling tests. For a stacked hypothesis, a stack matches each closing delimiter with the opening delimiter on top of the stack if it has the same name, and each match is a candidate interval, such as \<{(count > 100)}> and \<(count)> in the running example. Admitted intervals divide the input into \emph{regions}, each directly inside one interval. Within each region, a symmetric hypothesis pairs its occurrences in order: the first with the second, the third with the fourth, and so on (\Cref{app:pairs}). In \<f(\sq{}a\sq{}); g("it\sq{}s"); h(\sq{}b\sq{});>, the apostrophe in \<it\sq{}s> is alone inside \<g>'s admitted parentheses, so it pairs with nothing, while over the input it would pair with the quote opening \<\sq{}b\sq{}>. Candidate intervals need not be spans: family matching can leave stray delimiters inside one, and in-order pairing ignores nested regions.

\noindent\textbf{Ranking.} In the running example, the two \<(> and two \<)> occur in balanced order, which only two of their six orders achieve. Delimiters that the failure needs matched are balanced like this far more often than chance. Each hypothesis therefore gets a \emph{prior} $p$, the \emph{surprise} of its balance, meaning the negative log probability that a random shuffle of its occurrences leaves no more closing delimiters unmatched than the input does. For the parentheses, $p=\log3$. A family adds the surprise of its names, and a symmetric hypothesis counts the regions holding an odd number of its occurrences (\Cref{app:pairs}). Every prior is capped at $\log(1+|U|)$, and only hypotheses with at least two candidate intervals are eligible. Stacked hypotheses rank by $p$, and symmetric ones by $p+\log(1+|U|)$, which favors tokens with more sites to test. The prior only orders hypotheses and sets each one's testing allowance $\pi=2+\lfloor p\rfloor$, so that more surprising hypotheses get more tries.

\noindent\textbf{Screens.} Before spending coupling queries, \Cref{alg:pairs} \emph{screens} out any hypothesis that (1)~reuses an admitted delimiter, (2)~has more than $\pi$ intervals that cross admitted ones, or (3)~has no interval enclosing a non-blank token. These screens cost no queries. Screen (3) can discard a useful pair, since even an empty span may be worth unwrapping. A survivor must also be \emph{sensitive}: duplicating either delimiter of a representative pair must be rejected (\Cref{app:pairs}).

\noindent\textbf{Confirmation.} \textsc{Test} seeks two \emph{witnesses}, intervals that couple, among the intervals that cross no admitted one. It tries intervals of varied contexts and lengths, giving up after $\pi$ intervals without a witness, or $2\pi$ in all (\Cref{alg:pair-test}). On each interval, \textsc{Coupled} tries the rows of \eqref{eq:coupling} in order and succeeds at the first that couples, since a genuine span can fail some rows. Four consecutive failed hypotheses stop the current kind, and screened hypotheses do not count toward that streak.

Once a hypothesis is admitted, its pairs join $P$, and its intervals that cross no admitted one are admitted too, tested or not. A family's two witnesses may come from one name, so its other names are admitted by extrapolation, and the newly admitted intervals can suppress later hypotheses.

\subsection{Query Bounds}\label{sec:bounds}

Every stage's query count is bounded in the input size $N$. Neutral-word discovery makes at most 300 queries. Exhaustive tokenization makes $N-1$, and prediction and pooling usually make fewer, with $O(N)$ queries even in the worst case at the fixed thresholds (\Cref{app:tokenization}). Pair learning tests at most $K$ hypotheses of each kind, each with at most $2+20\pi$ queries, and $\pi\le2+\ln(1+N)$, so it makes $O(K\log N)$ queries (\Cref{app:pairs}). Laminar reduction makes $O(N^2)$ (\Cref{sec:reduction}), and shrinking, one \drdd{} pass per distinct token, makes at most quadratically many in the tokens' total length. The whole pipeline thus makes $O(N^2)$ queries in the worst case, the order of \drdd{} itself. Candidate generation, ranking, prediction, and tree construction make none.

\section{Evaluation}\label{sec:evaluation}

We evaluate \autodidact{} on \inputs{} failing inputs from \formats{} format families against byte-level \drdd{}, HDD, and C-Reduce. The most important question for a new reducer is whether it improves upon the baseline. Hence, we ask:

\noindent\textbf{RQ1: How does \autodidact{} compare with byte-level \drdd{} in output size and calls?}

With inferred structure, \autodidact{} can tackle structured inputs. But how good is that structure? Is it competitive with reducers that have format knowledge? Hence, we ask:

\noindent\textbf{RQ2: How does \autodidact{} compare with the format-aware HDD and C-Reduce?}

Finally, \autodidact{} rests on structure learned from the oracle alone, so we ask what that inference recovers and what it costs.

\noindent\textbf{RQ3: What structure does inference recover from the failing input and its oracle alone, and at what cost?} An answer tells us whether a failure oracle can stand in for the lexer and grammar that structural reducers need.

\subsection{Inputs and Oracles}

To cover a range of formats, we evaluate on \inputs{} failing inputs from \formats{} format families (\Cref{tab:benchmark}): \textInputs{} text-based and \binaryInputs{} binary, of \smallestInput{} to \largestInput{} bytes. Of these, \reproducers{} reproduce failures, and \controls{} are controls that must preserve a program's behavior.

\begin{table}[tp]
	\caption{Benchmark families and neutral-word discovery outcomes}
	\label{tab:benchmark}
	\tablestyle
	\begin{minipage}[t]{\halftable}
		\begin{tabularx}{\linewidth}{|l|N|r@{\,--\,}l|L|N|}
	\hline
	\textbf{Family} & \textbf{\#} & \multicolumn{2}{c|}{\textbf{Size}} & \textbf{Oracle} & \textbf{Word} \\
	\hline
	\subtablegap
	\multicolumn{6}{l}{\textit{Text-based inputs}} \\
	\hline
	c & 10 & 3.0 & 121.7\,KiB & ICE, control & \textvisiblespace{} \\
	crashjs & 10 & 644\,B & 1.3\,KiB & API exception & \textvisiblespace{} \\
	graphviz & 2 & 2.1 & 161.7\,KiB & sanitizer & \textvisiblespace{} \\
	lua & 2 & 889\,B & 1.2\,KiB & assertion & \textvisiblespace{} \\
	pdf & 5 & 4.4 & 15.2\,KiB & Magma fault & \textvisiblespace{} \\
	python & 5 & 212\,B & 1.1\,KiB & interpreter failure & \textvisiblespace{} \\
	smt & 10 & 2.4 & 218.9\,KiB & assertion & \textvisiblespace{} \\
	sql & 10 & 51 & 538\,B & error, wrong result & \textvisiblespace{} \\
	xml & 10 & 3.7 & 107.0\,KiB & differential & \textvisiblespace{} \\
	yaml & 10 & 16 & 177\,B & parse control & \textvisiblespace{} \\
	\hline
	\textbf{All} & \textbf{74} & \multicolumn{2}{c|}{} &  &  \\
	\hline
\end{tabularx}

	\end{minipage}\hfill
	\begin{minipage}[t]{\halftable}
		\begin{tabularx}{\linewidth}{|l|N|r@{\,--\,}l|L|N|}
	\hline
	\textbf{Family} & \textbf{\#} & \multicolumn{2}{c|}{\textbf{Size}} & \textbf{Oracle} & \textbf{Word} \\
	\hline
	\subtablegap
	\multicolumn{6}{l}{\textit{Binary inputs}} \\
	\hline
	asn1 & 6 & 500\,B & 1.2\,KiB & Magma fault & 10\,B \\
	binutils & 10 & 683\,B & 21.0\,KiB & signal, allocator & -- \\
	ffmpeg & 10 & 468\,B & 150.7\,KiB & sanitizer & -- \\
	fonts & 8 & 200\,B & 65.0\,KiB & sanitizer & -- \\
	parquet & 4 & 66.8 & 110.3\,KiB & sanitizer & -- \\
	pcap & 3 & 73 & 205\,B & sanitizer & -- \\
	png & 2 & 218 & 708\,B & Magma fault & -- \\
	tiff & 8 & 409\,B & 9.4\,KiB & Magma fault & -- \\
	 &  & \multicolumn{2}{c|}{} &  &  \\
	 &  & \multicolumn{2}{c|}{} &  &  \\
	\hline
	\textbf{All} & \textbf{51} & \multicolumn{2}{c|}{} &  &  \\
	\hline
\end{tabularx}

	\end{minipage}
	\tablenote{\# counts a family's inputs,
		and the benchmark has \inputs{} in all.
		\emph{ICE} denotes an internal compiler error,
		and \emph{control} denotes behavior preservation.
		In the word column, \textvisiblespace{} is a space,
		10\,B is the sequence returned on one ASN.1 input,
		and -- denotes abstention.}
\end{table}

The inputs come from existing collections and bug reports. C, JavaScript, Python, and SMT-LIB inputs come from RedBench~\cite{kreutzer2021redbench}, CrashJS~\cite{oliver2024crashjs}, GTR~\cite{herfert2017gtr}, and ddSMT~\cite{kremer2021ddsmtartifact}. Binutils, FFmpeg, font, Parquet, and libpcap inputs come from upstream reports and the OSS-Fuzz~\cite{serebryany2017ossfuzz} and FuzzBench~\cite{metzman2021fuzzbench} archives, Graphviz and LuaJIT from upstream reports, SQL from upstream reports and SQLFlex~\cite{an2026sqlflexartifact}, XML from~\cite{zhang2025ddartifact}, and ASN.1, PDF, PNG, and TIFF from Magma's proofs of concept~\cite{hazimeh2020magma}. We keep at most ten inputs per family, spread across failure groups and sizes, stratifying JavaScript by project and C compiler failures by crashing source file, after dropping four near-duplicates: inputs of one failure group whose normalized compression distance~\cite{cilibrasi2005clustering} is at most 0.12.

Each oracle fixes its target versions and accepts a candidate that reproduces the recorded outcome. For most families, this is a signal, sanitizer report, assertion, diagnostic, or stack signature. For ASN.1, PDF, PNG, and TIFF, it is a recorded Magma~\cite{hazimeh2020magma} fault, and for SQL, an internal error or wrong result. For XML, BaseX must disagree with Saxon on a fixed query at affected revisions and agree at fixed ones. The controls must compile two Csmith~\cite{yang2011csmith} programs with sanitizers, run them, and keep their checksums, or keep the libyaml event streams of ten YAML test-suite inputs~\cite{yaml2022testsuite}. All reducers share each input's oracle, which rejects timeouts. Inference sees only the oracle's verdicts on raw bytes. Every original input passes its oracle before a run, and every output after.

\subsection{Comparators}

We compare \autodidact{}, with one set of parameters for every family, against three reducers: byte-level \drdd{}, HDD, and C-Reduce. Where discovery abstains, it reduces exactly as \drdd{} does, so RQ1 and RQ2 use the \worded{} inputs with a returned word, from \wordedGroups{} failure groups.

\noindent\textbf{HDD.} We run a pinned Picireny~\cite{picireny,hodovan2016modernizing} to a fixpoint. It implements HDD* with \ddmin{} at each level, caching, pruning, and hoisting~\cite{vince2021hoisting}, over parse trees from grammars-v4 grammars~\cite{grammarsv4} under ANTLR 4.13.2. A grammar exists for \hddInputs{} inputs from \hddFamilies{} families: C, JavaScript, Graphviz, Lua, Python, XML, and the SQLite and SMT-LIB~2 inputs of SQL and SMT. The C grammar alone is \grammarBytes{} bytes. We keep C's and Python's hidden tokens, which preserve directives and indentation, and patch Python's synthetic token text so that unparsing preserves the source.

\noindent\textbf{C-Reduce.} We run a pinned C-Reduce~\cite{regehr2012test} as \texttt{creduce -{}-not-c -{}-n 1 -{}-tidy} on all \creduceInputs{} returned-word inputs, the C inputs included, so that every format gets the same passes, and let it run to its own fixpoint. This mode drops the Clang-based passes and keeps generic ones over lines, C tokens, and a fixed list of bracket pairs, plus integer reduction and peephole edits.

\subsection{Measurement}

\noindent\textbf{Output size.} We count non-whitespace bytes, since HDD spaces the tokens it unparses and C-Reduce reformats. For \autodidact{} against \drdd{}, both of which only delete, this changes the mean size ratio by under one percent.

\noindent\textbf{Calls.} We count oracle executions. Every reducer's oracle caches recent verdicts, and the counts exclude rechecks of inputs whose verdict is known: the checks of original and final inputs, C-Reduce's initial sanity check, and Picire's assertion at the start of each \ddmin{} round.

\noindent\textbf{Wall time.} Measured on one core of an AMD Ryzen AI 9 HX PRO 370 under Fedora~44, it covers inference or reduction, oracle execution, and each stage's startup and shutdown. HDD's includes generating its parser. Since setup costs differ across tools, comparisons emphasize calls.

\section{Results}\label{sec:results}

We answer the three questions in order, charging inference in full throughout. \Cref{tab:comparators} summarizes the comparisons of RQ1 and RQ2.

\begin{table}[tp]
	\caption{Reduction results for the \worded{} inputs with a returned word}
	\label{tab:results}
	\tablestyle
	{\tablemedium\renewcommand{\arraystretch}{0.85}\begin{tabularx}{\linewidth}{|L|N|N|N|N|N|N|N|N|N|N|N|N|N|}
	\hline
	\multirow{2}{*}{\textbf{Case}} & \multirow{2}{*}{\textbf{Input}} & \multicolumn{3}{c|}{\textbf{\autodidact{}}} & \multicolumn{3}{c|}{\textbf{\drdd{}}} & \multicolumn{3}{c|}{\textbf{HDD}} & \multicolumn{3}{c|}{\textbf{C-Reduce}} \\
	 & & \emph{Output} & \emph{Calls} & \emph{Time} & \emph{Output} & \emph{Calls} & \emph{Time} & \emph{Output} & \emph{Calls} & \emph{Time} & \emph{Output} & \emph{Calls} & \emph{Time} \\
	\hline
	\subtablegap
	\multicolumn{14}{l}{\textit{asn1}} \\
	\hline
	\case{asn1-aah055-nested} & 580 & \cellcolor[rgb]{0.999,0.987,0.852}112 & \cellcolor[rgb]{0.990,0.962,0.817}623 & 33\,s & \cellcolor[rgb]{0.990,0.955,0.809}164 & \cellcolor[rgb]{0.999,0.987,0.852}455 & 17\,s & -- & -- & -- & \cellcolor[rgb]{1.000,0.786,0.682}579 & \cellcolor[rgb]{0.998,0.743,0.665}3,280 & 4\,m \\
	\hline
	\subtablegap
	\multicolumn{14}{l}{\textit{c}} \\
	\hline
	\case{02} & 3,623 & \cellcolor[rgb]{0.993,0.978,0.837}457 & \cellcolor[rgb]{0.994,0.979,0.838}3,032 & 7\,m & \cellcolor[rgb]{0.993,0.914,0.768}824 & \cellcolor[rgb]{0.999,0.987,0.852}2,679 & 4\,m & \cellcolor[rgb]{0.992,0.974,0.831}481 & \cellcolor[rgb]{0.992,0.921,0.775}5,219 & 24\,m & \cellcolor[rgb]{0.999,0.987,0.852}399 & \cellcolor[rgb]{1.000,0.786,0.682}13,856 & 23\,m \\
	\case{10} & 3,071 & \cellcolor[rgb]{0.991,0.971,0.827}281 & \cellcolor[rgb]{0.990,0.960,0.814}2,126 & 4\,m & \cellcolor[rgb]{0.995,0.896,0.753}528 & \cellcolor[rgb]{0.992,0.972,0.829}1,859 & 3\,m & \cellcolor[rgb]{0.992,0.972,0.828}277 & \cellcolor[rgb]{0.999,0.987,0.852}1,515 & 3\,m & \cellcolor[rgb]{0.999,0.987,0.852}224 & \cellcolor[rgb]{0.998,0.869,0.732}4,319 & 7\,m \\
	\case{tree-ssa-structalias\_c\_4773\_96K} & 97,658 & \cellcolor[rgb]{0.993,0.978,0.837}151 & \cellcolor[rgb]{0.990,0.946,0.799}3,023 & 3\,m & \cellcolor[rgb]{0.649,0.502,0.727}45,061 & \cellcolor[rgb]{0.764,0.533,0.735}232,803 & 2.0\,h & \cellcolor[rgb]{0.990,0.961,0.816}182 & \cellcolor[rgb]{0.999,0.987,0.852}1,902 & 3\,m & \cellcolor[rgb]{0.999,0.987,0.852}132 & \cellcolor[rgb]{1.000,0.812,0.695}8,174 & 3\,m \\
	\case{tree-vect-transform\_c\_6496\_55K} & 55,923 & \cellcolor[rgb]{0.993,0.977,0.835}114 & \cellcolor[rgb]{0.999,0.987,0.852}5,658 & 4\,m & \cellcolor[rgb]{0.747,0.527,0.738}13,940 & \cellcolor[rgb]{0.925,0.615,0.671}146,979 & 1.6\,h & \cellcolor[rgb]{0.994,0.979,0.838}111 & \cellcolor[rgb]{0.991,0.971,0.828}7,046 & 11\,m & \cellcolor[rgb]{0.999,0.987,0.852}98 & \cellcolor[rgb]{0.996,0.984,0.846}5,993 & 3\,m \\
	\case{fold-const\_c\_8943\_23K} & 23,171 & \cellcolor[rgb]{0.999,0.987,0.852}101 & \cellcolor[rgb]{0.993,0.913,0.767}2,447 & 74\,s & \cellcolor[rgb]{0.792,0.542,0.728}9,620 & \cellcolor[rgb]{0.895,0.592,0.685}42,493 & 12\,m & \cellcolor[rgb]{0.990,0.952,0.805}152 & \cellcolor[rgb]{0.999,0.987,0.852}1,178 & 2\,m & \cellcolor[rgb]{0.991,0.971,0.827}126 & \cellcolor[rgb]{1.000,0.793,0.685}5,791 & 2\,m \\
	\case{recog\_c\_1990\_63K} & 63,509 & \cellcolor[rgb]{0.993,0.910,0.765}138 & \cellcolor[rgb]{1.000,0.792,0.685}2,533 & 2\,m & \cellcolor[rgb]{0.610,0.494,0.706}31,646 & \cellcolor[rgb]{0.661,0.505,0.731}156,700 & 61\,m & \cellcolor[rgb]{0.990,0.942,0.795}107 & \cellcolor[rgb]{0.999,0.987,0.852}513 & 78\,s & \cellcolor[rgb]{0.999,0.987,0.852}65 & \cellcolor[rgb]{1.000,0.779,0.679}2,795 & 70\,s \\
	\case{simplify-rtx\_c\_4922\_43K} & 43,720 & \cellcolor[rgb]{0.991,0.933,0.786}135 & \cellcolor[rgb]{0.997,0.876,0.737}3,029 & 91\,s & \cellcolor[rgb]{0.694,0.513,0.739}17,308 & \cellcolor[rgb]{0.844,0.564,0.708}66,093 & 25\,m & \cellcolor[rgb]{0.992,0.923,0.776}146 & \cellcolor[rgb]{0.999,0.987,0.852}1,116 & 2\,m & \cellcolor[rgb]{0.999,0.987,0.852}76 & \cellcolor[rgb]{1.000,0.818,0.698}4,586 & 2\,m \\
	\case{gimple\_c\_2343\_122K} & 124,593 & \cellcolor[rgb]{0.990,0.948,0.801}142 & \cellcolor[rgb]{0.999,0.987,0.852}10,302 & 13\,m & \cellcolor[rgb]{0.649,0.502,0.727}31,151 & \cellcolor[rgb]{0.920,0.610,0.673}284,208 & 3.2\,h & \cellcolor[rgb]{0.993,0.909,0.764}194 & \cellcolor[rgb]{0.990,0.960,0.815}14,342 & 42\,m & \cellcolor[rgb]{0.999,0.987,0.852}91 & \cellcolor[rgb]{0.992,0.973,0.830}12,537 & 8\,m \\
	\case{emit-rtl\_c\_1950\_75K} & 76,263 & \cellcolor[rgb]{0.993,0.978,0.837}155 & \cellcolor[rgb]{0.991,0.970,0.826}2,690 & 3\,m & \cellcolor[rgb]{0.678,0.509,0.736}35,434 & \cellcolor[rgb]{0.800,0.545,0.725}188,661 & 1.7\,h & \cellcolor[rgb]{0.990,0.962,0.817}185 & \cellcolor[rgb]{0.999,0.987,0.852}2,134 & 6\,m & \cellcolor[rgb]{0.999,0.987,0.852}135 & \cellcolor[rgb]{0.999,0.860,0.725}6,507 & 3\,m \\
	\case{ira-emit\_c\_371\_17K} & 16,424 & \cellcolor[rgb]{0.999,0.987,0.852}199 & \cellcolor[rgb]{0.999,0.987,0.852}3,110 & 77\,s & \cellcolor[rgb]{0.873,0.579,0.695}8,914 & \cellcolor[rgb]{0.978,0.681,0.655}38,223 & 11\,m & \cellcolor[rgb]{0.990,0.947,0.799}315 & \cellcolor[rgb]{0.994,0.980,0.840}3,467 & 8\,m & \cellcolor[rgb]{0.993,0.978,0.837}229 & \cellcolor[rgb]{1.000,0.792,0.685}15,364 & 6\,m \\
	\hline
	\subtablegap
	\multicolumn{14}{l}{\textit{crashjs}} \\
	\hline
	\case{7} & 1,012 & \cellcolor[rgb]{0.991,0.970,0.826}159 & \cellcolor[rgb]{0.999,0.745,0.666}1,782 & 5\,m & \cellcolor[rgb]{0.997,0.880,0.740}333 & \cellcolor[rgb]{0.996,0.733,0.662}1,966 & 6\,m & \cellcolor[rgb]{0.999,0.987,0.852}126 & \cellcolor[rgb]{0.999,0.987,0.852}252 & 2\,m & \cellcolor[rgb]{0.998,0.987,0.851}127 & \cellcolor[rgb]{0.996,0.730,0.661}2,016 & 6\,m \\
	\case{10} & 830 & \cellcolor[rgb]{0.999,0.987,0.852}113 & \cellcolor[rgb]{1.000,0.823,0.701}1,441 & 5\,m & \cellcolor[rgb]{0.999,0.865,0.729}332 & \cellcolor[rgb]{1.000,0.762,0.672}2,246 & 7\,m & \cellcolor[rgb]{0.990,0.965,0.819}151 & \cellcolor[rgb]{0.999,0.987,0.852}363 & 2\,m & \cellcolor[rgb]{0.992,0.975,0.833}134 & \cellcolor[rgb]{1.000,0.815,0.697}1,524 & 5\,m \\
	\case{commander-4} & 778 & \cellcolor[rgb]{0.999,0.987,0.852}166 & \cellcolor[rgb]{0.991,0.925,0.778}1,434 & 4\,m & \cellcolor[rgb]{0.991,0.925,0.778}313 & \cellcolor[rgb]{0.990,0.960,0.814}1,062 & 3\,m & \cellcolor[rgb]{0.990,0.953,0.806}249 & \cellcolor[rgb]{0.999,0.987,0.852}757 & 4\,m & \cellcolor[rgb]{0.991,0.971,0.827}208 & \cellcolor[rgb]{0.999,0.750,0.667}5,163 & 15\,m \\
	\case{commander-31} & 1,055 & \cellcolor[rgb]{0.999,0.987,0.852}156 & \cellcolor[rgb]{1.000,0.841,0.713}1,371 & 4\,m & \cellcolor[rgb]{0.991,0.924,0.777}297 & \cellcolor[rgb]{0.999,0.862,0.727}1,185 & 3\,m & \cellcolor[rgb]{0.990,0.964,0.819}209 & \cellcolor[rgb]{0.999,0.987,0.852}394 & 2\,m & \cellcolor[rgb]{0.993,0.976,0.834}184 & \cellcolor[rgb]{0.993,0.718,0.659}3,473 & 10\,m \\
	\case{javascript-algorithms-5} & 745 & \cellcolor[rgb]{0.999,0.987,0.852}220 & \cellcolor[rgb]{0.996,0.885,0.744}1,535 & 5\,m & \cellcolor[rgb]{0.994,0.902,0.757}497 & \cellcolor[rgb]{0.997,0.876,0.737}1,637 & 5\,m & \cellcolor[rgb]{0.990,0.945,0.797}354 & \cellcolor[rgb]{0.999,0.987,0.852}602 & 3\,m & \cellcolor[rgb]{0.995,0.982,0.842}240 & \cellcolor[rgb]{1.000,0.806,0.692}2,690 & 8\,m \\
	\case{javascript-algorithms-14} & 814 & \cellcolor[rgb]{0.991,0.967,0.823}252 & \cellcolor[rgb]{1.000,0.824,0.702}1,233 & 4\,m & \cellcolor[rgb]{0.990,0.938,0.790}331 & \cellcolor[rgb]{1.000,0.847,0.717}1,045 & 4\,m & \cellcolor[rgb]{0.996,0.984,0.846}204 & \cellcolor[rgb]{0.999,0.987,0.852}313 & 2\,m & \cellcolor[rgb]{0.999,0.987,0.852}194 & \cellcolor[rgb]{1.000,0.765,0.673}1,897 & 6\,m \\
	\case{moment-1} & 800 & \cellcolor[rgb]{0.990,0.960,0.815}181 & \cellcolor[rgb]{1.000,0.830,0.706}1,521 & 9\,m & \cellcolor[rgb]{1.000,0.847,0.717}434 & \cellcolor[rgb]{1.000,0.815,0.697}1,694 & 11\,m & \cellcolor[rgb]{0.999,0.987,0.852}130 & \cellcolor[rgb]{0.999,0.987,0.852}405 & 5\,m & \cellcolor[rgb]{0.994,0.981,0.841}144 & \cellcolor[rgb]{1.000,0.796,0.687}1,943 & 7\,m \\
	\case{moment-17} & 1,340 & \cellcolor[rgb]{0.999,0.987,0.852}188 & \cellcolor[rgb]{0.996,0.888,0.746}1,768 & 10\,m & \cellcolor[rgb]{0.996,0.884,0.743}483 & \cellcolor[rgb]{0.998,0.871,0.733}2,000 & 12\,m & \cellcolor[rgb]{0.993,0.976,0.834}222 & \cellcolor[rgb]{0.999,0.987,0.852}707 & 7\,m & \cellcolor[rgb]{0.995,0.981,0.841}207 & \cellcolor[rgb]{1.000,0.853,0.720}2,267 & 8\,m \\
	\case{express-1} & 644 & \cellcolor[rgb]{0.999,0.987,0.852}98 & \cellcolor[rgb]{1.000,0.759,0.670}1,182 & 4\,m & \cellcolor[rgb]{0.993,0.909,0.763}210 & \cellcolor[rgb]{1.000,0.758,0.670}1,185 & 4\,m & \cellcolor[rgb]{0.991,0.971,0.828}122 & \cellcolor[rgb]{0.999,0.987,0.852}186 & 87\,s & \cellcolor[rgb]{0.995,0.982,0.842}107 & \cellcolor[rgb]{0.987,0.702,0.656}1,888 & 6\,m \\
	\case{express-48} & 903 & \cellcolor[rgb]{0.999,0.987,0.852}117 & \cellcolor[rgb]{1.000,0.798,0.687}1,336 & 4\,m & \cellcolor[rgb]{0.990,0.949,0.802}181 & \cellcolor[rgb]{1.000,0.852,0.720}904 & 3\,m & \cellcolor[rgb]{0.991,0.926,0.779}220 & \cellcolor[rgb]{0.999,0.987,0.852}281 & 2\,m & \cellcolor[rgb]{0.991,0.930,0.783}212 & \cellcolor[rgb]{0.987,0.701,0.656}2,872 & 9\,m \\
	\hline
	\subtablegap
	\multicolumn{14}{l}{\textit{graphviz}} \\
	\hline
	\case{issue-1436} & 165,583 & \cellcolor[rgb]{0.874,0.579,0.695}2,270 & \cellcolor[rgb]{0.986,0.698,0.656}57,940 & 3.2\,h & \cellcolor[rgb]{0.682,0.510,0.737}12,949 & \cellcolor[rgb]{0.928,0.618,0.669}137,314 & 4.2\,h & \cellcolor[rgb]{0.907,0.600,0.679}1,614 & \cellcolor[rgb]{0.999,0.987,0.852}5,512 & 16\,m & \cellcolor[rgb]{0.999,0.987,0.852}51 & \cellcolor[rgb]{0.992,0.972,0.829}6,781 & 10\,m \\
	\case{issue-2723} & 2,141 & \cellcolor[rgb]{0.996,0.984,0.846}90 & \cellcolor[rgb]{0.990,0.960,0.814}1,011 & 47\,s & \cellcolor[rgb]{0.999,0.861,0.726}257 & \cellcolor[rgb]{0.992,0.916,0.770}1,461 & 55\,s & \cellcolor[rgb]{0.997,0.985,0.847}89 & \cellcolor[rgb]{0.999,0.987,0.852}722 & 2\,m & \cellcolor[rgb]{0.999,0.987,0.852}85 & \cellcolor[rgb]{0.998,0.870,0.733}2,048 & 71\,s \\
	\hline
	\subtablegap
	\multicolumn{14}{l}{\textit{lua}} \\
	\hline
	\case{luajit-913} & 1,248 & \cellcolor[rgb]{0.991,0.971,0.828}72 & \cellcolor[rgb]{1.000,0.811,0.695}1,325 & 52\,s & \cellcolor[rgb]{1.000,0.777,0.678}320 & \cellcolor[rgb]{1.000,0.760,0.671}1,934 & 28\,s & \cellcolor[rgb]{0.992,0.920,0.774}114 & \cellcolor[rgb]{0.999,0.987,0.852}308 & 59\,s & \cellcolor[rgb]{0.999,0.987,0.852}58 & \cellcolor[rgb]{1.000,0.770,0.675}1,789 & 49\,s \\
	\case{luajit-1116} & 889 & \cellcolor[rgb]{0.996,0.984,0.846}377 & \cellcolor[rgb]{0.992,0.975,0.832}2,385 & 58\,s & \cellcolor[rgb]{0.996,0.984,0.846}377 & \cellcolor[rgb]{0.999,0.987,0.852}1,991 & 42\,s & \cellcolor[rgb]{0.999,0.987,0.852}357 & \cellcolor[rgb]{0.991,0.933,0.786}3,529 & 8\,m & \cellcolor[rgb]{0.995,0.982,0.843}389 & \cellcolor[rgb]{1.000,0.789,0.683}10,087 & 3\,m \\
	\hline
	\subtablegap
	\multicolumn{14}{l}{\textit{pdf}} \\
	\hline
	\case{pdf-fuzzer-aah052} & 15,421 & \cellcolor[rgb]{0.997,0.985,0.847}118 & \cellcolor[rgb]{0.993,0.914,0.768}1,419 & 2\,m & \cellcolor[rgb]{0.997,0.985,0.847}118 & \cellcolor[rgb]{0.999,0.987,0.852}688 & 33\,s & -- & -- & -- & \cellcolor[rgb]{0.999,0.987,0.852}113 & \cellcolor[rgb]{1.000,0.807,0.692}3,053 & 2\,m \\
	\case{pdf-fuzzer-jch207} & 15,505 & \cellcolor[rgb]{0.999,0.987,0.852}7 & \cellcolor[rgb]{1.000,0.765,0.673}672 & 29\,s & \cellcolor[rgb]{0.999,0.987,0.852}7 & \cellcolor[rgb]{0.999,0.987,0.852}111 & 3.5\,s & -- & -- & -- & \cellcolor[rgb]{0.960,0.653,0.658}115 & \cellcolor[rgb]{0.994,0.904,0.760}246 & 15\,s \\
	\case{pdfimages-jch207-kw6} & 4,543 & \cellcolor[rgb]{0.999,0.987,0.852}3 & \cellcolor[rgb]{0.956,0.647,0.659}1,107 & 24\,s & \cellcolor[rgb]{0.999,0.987,0.852}3 & \cellcolor[rgb]{0.999,0.987,0.852}63 & 1.1\,s & -- & -- & -- & \cellcolor[rgb]{0.999,0.987,0.852}3 & \cellcolor[rgb]{0.993,0.721,0.659}541 & 20\,s \\
	\case{pdftoppm-aah052} & 15,495 & \cellcolor[rgb]{0.991,0.971,0.827}139 & \cellcolor[rgb]{0.994,0.902,0.758}1,576 & 86\,s & \cellcolor[rgb]{0.995,0.982,0.843}120 & \cellcolor[rgb]{0.999,0.987,0.852}701 & 25\,s & -- & -- & -- & \cellcolor[rgb]{0.999,0.987,0.852}111 & \cellcolor[rgb]{1.000,0.827,0.704}2,702 & 2\,m \\
	\case{pdftoppm-jch207} & 15,532 & \cellcolor[rgb]{0.677,0.509,0.736}794 & \cellcolor[rgb]{0.990,0.939,0.791}4,733 & 2\,m & \cellcolor[rgb]{0.677,0.509,0.736}799 & \cellcolor[rgb]{0.990,0.959,0.813}3,929 & 2\,m & -- & -- & -- & \cellcolor[rgb]{0.999,0.987,0.852}3 & \cellcolor[rgb]{0.999,0.987,0.852}2,794 & 2\,m \\
	\hline
	\subtablegap
	\multicolumn{14}{l}{\textit{python}} \\
	\hline
	\case{ackermann} & 483 & \cellcolor[rgb]{0.990,0.947,0.800}123 & \cellcolor[rgb]{0.997,0.985,0.848}1,113 & 2\,m & \cellcolor[rgb]{0.998,0.874,0.735}216 & \cellcolor[rgb]{0.999,0.987,0.852}1,073 & 86\,s & \cellcolor[rgb]{0.991,0.930,0.783}142 & \cellcolor[rgb]{0.997,0.985,0.847}1,117 & 5\,m & \cellcolor[rgb]{0.999,0.987,0.852}78 & \cellcolor[rgb]{0.997,0.883,0.742}2,784 & 4\,m \\
	\case{alloc} & 1,125 & \cellcolor[rgb]{0.990,0.961,0.816}304 & \cellcolor[rgb]{0.991,0.969,0.825}2,373 & 3\,m & \cellcolor[rgb]{0.993,0.910,0.765}466 & \cellcolor[rgb]{0.999,0.987,0.852}1,856 & 2\,m & \cellcolor[rgb]{0.992,0.922,0.775}426 & \cellcolor[rgb]{0.993,0.978,0.837}2,124 & 7\,m & \cellcolor[rgb]{0.999,0.987,0.852}220 & \cellcolor[rgb]{1.000,0.816,0.697}7,729 & 10\,m \\
	\case{dict} & 265 & \cellcolor[rgb]{0.990,0.955,0.808}135 & \cellcolor[rgb]{0.990,0.945,0.797}1,232 & 90\,s & \cellcolor[rgb]{0.990,0.945,0.797}148 & \cellcolor[rgb]{0.999,0.987,0.852}765 & 54\,s & \cellcolor[rgb]{0.990,0.943,0.796}150 & \cellcolor[rgb]{0.990,0.958,0.812}1,087 & 4\,m & \cellcolor[rgb]{0.999,0.987,0.852}92 & \cellcolor[rgb]{1.000,0.804,0.691}3,476 & 5\,m \\
	\case{mroref} & 383 & \cellcolor[rgb]{0.999,0.987,0.852}170 & \cellcolor[rgb]{0.994,0.980,0.839}1,052 & 83\,s & \cellcolor[rgb]{0.991,0.971,0.828}211 & \cellcolor[rgb]{0.999,0.987,0.852}935 & 66\,s & \cellcolor[rgb]{0.991,0.966,0.821}225 & \cellcolor[rgb]{0.995,0.982,0.843}1,014 & 4\,m & \cellcolor[rgb]{0.996,0.983,0.844}182 & \cellcolor[rgb]{1.000,0.773,0.676}5,328 & 7\,m \\
	\case{recursion} & 212 & \cellcolor[rgb]{0.999,0.987,0.852}59 & \cellcolor[rgb]{0.991,0.933,0.786}815 & 64\,s & \cellcolor[rgb]{0.992,0.923,0.777}113 & \cellcolor[rgb]{0.997,0.984,0.847}484 & 34\,s & \cellcolor[rgb]{0.991,0.971,0.827}74 & \cellcolor[rgb]{0.999,0.987,0.852}461 & 2\,m & \cellcolor[rgb]{0.997,0.984,0.846}62 & \cellcolor[rgb]{0.998,0.870,0.732}1,314 & 2\,m \\
	\hline
	\subtablegap
	\multicolumn{14}{l}{\textit{smt}} \\
	\hline
	\case{btor2-bug-16966-2630} & 2,435 & \cellcolor[rgb]{0.992,0.974,0.832}862 & \cellcolor[rgb]{0.995,0.982,0.842}8,632 & 44\,s & \cellcolor[rgb]{0.998,0.874,0.735}1,991 & \cellcolor[rgb]{0.999,0.987,0.852}7,914 & 29\,s & -- & -- & -- & \cellcolor[rgb]{0.999,0.987,0.852}719 & \cellcolor[rgb]{1.000,0.799,0.688}37,289 & 11\,m \\
	\case{btor2-bug-21435-401} & 3,546 & \cellcolor[rgb]{0.999,0.987,0.852}1,058 & \cellcolor[rgb]{0.999,0.987,0.852}9,564 & 51\,s & \cellcolor[rgb]{0.997,0.876,0.737}2,887 & \cellcolor[rgb]{0.992,0.975,0.832}11,445 & 44\,s & -- & -- & -- & \cellcolor[rgb]{0.994,0.980,0.840}1,185 & \cellcolor[rgb]{1.000,0.756,0.669}62,063 & 18\,m \\
	\case{btor2-bug-16965-103} & 9,773 & \cellcolor[rgb]{0.990,0.948,0.801}1,495 & \cellcolor[rgb]{0.999,0.987,0.852}31,307 & 3\,m & \cellcolor[rgb]{0.996,0.731,0.662}7,591 & \cellcolor[rgb]{0.992,0.972,0.829}38,428 & 3\,m & -- & -- & -- & \cellcolor[rgb]{0.999,0.987,0.852}957 & \cellcolor[rgb]{0.999,0.862,0.727}94,137 & 29\,m \\
	\case{btor2-bug-19025-698} & 63,795 & \cellcolor[rgb]{0.990,0.957,0.811}2,772 & \cellcolor[rgb]{0.999,0.987,0.852}185,324 & 56\,m & \cellcolor[rgb]{0.921,0.612,0.673}52,001 & \cellcolor[rgb]{0.990,0.949,0.802}287,101 & 54\,m & -- & -- & -- & \cellcolor[rgb]{0.999,0.987,0.852}1,925 & \cellcolor[rgb]{0.997,0.880,0.740}491,145 & 5.3\,h \\
	\case{btor4-bug-29933-6674} & 95,337 & \cellcolor[rgb]{0.993,0.913,0.767}1,364 & \cellcolor[rgb]{0.999,0.987,0.852}30,525 & 7\,m & \cellcolor[rgb]{0.959,0.651,0.658}11,027 & \cellcolor[rgb]{0.996,0.884,0.743}78,179 & 7\,m & -- & -- & -- & \cellcolor[rgb]{0.999,0.987,0.852}655 & \cellcolor[rgb]{0.999,0.862,0.726}92,026 & 26\,m \\
	\case{btor4-bug-1428-13899} & 162,928 & \cellcolor[rgb]{0.991,0.935,0.787}712 & \cellcolor[rgb]{0.999,0.987,0.852}144,595 & 67\,m & \cellcolor[rgb]{0.740,0.525,0.739}61,394 & \cellcolor[rgb]{0.999,0.859,0.725}444,404 & 1.6\,h & -- & -- & -- & \cellcolor[rgb]{0.999,0.987,0.852}407 & \cellcolor[rgb]{0.991,0.933,0.786}255,725 & 1.7\,h \\
	\case{btor4-bug-29800-10218} & 224,151 & \cellcolor[rgb]{0.990,0.947,0.800}708 & \cellcolor[rgb]{0.999,0.987,0.852}97,583 & 43\,m & \cellcolor[rgb]{0.822,0.554,0.717}32,617 & \cellcolor[rgb]{0.990,0.943,0.796}158,747 & 23\,m & -- & -- & -- & \cellcolor[rgb]{0.999,0.987,0.852}449 & \cellcolor[rgb]{0.991,0.935,0.788}170,158 & 58\,m \\
	\case{smtlib384a03} & 43,696 & \cellcolor[rgb]{0.995,0.981,0.842}172 & \cellcolor[rgb]{0.994,0.904,0.759}2,097 & 3\,m & \cellcolor[rgb]{0.912,0.604,0.677}4,723 & \cellcolor[rgb]{0.951,0.640,0.661}17,826 & 16\,m & \cellcolor[rgb]{0.990,0.958,0.812}223 & \cellcolor[rgb]{0.999,0.987,0.852}945 & 3\,m & \cellcolor[rgb]{0.999,0.987,0.852}157 & \cellcolor[rgb]{0.991,0.713,0.658}8,742 & 8\,m \\
	\case{smtlib46f14a} & 43,323 & \cellcolor[rgb]{0.995,0.981,0.842}172 & \cellcolor[rgb]{0.994,0.902,0.758}2,126 & 3\,m & \cellcolor[rgb]{0.928,0.617,0.670}3,943 & \cellcolor[rgb]{0.965,0.659,0.657}14,513 & 13\,m & \cellcolor[rgb]{0.990,0.958,0.812}223 & \cellcolor[rgb]{0.999,0.987,0.852}945 & 3\,m & \cellcolor[rgb]{0.999,0.987,0.852}157 & \cellcolor[rgb]{0.991,0.713,0.658}8,742 & 8\,m \\
	\case{smtlibf957ea} & 43,937 & \cellcolor[rgb]{0.999,0.987,0.852}151 & \cellcolor[rgb]{0.994,0.905,0.760}1,841 & 3\,m & \cellcolor[rgb]{0.933,0.622,0.667}3,556 & \cellcolor[rgb]{0.972,0.670,0.656}11,455 & 9\,m & \cellcolor[rgb]{0.990,0.949,0.802}234 & \cellcolor[rgb]{0.999,0.987,0.852}834 & 3\,m & \cellcolor[rgb]{0.997,0.985,0.848}157 & \cellcolor[rgb]{0.986,0.698,0.656}8,763 & 8\,m \\
	\hline
	\subtablegap
	\multicolumn{14}{l}{\textit{sql}} \\
	\hline
	\case{postgresql-18398} & 277 & \cellcolor[rgb]{0.991,0.971,0.827}126 & \cellcolor[rgb]{0.996,0.984,0.845}797 & 21\,s & \cellcolor[rgb]{0.993,0.908,0.763}218 & \cellcolor[rgb]{0.999,0.987,0.852}751 & 18\,s & -- & -- & -- & \cellcolor[rgb]{0.999,0.987,0.852}101 & \cellcolor[rgb]{0.996,0.889,0.747}1,866 & 68\,s \\
	\case{mysql-112460} & 538 & \cellcolor[rgb]{0.996,0.984,0.846}208 & \cellcolor[rgb]{0.994,0.980,0.840}1,486 & 2\,m & \cellcolor[rgb]{0.991,0.927,0.780}368 & \cellcolor[rgb]{0.999,0.987,0.852}1,333 & 64\,s & -- & -- & -- & \cellcolor[rgb]{0.999,0.987,0.852}197 & \cellcolor[rgb]{1.000,0.823,0.701}5,291 & 5\,m \\
	\case{sqlite-b7914e5795} & 503 & \cellcolor[rgb]{0.991,0.966,0.821}233 & \cellcolor[rgb]{0.997,0.876,0.737}2,300 & 32\,s & \cellcolor[rgb]{0.993,0.910,0.765}375 & \cellcolor[rgb]{0.990,0.962,0.816}1,164 & 16\,s & \cellcolor[rgb]{0.992,0.975,0.833}210 & \cellcolor[rgb]{0.999,0.987,0.852}846 & 18\,s & \cellcolor[rgb]{0.999,0.987,0.852}177 & \cellcolor[rgb]{1.000,0.751,0.668}5,689 & 3\,m \\
	\case{sqlflex-sqlite-79} & 128 & \cellcolor[rgb]{0.991,0.929,0.782}64 & \cellcolor[rgb]{1.000,0.784,0.681}579 & 7.9\,s & \cellcolor[rgb]{0.998,0.868,0.731}101 & \cellcolor[rgb]{0.998,0.871,0.733}310 & 4.1\,s & \cellcolor[rgb]{0.990,0.939,0.792}59 & \cellcolor[rgb]{0.999,0.987,0.852}110 & 5.3\,s & \cellcolor[rgb]{0.999,0.987,0.852}35 & \cellcolor[rgb]{0.998,0.744,0.665}786 & 24\,s \\
	\case{sqlflex-sqlite-165} & 163 & \cellcolor[rgb]{0.995,0.982,0.843}38 & \cellcolor[rgb]{0.995,0.726,0.660}504 & 6.9\,s & \cellcolor[rgb]{1.000,0.827,0.704}135 & \cellcolor[rgb]{1.000,0.759,0.670}387 & 5.2\,s & \cellcolor[rgb]{0.993,0.976,0.835}41 & \cellcolor[rgb]{0.999,0.987,0.852}61 & 4.5\,s & \cellcolor[rgb]{0.999,0.987,0.852}35 & \cellcolor[rgb]{0.973,0.672,0.656}824 & 25\,s \\
	\case{sqlflex-sqlite-256} & 104 & \cellcolor[rgb]{0.990,0.939,0.792}69 & \cellcolor[rgb]{1.000,0.808,0.693}554 & 7.7\,s & \cellcolor[rgb]{0.994,0.906,0.761}90 & \cellcolor[rgb]{0.990,0.936,0.789}218 & 3.0\,s & \cellcolor[rgb]{0.995,0.981,0.842}45 & \cellcolor[rgb]{0.999,0.987,0.852}126 & 5.8\,s & \cellcolor[rgb]{0.999,0.987,0.852}41 & \cellcolor[rgb]{1.000,0.760,0.671}791 & 25\,s \\
	\case{sqlflex-sqlite-622} & 53 & \cellcolor[rgb]{0.997,0.986,0.849}36 & \cellcolor[rgb]{0.998,0.739,0.664}432 & 5.9\,s & \cellcolor[rgb]{0.992,0.974,0.832}42 & \cellcolor[rgb]{0.993,0.915,0.769}119 & 1.6\,s & \cellcolor[rgb]{0.994,0.980,0.840}39 & \cellcolor[rgb]{0.999,0.987,0.852}58 & 4.4\,s & \cellcolor[rgb]{0.999,0.987,0.852}35 & \cellcolor[rgb]{0.998,0.741,0.665}422 & 15\,s \\
	\case{sqlflex-sqlite-686} & 179 & \cellcolor[rgb]{0.990,0.951,0.804}58 & \cellcolor[rgb]{1.000,0.773,0.676}640 & 8.7\,s & \cellcolor[rgb]{0.998,0.870,0.732}108 & \cellcolor[rgb]{1.000,0.832,0.707}417 & 5.6\,s & \cellcolor[rgb]{0.990,0.941,0.794}63 & \cellcolor[rgb]{0.999,0.987,0.852}112 & 5.4\,s & \cellcolor[rgb]{0.999,0.987,0.852}38 & \cellcolor[rgb]{1.000,0.811,0.695}483 & 17\,s \\
	\case{sqlflex-sqlite-808} & 51 & \cellcolor[rgb]{0.992,0.974,0.832}36 & \cellcolor[rgb]{0.998,0.743,0.665}430 & 5.9\,s & \cellcolor[rgb]{0.990,0.960,0.814}42 & \cellcolor[rgb]{0.992,0.921,0.774}117 & 1.6\,s & \cellcolor[rgb]{0.991,0.967,0.823}39 & \cellcolor[rgb]{0.999,0.987,0.852}60 & 4.5\,s & \cellcolor[rgb]{0.999,0.987,0.852}30 & \cellcolor[rgb]{0.999,0.746,0.666}420 & 15\,s \\
	\case{sqlflex-sqlite-909} & 144 & \cellcolor[rgb]{0.993,0.977,0.835}83 & \cellcolor[rgb]{1.000,0.795,0.686}691 & 9.5\,s & \cellcolor[rgb]{0.990,0.953,0.806}106 & \cellcolor[rgb]{0.999,0.866,0.729}419 & 5.6\,s & \cellcolor[rgb]{0.999,0.987,0.852}71 & \cellcolor[rgb]{0.999,0.987,0.852}143 & 5.9\,s & \cellcolor[rgb]{0.995,0.981,0.842}78 & \cellcolor[rgb]{0.985,0.696,0.656}1,534 & 43\,s \\
	\hline
\end{tabularx} 
}
	\tablenote{Sizes are in bytes, without whitespace.
		\autodidact{}'s calls and time include its inference,
		by stage in \Cref{tab:stages},
		and HDD's and C-Reduce's exclude validation queries.
		Dashes mark inputs a reducer did not run on.
		Each output and call count is shaded by its ratio
		to the smallest in its row: \tikz[baseline=-0.7pt]{\fill[color={rgb,1:red,0.996;green,0.983;blue,0.844}] (0.000,0) rectangle (0.067,0.12); \fill[color={rgb,1:red,0.991;green,0.971;blue,0.828}] (0.067,0) rectangle (0.133,0.12); \fill[color={rgb,1:red,0.990;green,0.957;blue,0.811}] (0.133,0) rectangle (0.200,0.12); \fill[color={rgb,1:red,0.990;green,0.941;blue,0.794}] (0.200,0) rectangle (0.267,0.12); \fill[color={rgb,1:red,0.992;green,0.924;blue,0.777}] (0.267,0) rectangle (0.333,0.12); \fill[color={rgb,1:red,0.994;green,0.905;blue,0.760}] (0.333,0) rectangle (0.400,0.12); \fill[color={rgb,1:red,0.996;green,0.885;blue,0.744}] (0.400,0) rectangle (0.467,0.12); \fill[color={rgb,1:red,0.999;green,0.865;blue,0.729}] (0.467,0) rectangle (0.533,0.12); \fill[color={rgb,1:red,1.000;green,0.845;blue,0.715}] (0.533,0) rectangle (0.600,0.12); \fill[color={rgb,1:red,1.000;green,0.824;blue,0.702}] (0.600,0) rectangle (0.667,0.12); \fill[color={rgb,1:red,1.000;green,0.804;blue,0.691}] (0.667,0) rectangle (0.733,0.12); \fill[color={rgb,1:red,1.000;green,0.784;blue,0.681}] (0.733,0) rectangle (0.800,0.12); \fill[color={rgb,1:red,1.000;green,0.765;blue,0.673}] (0.800,0) rectangle (0.867,0.12); \fill[color={rgb,1:red,0.999;green,0.747;blue,0.666}] (0.867,0) rectangle (0.933,0.12); \fill[color={rgb,1:red,0.995;green,0.729;blue,0.661}] (0.933,0) rectangle (1.000,0.12); \fill[color={rgb,1:red,0.991;green,0.712;blue,0.658}] (1.000,0) rectangle (1.067,0.12); \fill[color={rgb,1:red,0.985;green,0.696;blue,0.656}] (1.067,0) rectangle (1.133,0.12); \fill[color={rgb,1:red,0.978;green,0.680;blue,0.655}] (1.133,0) rectangle (1.200,0.12); \fill[color={rgb,1:red,0.969;green,0.666;blue,0.656}] (1.200,0) rectangle (1.267,0.12); \fill[color={rgb,1:red,0.960;green,0.652;blue,0.658}] (1.267,0) rectangle (1.333,0.12); \fill[color={rgb,1:red,0.950;green,0.639;blue,0.661}] (1.333,0) rectangle (1.400,0.12); \fill[color={rgb,1:red,0.939;green,0.627;blue,0.665}] (1.400,0) rectangle (1.467,0.12); \fill[color={rgb,1:red,0.927;green,0.616;blue,0.670}] (1.467,0) rectangle (1.533,0.12); \fill[color={rgb,1:red,0.914;green,0.605;blue,0.676}] (1.533,0) rectangle (1.600,0.12); \fill[color={rgb,1:red,0.900;green,0.596;blue,0.682}] (1.600,0) rectangle (1.667,0.12); \fill[color={rgb,1:red,0.886;green,0.587;blue,0.689}] (1.667,0) rectangle (1.733,0.12); \fill[color={rgb,1:red,0.872;green,0.578;blue,0.696}] (1.733,0) rectangle (1.800,0.12); \fill[color={rgb,1:red,0.857;green,0.570;blue,0.702}] (1.800,0) rectangle (1.867,0.12); \fill[color={rgb,1:red,0.842;green,0.563;blue,0.709}] (1.867,0) rectangle (1.933,0.12); \fill[color={rgb,1:red,0.826;green,0.556;blue,0.716}] (1.933,0) rectangle (2.000,0.12); \fill[color={rgb,1:red,0.810;green,0.549;blue,0.721}] (2.000,0) rectangle (2.067,0.12); \fill[color={rgb,1:red,0.794;green,0.543;blue,0.727}] (2.067,0) rectangle (2.133,0.12); \fill[color={rgb,1:red,0.778;green,0.538;blue,0.731}] (2.133,0) rectangle (2.200,0.12); \fill[color={rgb,1:red,0.762;green,0.532;blue,0.735}] (2.200,0) rectangle (2.267,0.12); \fill[color={rgb,1:red,0.746;green,0.527;blue,0.738}] (2.267,0) rectangle (2.333,0.12); \fill[color={rgb,1:red,0.730;green,0.522;blue,0.740}] (2.333,0) rectangle (2.400,0.12); \fill[color={rgb,1:red,0.714;green,0.518;blue,0.740}] (2.400,0) rectangle (2.467,0.12); \fill[color={rgb,1:red,0.698;green,0.514;blue,0.739}] (2.467,0) rectangle (2.533,0.12); \fill[color={rgb,1:red,0.682;green,0.510;blue,0.737}] (2.533,0) rectangle (2.600,0.12); \fill[color={rgb,1:red,0.666;green,0.506;blue,0.733}] (2.600,0) rectangle (2.667,0.12); \fill[color={rgb,1:red,0.650;green,0.502;blue,0.727}] (2.667,0) rectangle (2.733,0.12); \fill[color={rgb,1:red,0.634;green,0.499;blue,0.720}] (2.733,0) rectangle (2.800,0.12); \fill[color={rgb,1:red,0.618;green,0.496;blue,0.711}] (2.800,0) rectangle (2.867,0.12); \fill[color={rgb,1:red,0.602;green,0.493;blue,0.701}] (2.867,0) rectangle (2.933,0.12); \fill[color={rgb,1:red,0.587;green,0.490;blue,0.689}] (2.933,0) rectangle (3.000,0.12); \fill[color={rgb,1:red,0.571;green,0.487;blue,0.676}] (3.000,0) rectangle (3.067,0.12); \fill[color={rgb,1:red,0.556;green,0.485;blue,0.661}] (3.067,0) rectangle (3.133,0.12); \fill[color={rgb,1:red,0.541;green,0.482;blue,0.644}] (3.133,0) rectangle (3.200,0.12); \node[below, inner sep=0.5pt, font=\tiny] at (0.000,0) {1$\times$}; \node[below, inner sep=0.5pt, font=\tiny] at (1.067,0) {10$\times$}; \node[below, inner sep=0.5pt, font=\tiny] at (2.133,0) {100$\times$}; \node[below, inner sep=0.5pt, font=\tiny] at (3.200,0) {1000$\times$};} 
}
\end{table}

\begin{table}[tp]
	\continuedcaption{tab:results}{Reduction results for the \worded{} inputs with a returned word}
	\tablestyle
	{\tablemedium\begin{tabularx}{\linewidth}{|L|N|N|N|N|N|N|N|N|N|N|N|N|N|}
	\hline
	\multirow{2}{*}{\textbf{Case}} & \multirow{2}{*}{\textbf{Input}} & \multicolumn{3}{c|}{\textbf{\autodidact{}}} & \multicolumn{3}{c|}{\textbf{\drdd{}}} & \multicolumn{3}{c|}{\textbf{HDD}} & \multicolumn{3}{c|}{\textbf{C-Reduce}} \\
	 & & \emph{Output} & \emph{Calls} & \emph{Time} & \emph{Output} & \emph{Calls} & \emph{Time} & \emph{Output} & \emph{Calls} & \emph{Time} & \emph{Output} & \emph{Calls} & \emph{Time} \\
	\hline
	\subtablegap
	\multicolumn{14}{l}{\textit{xml}} \\
	\hline
	\case{1.2} & 3,974 & \cellcolor[rgb]{0.994,0.901,0.757}66 & \cellcolor[rgb]{0.972,0.670,0.656}1,322 & 86\,s & \cellcolor[rgb]{0.892,0.590,0.686}1,079 & \cellcolor[rgb]{0.878,0.581,0.693}4,110 & 78\,s & \cellcolor[rgb]{0.999,0.987,0.852}29 & \cellcolor[rgb]{0.999,0.987,0.852}96 & 15\,s & \cellcolor[rgb]{0.997,0.878,0.738}78 & \cellcolor[rgb]{0.812,0.550,0.721}7,633 & 3\,m \\
	\case{3.2} & 3,835 & \cellcolor[rgb]{0.991,0.970,0.826}149 & \cellcolor[rgb]{1.000,0.809,0.693}1,580 & 2\,m & \cellcolor[rgb]{0.993,0.719,0.659}1,035 & \cellcolor[rgb]{0.984,0.695,0.656}3,896 & 75\,s & \cellcolor[rgb]{0.999,0.987,0.852}118 & \cellcolor[rgb]{0.999,0.987,0.852}360 & 27\,s & \cellcolor[rgb]{0.979,0.684,0.655}1,416 & \cellcolor[rgb]{0.928,0.618,0.670}8,976 & 3\,m \\
	\case{4.1} & 3,933 & \cellcolor[rgb]{0.992,0.975,0.832}140 & \cellcolor[rgb]{1.000,0.826,0.703}2,308 & 2\,m & \cellcolor[rgb]{0.999,0.748,0.666}811 & \cellcolor[rgb]{1.000,0.769,0.674}3,491 & 81\,s & \cellcolor[rgb]{0.999,0.987,0.852}117 & \cellcolor[rgb]{0.999,0.987,0.852}595 & 53\,s & \cellcolor[rgb]{0.991,0.929,0.782}214 & \cellcolor[rgb]{0.945,0.634,0.663}12,071 & 4\,m \\
	\case{5.2} & 4,052 & \cellcolor[rgb]{0.991,0.971,0.827}130 & \cellcolor[rgb]{0.994,0.722,0.660}1,571 & 95\,s & \cellcolor[rgb]{0.989,0.708,0.657}1,004 & \cellcolor[rgb]{0.941,0.630,0.664}3,919 & 78\,s & \cellcolor[rgb]{0.999,0.987,0.852}104 & \cellcolor[rgb]{0.999,0.987,0.852}184 & 21\,s & \cellcolor[rgb]{0.990,0.963,0.818}141 & \cellcolor[rgb]{0.845,0.564,0.707}10,737 & 3\,m \\
	\case{071d221-1} & 101,759 & \cellcolor[rgb]{0.999,0.987,0.852}15 & \cellcolor[rgb]{0.750,0.528,0.737}3,887 & 8\,m & \cellcolor[rgb]{0.534,0.481,0.635}19,966 & \cellcolor[rgb]{0.534,0.481,0.635}108,336 & 54\,m & \cellcolor[rgb]{0.999,0.987,0.852}15 & \cellcolor[rgb]{0.999,0.987,0.852}28 & 14\,s & \cellcolor[rgb]{0.534,0.481,0.635}28,960 & \cellcolor[rgb]{0.534,0.481,0.635}142,284 & 77\,m \\
	\case{2d4ec80-1} & 108,560 & \cellcolor[rgb]{0.990,0.964,0.819}63 & \cellcolor[rgb]{0.856,0.570,0.703}4,221 & 3\,m & \cellcolor[rgb]{0.625,0.497,0.715}19,980 & \cellcolor[rgb]{0.534,0.481,0.635}108,244 & 35\,m & \cellcolor[rgb]{0.999,0.987,0.852}47 & \cellcolor[rgb]{0.999,0.987,0.852}80 & 16\,s & \cellcolor[rgb]{0.628,0.498,0.717}19,437 & \cellcolor[rgb]{0.534,0.481,0.635}125,681 & 63\,m \\
	\case{327c8af-1} & 103,527 & \cellcolor[rgb]{0.991,0.971,0.827}60 & \cellcolor[rgb]{0.956,0.647,0.659}4,168 & 4\,m & \cellcolor[rgb]{0.629,0.498,0.718}19,576 & \cellcolor[rgb]{0.619,0.496,0.712}106,657 & 41\,m & \cellcolor[rgb]{0.999,0.987,0.852}48 & \cellcolor[rgb]{0.999,0.987,0.852}237 & 33\,s & \cellcolor[rgb]{0.795,0.544,0.727}4,426 & \cellcolor[rgb]{0.591,0.491,0.693}137,325 & 54\,m \\
	\case{4c99b96-1} & 104,930 & \cellcolor[rgb]{0.999,0.987,0.852}62 & \cellcolor[rgb]{0.965,0.659,0.657}3,993 & 14\,m & \cellcolor[rgb]{0.655,0.504,0.729}20,000 & \cellcolor[rgb]{0.627,0.498,0.717}108,471 & 57\,m & \cellcolor[rgb]{0.990,0.938,0.790}106 & \cellcolor[rgb]{0.999,0.987,0.852}260 & 48\,s & \cellcolor[rgb]{0.621,0.496,0.713}27,401 & \cellcolor[rgb]{0.606,0.494,0.704}131,567 & 82\,m \\
	\case{8ede045-1} & 106,398 & \cellcolor[rgb]{0.990,0.965,0.820}60 & \cellcolor[rgb]{0.956,0.647,0.659}4,065 & 3\,m & \cellcolor[rgb]{0.621,0.497,0.713}19,841 & \cellcolor[rgb]{0.614,0.495,0.709}108,616 & 36\,m & \cellcolor[rgb]{0.999,0.987,0.852}45 & \cellcolor[rgb]{0.999,0.987,0.852}232 & 32\,s & \cellcolor[rgb]{0.589,0.490,0.691}26,684 & \cellcolor[rgb]{0.545,0.483,0.648}209,083 & 1.8\,h \\
	\case{f053486-1} & 109,614 & \cellcolor[rgb]{0.994,0.905,0.760}33 & \cellcolor[rgb]{0.679,0.509,0.736}5,230 & 13\,m & \cellcolor[rgb]{0.534,0.481,0.635}19,381 & \cellcolor[rgb]{0.534,0.481,0.635}107,771 & 76\,m & \cellcolor[rgb]{0.999,0.987,0.852}15 & \cellcolor[rgb]{0.999,0.987,0.852}20 & 14\,s & \cellcolor[rgb]{0.534,0.481,0.635}28,116 & \cellcolor[rgb]{0.534,0.481,0.635}135,287 & 1.7\,h \\
	\hline
	\subtablegap
	\multicolumn{14}{l}{\textit{yaml}} \\
	\hline
	\case{9J7A} & 16 & \cellcolor[rgb]{0.999,0.987,0.852}11 & \cellcolor[rgb]{1.000,0.846,0.716}138 & 0.3\,s & \cellcolor[rgb]{0.999,0.987,0.852}11 & \cellcolor[rgb]{0.999,0.987,0.852}41 & 0.1\,s & -- & -- & -- & \cellcolor[rgb]{0.999,0.987,0.852}11 & \cellcolor[rgb]{0.998,0.986,0.849}42 & 2.1\,s \\
	\case{9FMG} & 36 & \cellcolor[rgb]{0.999,0.987,0.852}15 & \cellcolor[rgb]{1.000,0.804,0.691}396 & 0.8\,s & \cellcolor[rgb]{0.999,0.987,0.852}15 & \cellcolor[rgb]{0.999,0.987,0.852}87 & 0.2\,s & -- & -- & -- & \cellcolor[rgb]{0.999,0.987,0.852}15 & \cellcolor[rgb]{0.996,0.889,0.747}216 & 4.8\,s \\
	\case{G4RS} & 177 & \cellcolor[rgb]{0.999,0.987,0.852}153 & \cellcolor[rgb]{0.990,0.940,0.793}682 & 1.3\,s & \cellcolor[rgb]{0.999,0.987,0.852}153 & \cellcolor[rgb]{0.999,0.987,0.852}407 & 0.8\,s & -- & -- & -- & \cellcolor[rgb]{0.999,0.987,0.852}153 & \cellcolor[rgb]{0.997,0.985,0.848}423 & 7.5\,s \\
	\case{FBC9} & 148 & \cellcolor[rgb]{0.999,0.987,0.852}130 & \cellcolor[rgb]{0.990,0.941,0.793}702 & 1.4\,s & \cellcolor[rgb]{0.999,0.987,0.852}130 & \cellcolor[rgb]{0.999,0.987,0.852}421 & 0.8\,s & -- & -- & -- & \cellcolor[rgb]{0.999,0.987,0.852}130 & \cellcolor[rgb]{0.995,0.891,0.749}1,029 & 16\,s \\
	\case{6JQW} & 44 & \cellcolor[rgb]{0.999,0.987,0.852}21 & \cellcolor[rgb]{0.996,0.732,0.662}384 & 0.7\,s & \cellcolor[rgb]{0.999,0.987,0.852}21 & \cellcolor[rgb]{0.999,0.987,0.852}49 & 0.1\,s & -- & -- & -- & \cellcolor[rgb]{0.999,0.987,0.852}21 & \cellcolor[rgb]{1.000,0.830,0.705}185 & 4.9\,s \\
	\case{96L6} & 63 & \cellcolor[rgb]{0.999,0.987,0.852}46 & \cellcolor[rgb]{1.000,0.805,0.691}415 & 0.8\,s & \cellcolor[rgb]{0.999,0.987,0.852}46 & \cellcolor[rgb]{0.999,0.987,0.852}92 & 0.2\,s & -- & -- & -- & \cellcolor[rgb]{0.999,0.987,0.852}46 & \cellcolor[rgb]{0.996,0.887,0.745}232 & 5.2\,s \\
	\case{3GZX} & 108 & \cellcolor[rgb]{0.999,0.987,0.852}94 & \cellcolor[rgb]{0.992,0.920,0.773}535 & 1.0\,s & \cellcolor[rgb]{0.999,0.987,0.852}94 & \cellcolor[rgb]{0.999,0.987,0.852}272 & 0.5\,s & -- & -- & -- & \cellcolor[rgb]{0.999,0.987,0.852}94 & \cellcolor[rgb]{0.990,0.960,0.814}382 & 6.8\,s \\
	\case{ZH7C} & 16 & \cellcolor[rgb]{0.999,0.987,0.852}10 & \cellcolor[rgb]{0.999,0.858,0.724}136 & 0.3\,s & \cellcolor[rgb]{0.999,0.987,0.852}10 & \cellcolor[rgb]{0.999,0.987,0.852}44 & 0.1\,s & -- & -- & -- & \cellcolor[rgb]{0.999,0.987,0.852}10 & \cellcolor[rgb]{0.997,0.883,0.742}114 & 3.4\,s \\
	\case{6FWR} & 20 & \cellcolor[rgb]{0.999,0.987,0.852}9 & \cellcolor[rgb]{1.000,0.793,0.685}202 & 0.4\,s & \cellcolor[rgb]{0.999,0.987,0.852}9 & \cellcolor[rgb]{0.999,0.987,0.852}41 & 0.1\,s & -- & -- & -- & \cellcolor[rgb]{0.999,0.987,0.852}9 & \cellcolor[rgb]{0.990,0.941,0.794}68 & 3.8\,s \\
	\case{NAT4} & 86 & \cellcolor[rgb]{0.999,0.987,0.852}35 & \cellcolor[rgb]{0.999,0.859,0.724}582 & 1.1\,s & \cellcolor[rgb]{0.999,0.987,0.852}35 & \cellcolor[rgb]{0.999,0.987,0.852}189 & 0.4\,s & -- & -- & -- & \cellcolor[rgb]{0.999,0.987,0.852}35 & \cellcolor[rgb]{0.997,0.878,0.739}506 & 11\,s \\
	\hline
\end{tabularx} 
}
\end{table}

\subsection{RQ1: Comparison with Byte-Level Delta Debugging}\label{sec:rqValue}

\begin{table}[tp]
	\caption{Reducers relative to \drdd{}.
		On HDD's \hddInputs{} inputs,
		\autodidact{}'s outputs are \hddCohortAtomSizeSaving\% smaller than \drdd{}'s
		with \hddCohortAtomCallSaving\% fewer calls,
		\hddSizeRatio{}$\times$ HDD's at \hddCallRatio{}$\times$ the calls,
		and smaller than C-Reduce's
		with \hddCohortCreduceCallSaving\% fewer calls}
	\label{tab:comparators}
	\tablestyle
	\begin{tabularx}{\linewidth}{|L|N|N|N|N|N|N|}
	\hline
	\multirow{2}{*}{\textbf{Reducer}} & \multicolumn{4}{c|}{\textbf{All \hddInputs{} inputs}} & \multicolumn{2}{c|}{\textbf{Without XML}} \\
	 & \emph{Size} & \emph{Smaller/larger} & \emph{Calls} & \emph{Reduction calls} & \emph{Size} & \emph{Calls} \\
	\hline
	\autodidact{} & 0.10 & 49/0 & 0.37 & 0.15 & 0.18 & 0.51 \\
	C-Reduce (\texttt{-{}-not-c}) & 0.19 & 43/7 & 1.06 & -- & 0.15 & 0.94 \\
	HDD (Picireny) & 0.11 & 47/3 & 0.10 & -- & 0.21 & 0.20 \\
	\hline
\end{tabularx}

	\tablenote{Every row covers the same \hddInputs{} inputs,
		the returned-word inputs with a grammar that HDD reduces,
		so the reducers also compare with each other through \drdd{}.
		Ratios are geometric means of reducer / \drdd{}.
		Ratios below 1 beat \drdd{}.
		Size counts non-whitespace bytes.
		\emph{Smaller/larger} counts inputs
		on which the reducer's output is strictly smaller
		or larger than \drdd{}'s.
		\autodidact{}'s \emph{calls} include its inference,
		and its \emph{reduction calls} leave it out.
		The tools' calls exclude their validation queries.}
\end{table}

We compare \autodidact{} with byte-level \drdd{} on the \fixpointInputs{} inputs with a returned word, charging all inference to \autodidact{}. Against \drdd{}, \autodidact{} returns outputs \fixpointSizeSaving\% smaller with \fixpointCallSaving\% fewer calls on geometric mean, inference included. Its outputs are smaller on \fixpointSmaller{} of \fixpointInputs{} inputs, equal on \fixpointEqual{}, and larger on \fixpointLarger{}, and it uses fewer calls on \fixpointCheaper{}. These results include \autodidact{}'s final shrinking pass, which reduces \shrinkImproved{} of the \shrinkInputs{} laminar outputs further, by \shrinkAdditionalSaving\% on geometric mean, for \shrinkAdditionalCost\% more calls. \Cref{tab:results} gives each input's outputs, calls, and time, and \Cref{tab:stages} itemizes \autodidact{}'s by stage.

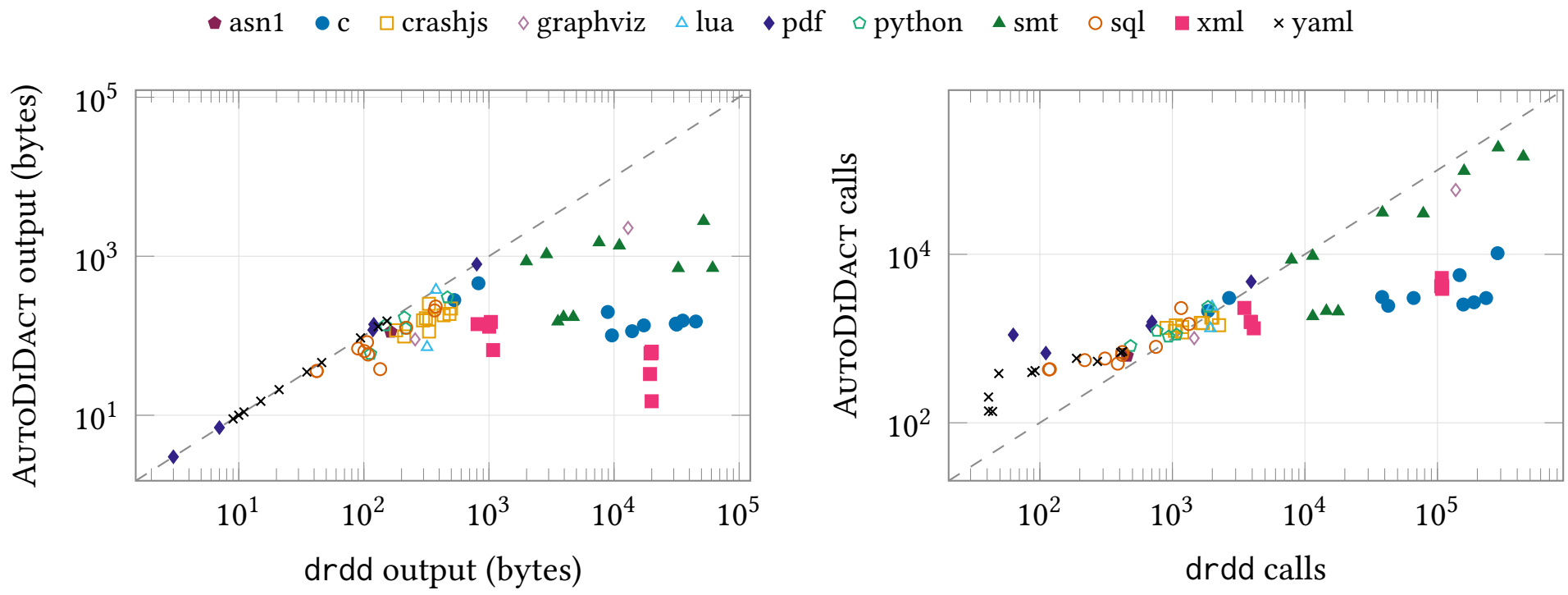
\begin{figure*}[tp]
	\centering
	\definecolor{family-asn1}{HTML}{882255}
\definecolor{family-c}{HTML}{0072B2}
\definecolor{family-crashjs}{HTML}{E69F00}
\definecolor{family-graphviz}{HTML}{B07AA1}
\definecolor{family-lua}{HTML}{33BBEE}
\definecolor{family-pdf}{HTML}{332288}
\definecolor{family-python}{HTML}{1BAF7A}
\definecolor{family-smt}{HTML}{117733}
\definecolor{family-sql}{HTML}{D55E00}
\definecolor{family-xml}{HTML}{EE3377}
\definecolor{family-yaml}{HTML}{000000}
\pgfplotslegendfromname{reduction-families}
\par\smallskip
\begin{tikzpicture}
\begin{groupplot}[
	group style={group size=2 by 1, horizontal sep=1.6cm, vertical sep=1.4cm},
	width=0.47\textwidth,
	height=0.34\textwidth,
	xmode=log,
	ymode=log,
	label style={font=\footnotesize},
	tick label style={font=\footnotesize},
	major grid style={line width=0.2pt, draw=black!12},
	grid=major,
	axis line style={draw=black!45},
	tick style={draw=black!45},
	legend columns=11,
	legend cell align=left,
	legend style={font=\scriptsize, draw=none, nodes={text height=1.2ex, text depth=0.3ex}, /tikz/every even column/.append style={column sep=5pt}},
]
\nextgroupplot[xlabel={\drdd{} output (bytes)}, ylabel={\autodidact{} output (bytes)}, xmin=1.5, xmax=122788, ymin=1.5, ymax=122788, legend to name=reduction-families]
\addplot[no marks, line width=0.4pt, black!45, dashed, forget plot] coordinates {
	(1.5, 1.50)
	(122788.0, 122788.00)
};
\addlegendimage{only marks, mark=pentagon*, mark size=1.4pt, color=family-asn1, mark options={line width=0.4pt}}
\addlegendentry{asn1}
\addlegendimage{only marks, mark=*, mark size=1.4pt, color=family-c, mark options={line width=0.4pt}}
\addlegendentry{c}
\addlegendimage{only marks, mark=square, mark size=1.4pt, color=family-crashjs, mark options={line width=0.4pt}}
\addlegendentry{crashjs}
\addlegendimage{only marks, mark=diamond, mark size=1.4pt, color=family-graphviz, mark options={line width=0.4pt}}
\addlegendentry{graphviz}
\addlegendimage{only marks, mark=triangle, mark size=1.4pt, color=family-lua, mark options={line width=0.4pt}}
\addlegendentry{lua}
\addlegendimage{only marks, mark=diamond*, mark size=1.4pt, color=family-pdf, mark options={line width=0.4pt}}
\addlegendentry{pdf}
\addlegendimage{only marks, mark=pentagon, mark size=1.4pt, color=family-python, mark options={line width=0.4pt}}
\addlegendentry{python}
\addlegendimage{only marks, mark=triangle*, mark size=1.4pt, color=family-smt, mark options={line width=0.4pt}}
\addlegendentry{smt}
\addlegendimage{only marks, mark=o, mark size=1.4pt, color=family-sql, mark options={line width=0.4pt}}
\addlegendentry{sql}
\addlegendimage{only marks, mark=square*, mark size=1.4pt, color=family-xml, mark options={line width=0.4pt}}
\addlegendentry{xml}
\addlegendimage{only marks, mark=x, mark size=1.4pt, color=family-yaml, mark options={line width=0.4pt}}
\addlegendentry{yaml}
\addplot[only marks, mark=pentagon*, mark size=1.4pt, color=family-asn1, forget plot, mark options={line width=0.4pt}] coordinates {
	(164, 112.00)
};
\addplot[only marks, mark=*, mark size=1.4pt, color=family-c, forget plot, mark options={line width=0.4pt}] coordinates {
	(824, 457.00)
	(528, 281.00)
	(45061, 151.00)
	(13940, 114.00)
	(9620, 101.00)
	(31646, 138.00)
	(17308, 135.00)
	(31151, 142.00)
	(35434, 155.00)
	(8914, 199.00)
};
\addplot[only marks, mark=square, mark size=1.4pt, color=family-crashjs, forget plot, mark options={line width=0.4pt}] coordinates {
	(333, 159.00)
	(332, 113.00)
	(313, 166.00)
	(297, 156.00)
	(497, 220.00)
	(331, 252.00)
	(434, 181.00)
	(483, 188.00)
	(210, 98.00)
	(181, 117.00)
};
\addplot[only marks, mark=diamond, mark size=1.4pt, color=family-graphviz, forget plot, mark options={line width=0.4pt}] coordinates {
	(12949, 2270.00)
	(257, 90.00)
};
\addplot[only marks, mark=triangle, mark size=1.4pt, color=family-lua, forget plot, mark options={line width=0.4pt}] coordinates {
	(320, 72.00)
	(377, 377.00)
};
\addplot[only marks, mark=diamond*, mark size=1.4pt, color=family-pdf, forget plot, mark options={line width=0.4pt}] coordinates {
	(118, 118.00)
	(7, 7.00)
	(3, 3.00)
	(120, 139.00)
	(799, 794.00)
};
\addplot[only marks, mark=pentagon, mark size=1.4pt, color=family-python, forget plot, mark options={line width=0.4pt}] coordinates {
	(216, 123.00)
	(466, 304.00)
	(148, 135.00)
	(211, 170.00)
	(113, 59.00)
};
\addplot[only marks, mark=triangle*, mark size=1.4pt, color=family-smt, forget plot, mark options={line width=0.4pt}] coordinates {
	(1991, 862.00)
	(2887, 1058.00)
	(7591, 1495.00)
	(52001, 2772.00)
	(11027, 1364.00)
	(61394, 712.00)
	(32617, 708.00)
	(4723, 172.00)
	(3943, 172.00)
	(3556, 151.00)
};
\addplot[only marks, mark=o, mark size=1.4pt, color=family-sql, forget plot, mark options={line width=0.4pt}] coordinates {
	(218, 126.00)
	(368, 208.00)
	(375, 233.00)
	(101, 64.00)
	(135, 38.00)
	(90, 69.00)
	(42, 36.00)
	(108, 58.00)
	(42, 36.00)
	(106, 83.00)
};
\addplot[only marks, mark=square*, mark size=1.4pt, color=family-xml, forget plot, mark options={line width=0.4pt}] coordinates {
	(1079, 66.00)
	(1035, 149.00)
	(811, 140.00)
	(1004, 130.00)
	(19966, 15.00)
	(19980, 63.00)
	(19576, 60.00)
	(20000, 62.00)
	(19841, 60.00)
	(19381, 33.00)
};
\addplot[only marks, mark=x, mark size=1.4pt, color=family-yaml, forget plot, mark options={line width=0.4pt}] coordinates {
	(11, 11.00)
	(15, 15.00)
	(153, 153.00)
	(130, 130.00)
	(21, 21.00)
	(46, 46.00)
	(94, 94.00)
	(10, 10.00)
	(9, 9.00)
	(35, 35.00)
};
\nextgroupplot[xlabel={\drdd{} calls}, ylabel={\autodidact{} calls}, xmin=20.5, xmax=888808, ymin=20.5, ymax=888808]
\addplot[no marks, line width=0.4pt, black!45, dashed, forget plot] coordinates {
	(20.5, 20.50)
	(888808.0, 888808.00)
};
\addplot[only marks, mark=pentagon*, mark size=1.4pt, color=family-asn1, forget plot, mark options={line width=0.4pt}] coordinates {
	(455, 623.00)
};
\addplot[only marks, mark=*, mark size=1.4pt, color=family-c, forget plot, mark options={line width=0.4pt}] coordinates {
	(2679, 3032.00)
	(1859, 2126.00)
	(232803, 3023.00)
	(146979, 5658.00)
	(42493, 2447.00)
	(156700, 2533.00)
	(66093, 3029.00)
	(284208, 10302.00)
	(188661, 2690.00)
	(38223, 3110.00)
};
\addplot[only marks, mark=square, mark size=1.4pt, color=family-crashjs, forget plot, mark options={line width=0.4pt}] coordinates {
	(1966, 1782.00)
	(2246, 1441.00)
	(1062, 1434.00)
	(1185, 1371.00)
	(1637, 1535.00)
	(1045, 1233.00)
	(1694, 1521.00)
	(2000, 1768.00)
	(1185, 1182.00)
	(904, 1336.00)
};
\addplot[only marks, mark=diamond, mark size=1.4pt, color=family-graphviz, forget plot, mark options={line width=0.4pt}] coordinates {
	(137314, 57940.00)
	(1461, 1011.00)
};
\addplot[only marks, mark=triangle, mark size=1.4pt, color=family-lua, forget plot, mark options={line width=0.4pt}] coordinates {
	(1934, 1325.00)
	(1991, 2385.00)
};
\addplot[only marks, mark=diamond*, mark size=1.4pt, color=family-pdf, forget plot, mark options={line width=0.4pt}] coordinates {
	(688, 1419.00)
	(111, 672.00)
	(63, 1107.00)
	(701, 1576.00)
	(3929, 4733.00)
};
\addplot[only marks, mark=pentagon, mark size=1.4pt, color=family-python, forget plot, mark options={line width=0.4pt}] coordinates {
	(1073, 1113.00)
	(1856, 2373.00)
	(765, 1232.00)
	(935, 1052.00)
	(484, 815.00)
};
\addplot[only marks, mark=triangle*, mark size=1.4pt, color=family-smt, forget plot, mark options={line width=0.4pt}] coordinates {
	(7914, 8632.00)
	(11445, 9564.00)
	(38428, 31307.00)
	(287101, 185324.00)
	(78179, 30525.00)
	(444404, 144595.00)
	(158747, 97583.00)
	(17826, 2097.00)
	(14513, 2126.00)
	(11455, 1841.00)
};
\addplot[only marks, mark=o, mark size=1.4pt, color=family-sql, forget plot, mark options={line width=0.4pt}] coordinates {
	(751, 797.00)
	(1333, 1486.00)
	(1164, 2300.00)
	(310, 579.00)
	(387, 504.00)
	(218, 554.00)
	(119, 432.00)
	(417, 640.00)
	(117, 430.00)
	(419, 691.00)
};
\addplot[only marks, mark=square*, mark size=1.4pt, color=family-xml, forget plot, mark options={line width=0.4pt}] coordinates {
	(4110, 1322.00)
	(3896, 1580.00)
	(3491, 2308.00)
	(3919, 1571.00)
	(108336, 3887.00)
	(108244, 4221.00)
	(106657, 4168.00)
	(108471, 3993.00)
	(108616, 4065.00)
	(107771, 5230.00)
};
\addplot[only marks, mark=x, mark size=1.4pt, color=family-yaml, forget plot, mark options={line width=0.4pt}] coordinates {
	(41, 138.00)
	(87, 396.00)
	(407, 682.00)
	(421, 702.00)
	(49, 384.00)
	(92, 415.00)
	(272, 535.00)
	(44, 136.00)
	(41, 202.00)
	(189, 582.00)
};
\end{groupplot}
\end{tikzpicture}
	\caption{\autodidact{}'s advantage over \drdd{} grows with input size:
		points fall further below the diagonal
		as \drdd{}'s outputs and calls grow.
		On the \fixpointInputs{} returned-word inputs,
		\autodidact{}'s output is smaller on \fixpointSmaller{}
		and larger on \fixpointLarger{}.
		It makes fewer calls on \fixpointLargeCheaper{}
		of the \fixpointLargeInputs{} inputs of at least 10\,KiB,
		but on only \fixpointSmallCheaper{}
		of the \fixpointSmallInputs{} below 1\,KiB.
		Left: output size in non-whitespace bytes.
		Right: calls, including all inference.
		Axes are logarithmic.
		Points below the dashed diagonal, where the two are equal,
		favor \autodidact{}, and colors and marks distinguish families.}
	\Description{Two logarithmic scatter plots compare AutoDiDact
		with byte-level delta debugging,
		by output size and by oracle calls including inference,
		one point per input, colored and marked by family.}
	\label{fig:reduction}
\end{figure*}

\noindent\textbf{Input size.} The savings grow with input size (\Cref{fig:reduction}). On the \fixpointLargeInputs{} inputs of at least 10\,KiB, outputs are \fixpointLargeSizeSaving\% smaller and calls \fixpointLargeCallSaving\% fewer, with \fixpointLargeSmaller{} outputs smaller and \fixpointLargeCheaper{} runs cheaper. Between 1 and 10\,KiB, outputs are \fixpointMidSizeSaving\% smaller for \fixpointMidCallSaving\% fewer calls. Below 1\,KiB, outputs are \fixpointSmallSizeSaving\% smaller, and calls rise \fixpointSmallCallRatio{}$\times$, with fewer calls on \fixpointSmallCheaper{} of \fixpointSmallInputs{} inputs. In wall time, \autodidact{} takes \fixpointLargeTimeRatio{}$\times$ \drdd{}'s time at 10\,KiB and above and \fixpointTimeRatio{}$\times$ overall, and longer on smaller inputs.

\noindent\textbf{1-minimality over atoms.} Laminar reduction's output is 1-minimal over nodes, and \drdd{}'s over atoms. Running \drdd{}'s closing atom scan on \autodidact{}'s output makes it 1-minimal over atoms too: the scan shrinks \minimalShrunk{} of \minimalInputs{} outputs, leaves outputs \minimalSizeRatio{}$\times$ \drdd{}'s on geometric mean, and adds \minimalCost\% to the calls, which stay \minimalCallSaving\% fewer than \drdd{}'s.

\begin{result}
	Against \drdd{}, \autodidact{}'s outputs are \fixpointSizeSaving\% smaller with \fixpointCallSaving\% fewer calls, inference included, and \fixpointLargeSizeSaving\% smaller with \fixpointLargeCallSaving\% fewer calls on inputs of at least 10\,KiB. Below 1\,KiB, it makes \fixpointSmallCallRatio{}$\times$ \drdd{}'s calls.
\end{result}

\subsection{RQ2: Inferred Structure against Format Knowledge}\label{sec:rqKnowledge}

\begin{figure*}[tp]
	\centering
	\input{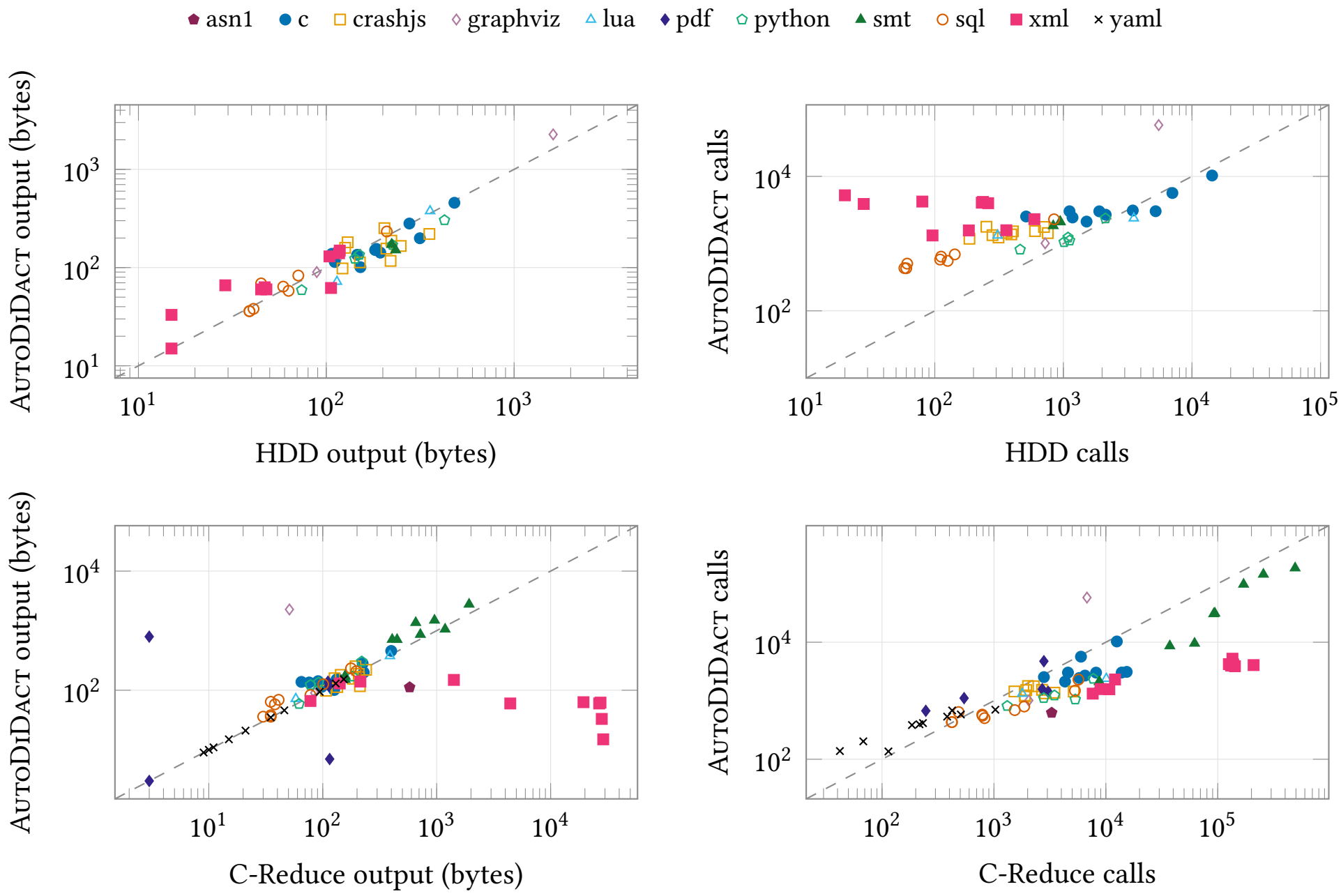}
	\caption{Without a grammar,
		\autodidact{} matches HDD's output sizes at more calls,
		and it needs fewer calls than C-Reduce.
		Top: against HDD on its \hddInputs{} inputs,
		size points straddle the diagonal, smaller on \hddSmaller{},
		and call points lie mostly above it, fewer on only \hddCheaper{}.
		Bottom: against C-Reduce's \texttt{-{}-not-c} mode
		on its \creduceInputs{} inputs,
		call points lie mostly below the diagonal, fewer on \creduceCheaper{},
		XML's size points lie far below it,
		and the other families' lie slightly above.
		Left: output size in non-whitespace bytes.
		Right: calls,
		\autodidact{}'s including all inference
		and HDD's and C-Reduce's excluding their validation queries.
		Axes are logarithmic.
		Points below the dashed diagonal, where the two are equal,
		favor \autodidact{}, and colors and marks distinguish families.}
	\Description{Four logarithmic scatter plots compare AutoDiDact
		with HDD and with C-Reduce,
		by output size and by oracle calls,
		one point per input, colored and marked by family.}
	\label{fig:comparison}
\end{figure*}

We compare \autodidact{} with HDD, which is given a grammar, and with C-Reduce's language-agnostic mode, which has a fixed list of bracket pairs and a C lexer built in (\Cref{tab:comparators} and \Cref{fig:comparison}).

\noindent\textbf{HDD.} On the \hddInputs{} inputs with a grammar, \autodidact{}'s outputs are \hddSizeRatio{}$\times$ HDD's on geometric mean, smaller on \hddSmaller{} inputs, with a median ratio of \hddMedianSizeRatio{}. \drdd{}'s outputs are \hddAtomSizeRatio{}$\times$ HDD's on the same inputs. \autodidact{} makes \hddCallRatio{}$\times$ HDD's calls, fewer on \hddCheaper{} inputs, and \hddReduceCallRatio{}$\times$ for reduction alone. On XML, \autodidact{} makes \hddXmlCallRatio{}$\times$ HDD's calls for outputs \hddXmlSizeRatio{}$\times$ as large. Without XML, its outputs are \hddNoXmlSizeRatio{}$\times$ HDD's at \hddNoXmlCallRatio{}$\times$ the calls, and without the \hddSyntaxInputs{} Python inputs that ANTLR parses only by recovering from syntax errors, \hddCleanSizeRatio{}$\times$ at \hddCleanCallRatio{}$\times$.

\noindent\textbf{C-Reduce.}\looseness=-1 Against C-Reduce's \texttt{-{}-not-c} mode, \autodidact{} uses \creduceCallSaving\% fewer calls, fewer on \creduceCheaper{} of \creduceInputs{}, or \creduceReduceCallRatio{}$\times$ C-Reduce's for reduction alone, and its outputs are \creduceSizeSaving\% smaller on geometric mean. On XML, C-Reduce leaves most large inputs at tens of kilobytes, and \autodidact{}'s are \creduceXmlSizeRatio{}$\times$ C-Reduce's. On other families, \autodidact{}'s outputs are \creduceNoXmlSizeRatio{}$\times$ C-Reduce's at \creduceNoXmlCallRatio{}$\times$ the calls. Overall, \autodidact{}'s are smaller on \creduceSmaller{}, equal on \creduceEqual{}, and larger on \creduceLarger{}, with a median ratio of \creduceMedianSizeRatio{}.

\noindent\textbf{Whitespace.} Counting whitespace, \autodidact{}'s outputs are \creduceRawSizeRatio{}$\times$ C-Reduce's and \hddRawSizeRatio{}$\times$ HDD's.

\begin{result}
	\autodidact{}'s outputs are \hddSizeRatio{}$\times$ HDD's at \hddCallRatio{}$\times$ HDD's calls, or \hddReduceCallRatio{}$\times$ excluding inference. Against C-Reduce's language-agnostic mode, it makes \creduceCallSaving\% fewer calls, and its outputs are \creduceSizeSaving\% smaller overall and \creduceNoXmlSizeRatio{}$\times$ C-Reduce's outside XML.
\end{result}

\subsection{RQ3: What Inference Recovers, at What Cost}\label{sec:rqRecovery}

We check the inferred tokens against real lexers and the pairs against each format's delimiters, and give each learner's cost.

\noindent\textbf{The neutral word.} Neutral-word discovery returns a single space on all \textInputs{} text-based inputs, across \textFormats{} families. It abstains on \abstained{} of the \binaryInputs{} binary inputs and returns a 10-byte sequence on one ASN.1 input. In a budget ablation on \budgetInputs{} inputs, every text input finds the space from \budgetSufficient{} calls on, and \budgetStable{} calls is the smallest budget that reproduces every main-run outcome.

Discovery spends nearly its whole budget whether it finds a word or abstains: \textDiscoveryCalls{} calls on text inputs and \binaryDiscoveryCalls{} on binary ones, taking \discoveryTime\,s, all on geometric mean. An abstention costs \abstentionLeast{}--\abstentionMost{} calls before \drdd{}, adding \abstentionPooled\% to the pooled calls of the abstaining inputs and \abstentionMean\% per input on geometric mean, and outweighing the reduction itself on \abstentionCostlier{} small inputs, by up to \abstentionWorst$\times$.

\begin{table}[tp]
	\caption{\autodidact{}'s inferred tokens
		against each family's standard lexer,
		on the \lexerInputs{} returned-word inputs with a lexer.
		\lexerWithin\% of inferred tokens lie within one lexical token}
	\label{tab:lexer}
	\tablestyle
	\begin{tabularx}{\linewidth}{|L|N|N|N|N|N|N|N|N|N|N|}
	\hline
	\multirow{2}{*}{\textbf{Family}} & \multirow{2}{*}{\textbf{Inputs}} & \multicolumn{2}{c|}{\textbf{Tokens}} & \multicolumn{4}{c|}{\textbf{Inferred tokens (\%)}} & \multicolumn{2}{c|}{\textbf{Crossing}} & \multirow{2}{*}{\textbf{Recall (\%)}} \\
	 & & \emph{Lexical} & \emph{Inferred} & \emph{Exact} & \emph{Space} & \emph{Quoted} & \emph{Other} & \emph{Merged} & \emph{Straddling} & \\
	\hline
	c & 10 & 287,893 & 320,842 & 88.3 & 9.3 & 0.6 & 1.8 & 13 & 0 & $>$99.9 \\
	crashjs & 10 & 2,526 & 4,342 & 51.5 & 9.3 & 23.6 & 15.4 & 11 & 0 & 99.6 \\
	graphviz & 2 & 25,489 & 131,760 & 13.2 & 39.1 & 46.3 & 1.4 & 3 & 10 & 99.9 \\
	lua & 2 & 1,026 & 1,341 & 70.6 & 18.7 & 1.2 & 9.5 & 0 & 0 & 100.0 \\
	python & 5 & 997 & 1,132 & 68.4 & 17.1 & 7.0 & 0.0 & 63 & 22 & 87.9 \\
	smt & 10 & 332,269 & 480,203 & 61.5 & $<$0.1 & $<$0.1 & 38.5 & 0 & 1 & $>$99.9 \\
	sql & 10 & 1,007 & 1,089 & 89.8 & 5.2 & 0.9 & 4.0 & 0 & 0 & 100.0 \\
	xml & 10 & 147,295 & 546,158 & 14.9 & 8.5 & 50.6 & 23.7 & 12,334 & 9 & 87.4 \\
	yaml & 10 & 191 & 214 & 54.7 & 4.7 & 23.4 & 5.1 & 24 & 2 & 80.7 \\
	\hline
	\textbf{All} & \textbf{69} & \textbf{798,693} & \textbf{1,487,081} & \textbf{45.9} & \textbf{8.7} & \textbf{22.9} & \textbf{21.7} & \textbf{12,448} & \textbf{44} & \textbf{97.7} \\
	\hline
\end{tabularx}

	\tablenote{\emph{Exact} is the share of a family's inferred tokens
		equal to a lexical token.
		\emph{Space}, \emph{Quoted}, and \emph{Other} are the shares
		strictly inside a lexical token
		that is a whitespace run, a quoted literal, or any other token.
		\emph{Merged} counts inferred tokens that join whole lexical tokens,
		and \emph{Straddling} those that cross a lexical boundary
		with an end inside a lexical token.
		\emph{Recall} is the share of the lexer's token boundaries
		that the inferred tokens recover.
		The lexers are the grammars-v4 ANTLR~4 lexers,
		except Python's \texttt{tokenize} module and PyYAML's scanner.
		Lexical tokens keep every byte: hidden tokens such as comments stay,
		and skipped text becomes
		whitespace runs, line breaks, and single characters.}
\end{table}

\noindent\textbf{Tokens.} On the \lexerInputs{} returned-word inputs from the \lexerFamilies{} families with a lexer (\Cref{tab:lexer}), \lexerWithin\% of inferred tokens lie within one lexical token, \lexerExact\% matching one exactly, and they recover \lexerRecall\% of the lexer's token boundaries. Of the inferred tokens strictly inside a lexical token, \lexerInsideQuoted\% lie in quoted literals, such as XML attribute values and Graphviz strings, and \lexerInsideSpace\% in whitespace runs. Most of the rest lie in SMT symbols and XML text. \lexerCrossings{} inferred tokens cross a lexical boundary, \lexerMergedShare\% of them joining whole lexical tokens and \lexerCrossShare\% in \lexerCrossFamilies{}.

\noindent\textbf{Tokenization cost.} Probing every internal gap alone makes up to one call per gap. With prediction and pooling, tokenization makes \probedGaps{} calls for \internalGaps{} gaps, \probedShare\% as many. This share falls with input size (\Cref{fig:probing}), from \smallShare\% below 1\,KiB to \largeShare\% at 10\,KiB and above. On the largest input, of \largestInput{} bytes, tokenization makes \largestTokenizeCalls{} calls, \largestTokenizeShare\% of its gaps.

\begin{figure}[tp]
	\centering
	\begin{minipage}[t]{0.48\linewidth}
		\centering
		\definecolor{stage-reduction}{HTML}{898781}
\definecolor{stage-discovery}{HTML}{1BAF7A}
\definecolor{stage-tokenization}{HTML}{2A78D6}
\definecolor{stage-pair-learning}{HTML}{EB6834}
\begin{tikzpicture}[baseline=(current axis.north)]
\begin{axis}[
	scale only axis,
	width=0.78\linewidth,
	height=0.68\linewidth,
	xmode=log,
	ymode=log,
	xmin=11,
	xmax=336226,
	ymin=0.67,
	ymax=336226,
	xlabel={input size (bytes)},
	ylabel={oracle calls},
	label style={font=\footnotesize},
	tick label style={font=\footnotesize},
	major grid style={line width=0.2pt, draw=black!12},
	axis line style={draw=black!45},
	tick style={draw=black!45},
	legend cell align=left,
	grid=major,
	legend columns=3,
	legend style={font=\scriptsize, draw=none, at={(0.5,1.03)}, anchor=south, nodes={text height=1.2ex, text depth=0.3ex}, /tikz/every even column/.append style={column sep=3pt}},
]
\addplot[no marks, line width=0.5pt, black!60, dashed] coordinates {
	(10.7, 9.67)
	(336226.5, 336225.50)
};
\addlegendentry{one call per gap}
\addplot[only marks, mark=diamond*, mark size=1.2pt, color=stage-reduction, mark options={line width=0.4pt}] coordinates {
	(16, 27.00)
	(16, 38.00)
	(20, 34.00)
	(36, 81.00)
	(44, 59.00)
	(51, 93.00)
	(53, 97.00)
	(63, 94.00)
	(86, 210.00)
	(104, 186.00)
	(108, 169.00)
	(128, 191.00)
	(144, 295.00)
	(148, 323.00)
	(163, 121.00)
	(177, 323.00)
	(179, 242.00)
	(212, 355.00)
	(265, 726.00)
	(277, 362.00)
	(383, 527.00)
	(483, 616.00)
	(503, 1776.00)
	(538, 942.00)
	(580, 210.00)
	(644, 483.00)
	(745, 820.00)
	(778, 696.00)
	(800, 802.00)
	(814, 626.00)
	(830, 694.00)
	(889, 1903.00)
	(903, 473.00)
	(1012, 998.00)
	(1055, 498.00)
	(1125, 1591.00)
	(1248, 441.00)
	(1340, 886.00)
	(2141, 483.00)
	(2435, 8149.00)
	(3071, 1128.00)
	(3546, 9045.00)
	(3623, 1967.00)
	(3835, 722.00)
	(3933, 1422.00)
	(3974, 451.00)
	(4052, 699.00)
	(4543, 106.00)
	(9773, 30782.00)
	(15421, 709.00)
	(15495, 749.00)
	(15505, 116.00)
	(15532, 3905.00)
	(16424, 2248.00)
	(23171, 1545.00)
	(43323, 1100.00)
	(43696, 1100.00)
	(43720, 2174.00)
	(43937, 795.00)
	(55923, 4181.00)
	(63509, 1644.00)
	(63795, 184753.00)
	(76263, 1592.00)
	(95337, 29649.00)
	(97658, 1900.00)
	(101759, 117.00)
	(103527, 557.00)
	(104930, 317.00)
	(106398, 645.00)
	(108560, 702.00)
	(109614, 190.00)
	(124593, 8943.00)
	(162928, 144025.00)
	(165583, 46196.00)
	(224151, 96722.00)
};
\addlegendentry{reduction}
\addplot[only marks, mark=square*, mark size=1.2pt, color=stage-discovery, mark options={line width=0.4pt}] coordinates {
	(16, 85.00)
	(16, 98.00)
	(20, 151.00)
	(36, 288.00)
	(44, 288.00)
	(51, 289.00)
	(53, 289.00)
	(63, 289.00)
	(86, 288.00)
	(104, 289.00)
	(108, 289.00)
	(128, 289.00)
	(144, 289.00)
	(148, 288.00)
	(163, 289.00)
	(177, 290.00)
	(179, 289.00)
	(212, 288.00)
	(265, 288.00)
	(277, 289.00)
	(383, 288.00)
	(483, 288.00)
	(503, 289.00)
	(538, 288.00)
	(580, 287.00)
	(644, 289.00)
	(745, 289.00)
	(778, 289.00)
	(800, 289.00)
	(814, 289.00)
	(830, 289.00)
	(889, 288.00)
	(903, 289.00)
	(1012, 289.00)
	(1055, 289.00)
	(1125, 288.00)
	(1248, 289.00)
	(1340, 289.00)
	(2141, 289.00)
	(2435, 289.00)
	(3071, 290.00)
	(3546, 289.00)
	(3623, 288.00)
	(3835, 289.00)
	(3933, 289.00)
	(3974, 289.00)
	(4052, 289.00)
	(4543, 291.00)
	(9773, 289.00)
	(15421, 289.00)
	(15495, 291.00)
	(15505, 289.00)
	(15532, 289.00)
	(16424, 289.00)
	(23171, 289.00)
	(43323, 289.00)
	(43696, 289.00)
	(43720, 289.00)
	(43937, 289.00)
	(55923, 288.00)
	(63509, 289.00)
	(63795, 291.00)
	(76263, 289.00)
	(95337, 289.00)
	(97658, 289.00)
	(101759, 289.00)
	(103527, 289.00)
	(104930, 289.00)
	(106398, 288.00)
	(108560, 289.00)
	(109614, 289.00)
	(124593, 288.00)
	(162928, 289.00)
	(165583, 289.00)
	(224151, 289.00)
};
\addlegendentry{discovery}
\addplot[only marks, mark=*, mark size=1.2pt, color=stage-tokenization, mark options={line width=0.4pt}] coordinates {
	(16, 11.00)
	(16, 13.00)
	(20, 17.00)
	(36, 17.00)
	(44, 37.00)
	(51, 39.00)
	(53, 35.00)
	(63, 32.00)
	(86, 23.00)
	(104, 59.00)
	(108, 50.00)
	(128, 74.00)
	(144, 83.00)
	(148, 82.00)
	(163, 74.00)
	(177, 60.00)
	(179, 95.00)
	(212, 109.00)
	(265, 129.00)
	(277, 98.00)
	(383, 194.00)
	(483, 174.00)
	(503, 150.00)
	(538, 155.00)
	(580, 125.00)
	(644, 341.00)
	(745, 342.00)
	(778, 371.00)
	(800, 325.00)
	(814, 247.00)
	(830, 314.00)
	(889, 96.00)
	(903, 425.00)
	(1012, 332.00)
	(1055, 473.00)
	(1125, 350.00)
	(1248, 444.00)
	(1340, 457.00)
	(2141, 120.00)
	(2435, 161.00)
	(3071, 514.00)
	(3546, 190.00)
	(3623, 606.00)
	(3835, 505.00)
	(3933, 533.00)
	(3974, 505.00)
	(4052, 501.00)
	(4543, 480.00)
	(9773, 220.00)
	(15421, 226.00)
	(15495, 344.00)
	(15505, 12.00)
	(15532, 313.00)
	(16424, 455.00)
	(23171, 494.00)
	(43323, 659.00)
	(43696, 657.00)
	(43720, 453.00)
	(43937, 679.00)
	(55923, 1072.00)
	(63509, 518.00)
	(63795, 265.00)
	(76263, 716.00)
	(95337, 445.00)
	(97658, 733.00)
	(101759, 3452.00)
	(103527, 3282.00)
	(104930, 3352.00)
	(106398, 3095.00)
	(108560, 3198.00)
	(109614, 4727.00)
	(124593, 988.00)
	(162928, 270.00)
	(165583, 11297.00)
	(224151, 454.00)
};
\addlegendentry{tokenization}
\addplot[only marks, mark=triangle*, mark size=1.2pt, color=stage-pair-learning, mark options={line width=0.4pt}] coordinates {
	(16, 2.00)
	(36, 10.00)
	(51, 9.00)
	(53, 11.00)
	(86, 61.00)
	(104, 20.00)
	(108, 27.00)
	(128, 25.00)
	(144, 24.00)
	(148, 9.00)
	(163, 20.00)
	(177, 9.00)
	(179, 14.00)
	(212, 63.00)
	(265, 89.00)
	(277, 48.00)
	(383, 43.00)
	(483, 35.00)
	(503, 85.00)
	(538, 101.00)
	(580, 1.00)
	(644, 69.00)
	(745, 84.00)
	(778, 78.00)
	(800, 105.00)
	(814, 71.00)
	(830, 144.00)
	(889, 98.00)
	(903, 149.00)
	(1012, 163.00)
	(1055, 111.00)
	(1125, 144.00)
	(1248, 151.00)
	(1340, 136.00)
	(2141, 119.00)
	(2435, 33.00)
	(3071, 194.00)
	(3546, 40.00)
	(3623, 171.00)
	(3835, 64.00)
	(3933, 64.00)
	(3974, 77.00)
	(4052, 82.00)
	(4543, 230.00)
	(9773, 16.00)
	(15421, 195.00)
	(15495, 192.00)
	(15505, 255.00)
	(15532, 226.00)
	(16424, 118.00)
	(23171, 119.00)
	(43323, 78.00)
	(43696, 51.00)
	(43720, 113.00)
	(43937, 78.00)
	(55923, 117.00)
	(63509, 82.00)
	(63795, 15.00)
	(76263, 93.00)
	(95337, 142.00)
	(97658, 101.00)
	(101759, 29.00)
	(103527, 40.00)
	(104930, 35.00)
	(106398, 37.00)
	(108560, 32.00)
	(109614, 24.00)
	(124593, 83.00)
	(162928, 11.00)
	(165583, 158.00)
	(224151, 118.00)
};
\addlegendentry{pair learning}
\end{axis}
\end{tikzpicture}
		\caption{Inference's share of calls falls with input size,
			from \inferenceSmallShare\% below 1\,KiB
			to \inferenceLargeShare\% at 10\,KiB and above.
			Discovery and pair learning make nearly constant calls,
			and tokenization's grow roughly with the square root of size.
			Reduction includes shrinking.
			The dashed line marks one call per gap.}
		\Description{A logarithmic scatter plot of oracle calls
			against input size for each stage:
			reduction, discovery, tokenization, and pair learning,
			with a dashed line at one call per internal gap.}
		\label{fig:costs}
	\end{minipage}\hfill
	\begin{minipage}[t]{0.48\linewidth}
		\centering
		\definecolor{family-asn1}{HTML}{882255}
\definecolor{family-c}{HTML}{0072B2}
\definecolor{family-crashjs}{HTML}{E69F00}
\definecolor{family-graphviz}{HTML}{B07AA1}
\definecolor{family-lua}{HTML}{33BBEE}
\definecolor{family-pdf}{HTML}{332288}
\definecolor{family-python}{HTML}{1BAF7A}
\definecolor{family-smt}{HTML}{117733}
\definecolor{family-sql}{HTML}{D55E00}
\definecolor{family-xml}{HTML}{EE3377}
\definecolor{family-yaml}{HTML}{000000}
\begin{tikzpicture}[baseline=(current axis.north)]
\begin{axis}[
	scale only axis,
	width=0.78\linewidth,
	height=0.68\linewidth,
	xmode=log,
	log basis x=10,
	ymin=0,
	ymax=100,
	xlabel={input size (bytes)},
	ylabel={calls per internal gap (\%)},
	label style={font=\footnotesize},
	tick label style={font=\footnotesize},
	major grid style={line width=0.2pt, draw=black!12},
	axis line style={draw=black!45},
	tick style={draw=black!45},
	legend cell align=left,
	ymajorgrids,
	legend columns=4,
	legend style={font=\scriptsize, draw=none, at={(0.5,1.03)}, anchor=south, nodes={text height=1.2ex, text depth=0.3ex}, /tikz/every even column/.append style={column sep=3pt}},
]
\addplot[no marks, line width=0.6pt, black!55, forget plot] coordinates {
	(16.0, 84.88)
	(18.6, 83.43)
	(21.7, 81.87)
	(25.2, 80.20)
	(29.3, 78.41)
	(34.1, 76.51)
	(39.7, 74.50)
	(46.2, 72.38)
	(53.8, 70.16)
	(62.6, 67.83)
	(72.8, 65.41)
	(84.7, 62.91)
	(98.6, 60.34)
	(114.7, 57.71)
	(133.5, 55.04)
	(155.4, 52.33)
	(180.8, 49.62)
	(210.4, 46.90)
	(244.8, 44.20)
	(284.9, 41.54)
	(331.5, 38.93)
	(385.7, 36.37)
	(448.8, 33.90)
	(522.3, 31.50)
	(607.7, 29.21)
	(707.2, 27.01)
	(822.9, 24.92)
	(957.5, 22.94)
	(1114.2, 21.07)
	(1296.6, 19.32)
	(1508.7, 17.68)
	(1755.6, 16.15)
	(2042.9, 14.74)
	(2377.1, 13.42)
	(2766.1, 12.21)
	(3218.7, 11.09)
	(3745.4, 10.06)
	(4358.3, 9.12)
	(5071.5, 8.26)
	(5901.4, 7.47)
	(6867.0, 6.75)
	(7990.7, 6.10)
	(9298.2, 5.50)
	(10819.7, 4.96)
	(12590.2, 4.48)
	(14650.4, 4.03)
	(17047.7, 3.63)
	(19837.2, 3.27)
	(23083.3, 2.94)
	(26860.5, 2.65)
	(31255.8, 2.38)
	(36370.2, 2.14)
	(42321.6, 1.93)
	(49246.9, 1.73)
	(57305.3, 1.55)
	(66682.4, 1.40)
	(77593.8, 1.25)
	(90290.8, 1.13)
	(105065.4, 1.01)
	(122257.6, 0.91)
	(142263.0, 0.82)
	(165542.0, 0.73)
	(192630.2, 0.66)
	(224151.0, 0.59)
};
\addplot[only marks, mark=pentagon*, mark size=1.4pt, color=family-asn1, mark options={line width=0.4pt}] coordinates {
	(580, 21.59)
};
\addlegendentry{asn1}
\addplot[only marks, mark=*, mark size=1.4pt, color=family-c, mark options={line width=0.4pt}] coordinates {
	(3071, 16.74)
	(3623, 16.73)
	(16424, 2.77)
	(23171, 2.13)
	(43720, 1.04)
	(55923, 1.92)
	(63509, 0.82)
	(76263, 0.94)
	(97658, 0.75)
	(124593, 0.79)
};
\addlegendentry{c}
\addplot[only marks, mark=square, mark size=1.4pt, color=family-crashjs, mark options={line width=0.4pt}] coordinates {
	(644, 53.03)
	(745, 45.97)
	(778, 47.75)
	(800, 40.68)
	(814, 30.38)
	(830, 37.88)
	(903, 47.12)
	(1012, 32.84)
	(1055, 44.88)
	(1340, 34.13)
};
\addlegendentry{crashjs}
\addplot[only marks, mark=diamond, mark size=1.4pt, color=family-graphviz, mark options={line width=0.4pt}] coordinates {
	(2141, 5.61)
	(165583, 6.82)
};
\addlegendentry{graphviz}
\addplot[only marks, mark=triangle, mark size=1.4pt, color=family-lua, mark options={line width=0.4pt}] coordinates {
	(889, 10.81)
	(1248, 35.61)
};
\addlegendentry{lua}
\addplot[only marks, mark=diamond*, mark size=1.4pt, color=family-pdf, mark options={line width=0.4pt}] coordinates {
	(4543, 10.57)
	(15421, 1.47)
	(15495, 2.22)
	(15505, 0.08)
	(15532, 2.02)
};
\addlegendentry{pdf}
\addplot[only marks, mark=pentagon, mark size=1.4pt, color=family-python, mark options={line width=0.4pt}] coordinates {
	(212, 51.66)
	(265, 48.86)
	(383, 50.79)
	(483, 36.10)
	(1125, 31.14)
};
\addlegendentry{python}
\addplot[only marks, mark=triangle*, mark size=1.4pt, color=family-smt, mark options={line width=0.4pt}] coordinates {
	(2435, 6.61)
	(3546, 5.36)
	(9773, 2.25)
	(43323, 1.52)
	(43696, 1.50)
	(43937, 1.55)
	(63795, 0.42)
	(95337, 0.47)
	(162928, 0.17)
	(224151, 0.20)
};
\addlegendentry{smt}
\addplot[only marks, mark=o, mark size=1.4pt, color=family-sql, mark options={line width=0.4pt}] coordinates {
	(51, 78.00)
	(53, 67.31)
	(104, 57.28)
	(128, 58.27)
	(144, 58.04)
	(163, 45.68)
	(179, 53.37)
	(277, 35.51)
	(503, 29.88)
	(538, 28.86)
};
\addlegendentry{sql}
\addplot[only marks, mark=square*, mark size=1.4pt, color=family-xml, mark options={line width=0.4pt}] coordinates {
	(3835, 13.17)
	(3933, 13.56)
	(3974, 12.71)
	(4052, 12.37)
	(101759, 3.39)
	(103527, 3.17)
	(104930, 3.19)
	(106398, 2.91)
	(108560, 2.95)
	(109614, 4.31)
};
\addlegendentry{xml}
\addplot[only marks, mark=x, mark size=1.4pt, color=family-yaml, mark options={line width=0.4pt}] coordinates {
	(16, 73.33)
	(16, 86.67)
	(20, 89.47)
	(36, 48.57)
	(44, 86.05)
	(63, 51.61)
	(86, 27.06)
	(108, 46.73)
	(148, 55.78)
	(177, 34.09)
};
\addlegendentry{yaml}
\end{axis}
\end{tikzpicture}
		\caption{Tokenization's calls per gap fall sharply with size,
			becoming negligible for large inputs.
			Calls fall from \smallShare\% of gaps below 1\,KiB
			to \largeShare\% at 10\,KiB and above.
			The plot gives these calls as a percentage of internal gaps
			on each of the \worded{} returned-word inputs,
			against input size,
			with a logistic fit in log size.}
		\Description{A scatter plot of tokenization calls per internal gap
			against input size,
			one point per input colored and marked by family,
			with a fitted curve falling toward zero.}
		\label{fig:probing}
	\end{minipage}
\end{figure}

\begin{table}[tp]
	\caption{The default tokenizer against exhaustive probing
		on \controlInputs{} selected inputs.
		It makes \controlCallShare\% of exhaustive probing's calls
		and agrees with it at \controlAgreement\% of gaps}
	\label{tab:prediction}
	\tablestyle
	\begin{tabularx}{\linewidth}{|l|L|N|N|N|N|N|N|N|}
	\hline
	\textbf{Family} & \textbf{Case} & \textbf{Gaps} & \textbf{Predicted (\%)} & \textbf{Calls} & \textbf{Exhaustive} & \textbf{Missed} & \textbf{Extra} & \textbf{Agreement (\%)} \\
	\hline
	asn1 & \case{asn1\ldots{}055-nested} & 579 & 72.9 & 125 & 529 & 20 & 0 & 96.55 \\
	c & \case{10} & 3,070 & 69.3 & 514 & 2,582 & 8 & 44 & 98.31 \\
	c & \case{simp\ldots{}c\_4922\_43K} & 43,719 & 98.8 & 453 & 36,099 & 0 & 40 & 99.91 \\
	crashjs & \case{java\ldots{}orithms-14} & 813 & 65.7 & 247 & 750 & 3 & 0 & 99.63 \\
	crashjs & \case{express-1} & 643 & 34.1 & 341 & 579 & 0 & 0 & 100.00 \\
	graphviz & \case{issue-2723} & 2,140 & 91.2 & 120 & 1,753 & 1 & 47 & 97.76 \\
	lua & \case{luajit-913} & 1,247 & 58.5 & 444 & 1,124 & 1 & 2 & 99.76 \\
	lua & \case{luajit-1116} & 888 & 83.9 & 96 & 602 & 0 & 0 & 100.00 \\
	pdf & \case{pdf-fuzzer-jch207} & 15,504 & 99.1 & 12 & 14,885 & 0 & 80 & 99.48 \\
	pdf & \case{pdfi\ldots{}jch207-kw6} & 4,542 & 74.2 & 480 & 4,079 & 0 & 24 & 99.47 \\
	python & \case{alloc} & 1,124 & 55.2 & 350 & 791 & 1 & 8 & 99.20 \\
	python & \case{recursion} & 211 & 28.9 & 109 & 144 & 0 & 12 & 94.31 \\
	smt & \case{btor\ldots{}16966-2630} & 2,434 & 92.5 & 161 & 2,113 & 0 & 0 & 100.00 \\
	smt & \case{btor\ldots{}-19025-698} & 63,794 & 99.5 & 265 & 56,083 & 0 & 0 & 100.00 \\
	sql & \case{mysql-112460} & 537 & 65.9 & 155 & 420 & 0 & 0 & 100.00 \\
	sql & \case{sqlf\ldots{}sqlite-808} & 50 & 4.0 & 39 & 45 & 0 & 0 & 100.00 \\
	xml & \case{4.1} & 3,932 & 72.4 & 533 & 3,731 & 0 & 0 & 100.00 \\
	xml & \case{8ede045-1} & 106,397 & 93.3 & 3,095 & 97,250 & 12 & 1 & 99.99 \\
	yaml & \case{6FWR} & 19 & 5.3 & 17 & 14 & 0 & 0 & 100.00 \\
	yaml & \case{NAT4} & 85 & 58.8 & 23 & 57 & 0 & 0 & 100.00 \\
	\hline
	\textbf{All} &  & \textbf{251,728} & \textbf{94.4} & \textbf{7,579} & \textbf{223,630} & \textbf{46} & \textbf{258} & \textbf{99.88} \\
	\hline
\end{tabularx}

	\tablenote{\emph{Gaps} counts internal offsets,
		and \emph{Predicted} is the share of them
		whose verdict the default tokenizer predicted without a query.
		\emph{Calls} and \emph{Exhaustive} count the oracle calls
		of the default tokenizer and of probing every gap alone.
		\emph{Missed} and \emph{Extra} count the boundaries
		the default tokenizer omits or adds relative to exhaustive probing,
		over all gaps,
		and \emph{Agreement} is the share of gaps on which both agree.
		The total pools gaps.}
\end{table}

\noindent\textbf{Prediction accuracy.} On \controlInputs{} inputs from \controlFamilies{} families, we also probed every gap alone with the same word (\Cref{tab:prediction}). In each family, we chose the inputs with the smallest and the largest share of calls among those estimated to take under an hour to probe exhaustively. Where only one qualified, we used it. On these inputs, the default tokenizer makes \controlCalls{} calls where exhaustive probing makes \controlExhaustive{}, \controlCallShare\% as many, and the two agree at \controlAgreement\% of the \controlGaps{} gaps, from \controlLowest\% to \controlHighest\% per input, and on every gap of the \controlErrorless{} inputs. Of the \controlDisagreements{} disagreements, \controlMispredicted{} are mispredictions, and the other \controlPooled{} lie at gaps settled by pooling. Most are extra boundaries, which exhaustive probing rejects: in a C preprocessor directive, between the adjacent minus signs of C decrements, in Graphviz's numeric identifiers, in Python indentation, and in PDF keywords.

\noindent\textbf{Pairs.} Pair learning admits \pairAdmitted{} pairs on \pairLearned{} of the \pairInputs{} inputs, nearly all the format's own delimiters. Apart from tags, \pairDelimiters{} of the \pairUntagged{} pairs are brackets and quotes: parentheses on every C, Python, SMT, and SQL input, braces, brackets, and quotes in C, JavaScript, and Graphviz, and quotes in Lua and Python. On XML, affix families recover \pairTags{} start- and end-tag pairs. The conditional operator \<?>\,\ldots\,\<:> is admitted on \pairConditional{} C inputs, and the \pairOther{} others are not delimiters of the format. Where no pair is admitted, as on ASN.1, PDF, and most YAML inputs, reduction proceeds over flat tokens.

Pair learning is the cheapest stage. It makes \pairCalls{} calls over all \worded{} inputs, a median of \pairMedian{} per input and at most \pairMost{}, and takes a median of \pairTime\,s per input.

\noindent\textbf{Where the calls go.} Inference makes \inferenceShare\% of \autodidact{}'s calls: tokenization \tokensShare\%, discovery \wordShare\%, and pair learning \pairsShare\%. Its share falls with input size (\Cref{fig:costs}), from \inferenceSmallShare\% below 1\,KiB to \inferenceLargeShare\% at 10\,KiB and above. Discovery's and pair learning's calls stay nearly constant, and tokenization's grow roughly with the square root of input size.

\begin{result}
	Discovery finds a space on every text input. Inferred tokens recover \lexerRecall\% of the lexer's token boundaries, and pair learning admits the formats' brackets, quotes, and XML tags. Inference makes \inferenceShare\% of \autodidact{}'s calls, and \inferenceLargeShare\% at 10\,KiB and above.
\end{result}

\section{Discussion}\label{sec:discussion}

\noindent\textbf{Competitive without a grammar.} \autodidact{} receives only what \drdd{} receives, a failing input and its oracle. From these alone, it returns outputs \fixpointSizeSaving\% smaller than \drdd{}'s with \fixpointCallSaving\% fewer calls, and this gain makes it competitive with grammar-aware reducers. On the inputs with a grammar, \drdd{}'s outputs are \hddAtomSizeRatio{}$\times$ HDD's, and \autodidact{}'s are \hddSizeRatio{}$\times$. Against C-Reduce, which builds in a C lexer and a list of bracket pairs, \autodidact{} makes \creduceCallSaving\% fewer calls for outputs \creduceSizeSaving\% smaller. The inferred structure is small: for the GCC input of \Cref{sec:introduction}, the word and pairs take \structureBytes{} bytes, against \grammarBytes{} bytes for the C grammar HDD needs. The price is calls against HDD, \hddCallRatio{}$\times$ HDD's in all and \hddReduceCallRatio{}$\times$ for reduction alone, so learning the structure accounts for most of the difference. The grammar helps HDD most on XML, whose tokens \autodidact{} must find by probing.

\noindent\textbf{Tolerating errors.} A grammar-aware reducer needs a grammar for the format and a parse of the input. HDD ran on only \hddInputs{} of the \fixpointInputs{} inputs, since no grammar was at hand for ASN.1, PDF, YAML, or the other SQL and SMT dialects, and ANTLR parsed \hddSyntaxInputs{} Python inputs only by recovering from syntax errors. Failing inputs are often malformed, and \autodidact{} parses nothing. A stray delimiter changes the span count by at most one (\Cref{sec:reduction}). \autodidact{} thus ran on all \fixpointInputs{} returned-word inputs, including a binary ASN.1 input. In this respect, inferred structure goes beyond a grammar.

\noindent\textbf{A minimizer for any fuzzer.} Fuzzers minimize the crashing inputs they find at the byte level, as \texttt{afl-tmin}~\cite{zalewski2014afl} and libFuzzer's \texttt{-minimize\_crash}~\cite{libfuzzer} do, because they cannot assume the target's input format. With inferred structure, a fuzzer can carry a general minimizer that works on any input. It needs nothing beyond the oracle the fuzzer already has, and where discovery abstains, it falls back to bytes after at most 300 calls.

\noindent\textbf{Inferred against built-in knowledge.} Because the structure comes from the oracle, it can also exceed a fixed description of the format: C-Reduce's bracket list cannot pair XML tags, but affix families learn them. On the other families, C-Reduce also makes changes that no deletion can, through integer reduction and peephole edits, and inferred tokens could guide such edits as they guide shrinking. Ignoring whitespace in these comparisons is conservative, since counting it favors \autodidact{} (\Cref{sec:rqKnowledge}).

\noindent\textbf{Why the savings over \drdd{} grow with size.} On large inputs, \drdd{} stops at large 1-minimal outputs, such as the GCC input's \showcaseAtomSize{} bytes. Deleting or unwrapping whole spans gets past them. Below 1\,KiB, \drdd{} needs only \smallAtomCalls{} calls on geometric mean, not much more than discovery's fixed budget, so inference rarely pays for itself. Inference queries also run on the whole input, so inference weighs more in wall time than in calls. With the closing atom scan, \autodidact{}'s calls stay below \drdd{}'s, so the comparison also holds when both outputs are 1-minimal over atoms.

\noindent\textbf{What the oracle reveals.} Inference sees only the structure that the failure makes visible. Inferred tokens split a lexical token where the failure does not need it whole, as \Cref{app:lexer} predicts: a whitespace run accepts the neutral word anywhere, and a comment may split into single bytes because insertions there do not matter. They cross lexical tokens where a lexer places boundaries that the neutral word cannot reach, such as after \<<> in an XML tag or before indentation in Python and YAML. Coupling likewise follows the failure: the conditional operator \<?>\,\ldots\,\<:> is admitted because the compiler accepts its halves only together. A genuine pair may yield no witness when every paired edit destroys the failure. An abstention likewise leaves open whether useful structure exists. Such partial structure is safe to use, since the oracle checks every edit during reduction: it guides reduction without having to be a correct lexer or parser. Where the evidence is insufficient, discovery abstains at bounded cost, and reduction falls back to bytes.

\noindent\textbf{Where inference is cheap.} Tokenization's share of calls falls with input size, consistent with larger inputs repeating more contexts. Pair learning would stay a small fraction of all calls even at ten times its cost, which leaves room for a learner that tests more hypotheses or other relations. The remaining overheads are discovery's fixed budget, which small inputs cannot repay, and tokenizing inputs with little repetition, such as the large XML inputs. Both belong to the current learners, which are one way to infer the targets of \Cref{sec:core}. Other learners could do better: a word learned on one input of a format could serve the next, and better prediction would cut tokenization further. How cheaply an oracle can reveal structure remains open.

\section{Related Work}\label{sec:relatedWork}

\noindent\textbf{Delta debugging.} \ddmin{}~\cite{zeller2002simplifying} tests subsets and complements of supplied units, \drdd{}~\cite{kumbhakern2026drdd} changes its restart schedule while keeping 1-minimality, and ProbDD~\cite{wang2021probabilistic} chooses deletions with a probabilistic model. Our inference supplies units and nesting to any such reducer.

\noindent\textbf{Reduction with format knowledge.} HDD~\cite{misherghi2006hdd} reduces a parse tree level by level, as Picireny~\cite{hodovan2016modernizing,picireny} does for ANTLR grammars, with hoisting~\cite{vince2021hoisting}. Perses~\cite{sun2018perses} uses grammar rules for syntax-preserving deletions and replacements, GTR~\cite{herfert2017gtr} learns tree transformations from a corpus, and T-Rec~\cite{xu2025trec} reduces within tokens using a language's lexical syntax. C-Reduce~\cite{regehr2012test} deletes or unwraps brackets from a fixed list among other passes, and Shrink Ray~\cite{shrinkray} pairs generic passes with format-specific ones. Each obtains units and hierarchy from a grammar, a lexer, a corpus, or built-in knowledge. We obtain them from the failing input and its oracle.

\noindent\textbf{LLM-assisted reduction.} LPR~\cite{zhang2024lpr} alternates a generic reducer with transformations that a large language model proposes, and Yang et al.~\cite{yang2025inputreduction} prompt a model to write a reducer for each failing input. Model queries add cost to every reduction, and the cost grows when a fuzzing campaign yields thousands of inputs. With \autodidact{} reducing them, a model's effort can go to selecting which inputs to reduce and which failures to investigate.

\noindent\textbf{Grammar inference.} GLADE~\cite{bastani2017glade} generalizes seed inputs with membership queries, Arvada~\cite{kulkarni2021arvada} and TreeVada~\cite{arefin2024treevada} merge nodes of trees over example inputs, Kedavra~\cite{li2024kedavra} infers grammars incrementally, and Crucio~\cite{li2026crucio} scales inference to C, confirming token boundaries by inserting a space after checking that the parser ignores spaces. V-Star~\cite{jia2024vstar} is closest to pair learning: it infers call and return tokens from nesting patterns and learns a visibly pushdown grammar. Our coupling tests share its coordinated edits and rejected controls without requiring a visibly pushdown language. These learners assume a parser oracle and positive examples and seek a grammar of the whole language. We have one failing input and a failure oracle, which tests failure preservation and may accept malformed inputs. We discover the separator, learn only the relations reduction uses, abstain when evidence is weak, and check every edit.

\section{Threats to Validity}\label{sec:threats}

\noindent\textbf{Internal validity.} All reducers use the same oracle for each input, and original and final inputs are checked, though a shared symptom need not identify the same fault. One run per input and reducer cannot separate nondeterminism from the methods' effects. Parameters are fixed across families but were chosen on these inputs, and the budget ablation covers selected inputs and one parameter.

\noindent\textbf{External validity.} The \inputs{} inputs cover \formats{} families, at most ten per family, largely from existing reproducer collections, some already reduced. They are not independent samples of defects, and other inputs may behave differently. Every text input yielded a space and one binary input another word, so evidence for other words and for binary formats is limited. The comparisons cover one configuration each of HDD and C-Reduce, on cohorts set by discovery and the available grammars, and do not isolate the effects of nesting, hoisting, unwrapping, shrinking, and traversal order.

\noindent\textbf{Construct validity.} Agreement with a lexer depends on its conventions for hidden tokens and skipped text, and agreement with exhaustive probing measures fidelity to probing on inputs chosen for affordable probing. Coupling witnesses establish only the contrasts tested. Oracle calls have unequal costs, and wall times come from one machine without repeated trials, so we report both. Byte counts may differ from debugging effort, and ignoring whitespace discounts it in Python and YAML. Per-input results, medians, and win counts expose variation that geometric means hide.

\section{Conclusion}\label{sec:conclusion}

The most widely used fuzzers, such as AFL~\cite{zalewski2014afl,fioraldi2020aflpp} and libFuzzer~\cite{libfuzzer}, know nothing of their targets' input formats. Yet they must reduce what they find, from corpus seeds to crashing inputs, and their minimizers have so far worked on bytes. This paper removes that limit. A neutral word and delimiter pairs, inferred from an input and its oracle alone, give structured reduction without a grammar. Reduction over them returns outputs \fixpointSizeSaving\% smaller than byte-level \drdd{}'s with \fixpointCallSaving\% fewer calls, matches grammar-based HDD in output size, and uses \creduceCallSaving\% fewer calls than C-Reduce's \texttt{-{}-not-c} mode. Because \autodidact{} parses nothing and checks every edit, it accepts whatever a fuzzer finds, malformed inputs included, with any property the fuzzer can check, from a crash to a seed's coverage. Grammar-agnostic fuzzers can now carry a structure-aware minimizer, for smaller seeds that are cheaper to run and smaller crash inputs that are quicker to diagnose.

\section*{AI Use}

We used Codex (OpenAI) and Claude Code (Anthropic) primarily to acquire and build benchmark targets, implement and validate oracle wrappers, and write experiment and analysis scripts, including table and figure generators. We also used them for localized implementation cleanup, documentation, debugging, and validation tests, to search for related literature, and to check the paper's algorithms, arguments, and numerical claims against the implementation and saved results.

\section*{Data-Availability Statement}

A replication package and prebuilt image are available at \href{https://doi.org/10.5281/zenodo.24164465}{doi:10.5281/zenodo.24164465}. The package contains the implementation source, benchmark inputs and oracles, experiment and analysis scripts, and saved results.

\bibliographystyle{ACM-Reference-Format}
\bibliography{refs}

\clearpage
\appendix
\counterwithin{table}{section} \counterwithin{figure}{section} \counterwithin{algorithm}{section} \counterwithin{equation}{section}

\section{Structural Arguments}\label{app:structure}

The following arguments support \Cref{sec:core}; \Cref{app:reduction,app:details,app:results} give reduction algorithms, inference details, and supplementary results.

\subsection{Tokens under a Lexer}\label{app:lexer}

Let $\test=F\circ L$ and let $\Lambda(S)$ be the internal lexical boundaries. Assume that $L$ erases arbitrary copies of $w$ inserted at any subset of $\Lambda(S)$, while inserting either control, $\kappa$ or $\kappa^{|w|}$, there is rejected. Erasure preserves $L(S)$, so every lexical boundary is selective and, when $\Lambda(S)\ne\varnothing$, $w$ is neutral with $G=\Lambda(S)$. By \eqref{eq:boundaries},
\begin{equation}\label{eq:failure-relative}
	B_w(S)=\Lambda(S)\cup\{\,g\notin\Lambda(S) : F(L(S\oplus_g w))=\ACCEPT\,\}.
\end{equation}
Exhaustive probing therefore refines the lexical partition; prediction and pooling need not. Erasure ensures composition at lexical boundaries, but not at extra boundaries within tokens, where insertions can interact.

\subsection{Delimiter Pairs under a Parser}\label{app:parser}

Assume that $S$ is matched for $(\opn,\cls)$ and the oracle rejects every unmatched input. The paired edits of \eqref{eq:coupling}, span deletion, and unwrapping preserve matchedness: for distinct delimiters, each replacement and the original span have equal opening and closing counts and no prefix with more closes than opens, preserving matchedness in the walk of \eqref{eq:matched}. For symmetric delimiters, each edit changes the count by an even number. Each control instead changes the total delimiter parity and is rejected. Coupling therefore reduces to acceptance of a paired edit; matchedness alone does not ensure that acceptance.

\section{Reduction Algorithms and Guarantees}\label{app:reduction}

Assume a deterministic Boolean oracle. Queries count requests before caching and hence also bound oracle executions.

\FloatBarrier
\subsection{Matching and Tree Construction}\label{app:matching}

\noindent\textbf{Stability.} For fixed admitted pairs, let $m(S)$ be the largest matching size. Deleting delimiter $i$ removes at most one pair from a largest matching; every matching of the shorter sequence is also valid before deletion. Hence $m(S)-1\le m(S\setminus\{i\})\le m(S)$. Insertion reverses this argument, so $k$ inserted or deleted delimiters change the size by at most $k$. Admitting another pair enlarges the feasible set and cannot decrease $m(S)$. These bounds concern span counts, not identities.

\noindent\textbf{Construction.} Some largest matching contains all pairs made by the stack pass of \Cref{alg:tree}. When it pairs adjacent unmatched delimiters, an optimal matching can add their pair if neither is used, or swap it for the existing pair if one is used. Pairing both elsewhere would cross unless a token serves both roles, so two-sided tokens are skipped; for a symmetric pair, the two partners can instead pair with each other. The pass is linear and complete on well-nested input without two-sided tokens. An $O(m^3)$ interval program matches the remaining delimiters $r_1,\ldots,r_m$. Let $\mu(i,j)$ be the largest matching size on $r_i,\ldots,r_{j-1}$, with $\mu(i,j)=0$ for $j-i\le1$; otherwise,
\begin{equation}\label{eq:nussinov}
	\mu(i,j)=\max\Bigl(\mu(i+1,j),\ \max_{\substack{i<k<j\\(t_{r_i},t_{r_k})\text{ admitted}}}1+\mu(i+1,k)+\mu(k+1,j)\Bigr),
\end{equation}
since $r_i$ is unpaired or paired with some $r_k$, separating the inside and following intervals. Traceback skips $r_i$ on ties; otherwise it takes the least maximizing $k$ and recurses inside and after that pair.

\begin{algorithm}[h]
	\caption{\textsc{Tree}: match delimiters into spans}
	\label{alg:tree}
	\small
	\begin{algorithmic}[1]
		\Require Tokens $t_1\cdots t_n$, admitted pairs $P$
		\Ensure The root of the span tree
		\Statex \textit{A \emph{two-sided} token both opens and closes pairs of $P$ and occurs in a nonsymmetric pair.}
		\State $M\gets\varnothing$, and $r\gets$ empty list \Comment{matched pairs, unmatched delimiters}
		\For{each position $i$ whose token is a delimiter of $P$, in order}
			\State Append $i$ to $r$
			\State \textbf{if} $r$ ends in $j,i$ with $(t_j,t_i)\in P$, and neither $t_j$ nor $t_i$ is two-sided, \textbf{then} add $(j,i)$ to $M$, and remove $j$ and $i$ from $r$
		\EndFor
		\State Complete $M$ on $r$ using \eqref{eq:nussinov} and its traceback
		\State Start a stack with a delimiterless root
		\For{each token position $i$, in order}
			\State \textbf{if} $i$ opens a pair of $M$ \textbf{then} create and push a child of the top node
			\State \textbf{else if} $i$ closes a pair of $M$ \textbf{then} close and pop the top node
			\State \textbf{else} append $i$ to the top node's items
		\EndFor
		\State \Return the root
	\end{algorithmic}
\end{algorithm}

\FloatBarrier
\subsection{Base Reducer}

\Cref{alg:drdd} gives \drdd{}~\cite{kumbhakern2026drdd} with a closing-scan budget $R$: laminar reduction uses $R=1$ per node, shrinking $R=\infty$.

\begin{algorithm}[h]
	\caption{\drdd{}: chunk sweeps and closing scans}
	\label{alg:drdd}
	\small
	\begin{algorithmic}[1]
		\Require Accepted unit sequence $U$, oracle $\test$, closing-scan budget $R$
		\Ensure An accepted subsequence, 1-minimal over units if $R=\infty$
		\State $s\gets\lfloor|U|/2\rfloor$
		\While{$s>1$}
			\State $\Call{Sweep}{s}$, and $s\gets\lfloor s/2\rfloor$
		\EndWhile
		\Repeat
			\State $\Call{Sweep}{1}$
		\Until{the sweep deletes nothing, or $R$ sweeps of size 1 have run}
		\State \Return $U$
		\Statex
		\Procedure{Sweep}{$s$}
			\State Partition $U$ into chunks $c_1,\ldots,c_k$ of $s$ units, the last possibly shorter
			\For{$i\gets1$ \textbf{to} $k$}
				\State \textbf{if} $\test(U\setminus c_i)=\ACCEPT$ \textbf{then} $U\gets U\setminus c_i$
			\EndFor
		\EndProcedure
	\end{algorithmic}
\end{algorithm}

\FloatBarrier
\subsection{Laminar Reduction}
Promoted spans join the current level in \Cref{alg:laminar} and are processed in their new positions.

\begin{algorithm}[h]
	\caption{Laminar reduction: reduce the span tree level by level}
	\label{alg:laminar}
	\small
	\begin{algorithmic}[1]
		\Require The root of the tree from \Cref{alg:tree}, oracle $\test$
		\Ensure The surviving tokens, in order
		\Statex \textit{Every query tests the whole input after one proposed replacement; accepted edits update the tree.}
		\State $\mathit{stop}\gets\bot$ \Comment{last node changed in the preceding traversal}
		\Repeat
			\State $\mathit{last}\gets\bot$, and $L\gets((\bot,\mathit{root}))$ \Comment{(parent, node) pairs}
			\While{$L$ is nonempty}
				\For{each $(q,v)\in L$ in order, including appended pairs}
					\State Reduce $v$'s items by \drdd{} with $R=1$
					\If{$q\ne\bot$ and a nearest same-pair descendant $u$, tried in order, is accepted in $v$'s place}
						\State Append $(q,u)$ to $L$ \Comment{first accepted hoist}
					\ElsIf{$q\ne\bot$ and unwrapping $v$ is accepted}
						\State Append $(q,u)$ to $L$ for each lifted span $u$
					\EndIf
					\State \textbf{if} processing $v$ changed the tree \textbf{then} $\mathit{last}\gets v$
					\State \textbf{if} $v=\mathit{stop}$ and $\mathit{last}=\bot$ \textbf{then return} the surviving tokens, in order
				\EndFor
				\State $L\gets[(v,u):(q,v)\in L,\ v\text{ survives},\ u\text{ is a span child of }v]$
			\EndWhile
			\State $\mathit{stop}\gets\mathit{last}$
		\Until{$\mathit{last}=\bot$}
		\State \Return the surviving tokens, in order
	\end{algorithmic}
\end{algorithm}

\noindent\textbf{Fixpoint.} Each accepted edit removes a nonempty token, so at most $n$ traversals change an input of $n$ tokens. An unchanged traversal rejects every single-item deletion, permitted hoist, and unwrapping against the final input, proving local minimality despite $R=1$. An early stop combines the unchanged prefix through $\mathit{stop}$ with the suffix checked after the preceding traversal's last change. Neither oracle monotonicity nor correct inferred structure is needed.

\noindent\textbf{Query cost.} With $R=1$, \drdd{} costs $O(k)$ queries for $k$ items. A traversal processes each node once, including promoted nodes; their item lists total $O(n)$ entries. Charge each hoist query to its offered descendant: only its nearest remaining ancestor of the same pair can offer it, and ancestors are processed first. Each descendant is thus offered at most once per traversal; each span also costs at most one unwrapping query. Each traversal costs $O(n)$, and at most $n$ changing traversals plus a final check cost $O(n^2)\subseteq O(N^2)$.

\subsection{Shrinking}

\begin{algorithm}[h]
	\caption{Shrinking: shorten each distinct token at all occurrences}
	\label{alg:shrink}
	\small
	\begin{algorithmic}[1]
		\Require Tokens $t_1\cdots t_n$ left by laminar reduction, oracle $\test$
		\Ensure The shrunk tokens
		\For{each distinct token $u$ with $|u|>1$, in order of first occurrence}
			\State $O\gets\{i:t_i=u\}$ \Comment{fixed while reducing $u$}
			\State Reduce $u$'s atoms by \drdd{} with $R=\infty$, testing each $u'$ at all positions $O$ in the current input and keeping accepted edits
		\EndFor
		\State \Return $t_1\cdots t_n$ without empty tokens
	\end{algorithmic}
\end{algorithm}

For initial distinct-token lengths $\ell_1,\ldots,\ell_d$, shrinking costs $O(\sum_i\ell_i^2)\subseteq O(N^2)$, since $\sum_i\ell_i\le N$. Later token edits may invalidate an earlier token's local fixpoint.

\FloatBarrier
\section{Inference Procedures}\label{app:details}

These subroutines complete \Cref{alg:discover,alg:tokenize,alg:pairs}. Discovery's budget counts executions, so cached verdicts cost no calls.

\subsection{Neutral-Word Discovery}\label{app:discovery}

\noindent\textbf{Candidate enumeration.} \textsc{Candidates} enumerates maximal repeats as internal intervals of a suffix array's longest-common-prefix tree whose occurrences have at least two distinct preceding atoms. Children give distinct following atoms; input ends count as distinct markers. Prefix doubling takes $O(N\log^2N)$ comparisons, Kasai's longest-common-prefix construction $O(N)$ time, and merging preceding-atom sets smaller into larger $O(N\log N)$ insertions, without queries.

\noindent\textbf{Runs and classes.} Maximal consecutive copies of $\rho(u)$ on either side of an offset form a \emph{run}; its two adjacent atoms, or end markers, are its \emph{surroundings}. Inserting $u$ at any copy boundary within a run gives the same input. A word's \emph{class} is $\lfloor\log_2|\rho(u)|\rfloor$.

\noindent\textbf{Schedule.} \textsc{Schedule} visits distinct offsets in this order:
\begin{enumerate}
	\item the starts of the leading candidate's first and last occurrences;
	\item both ends of the runs around each source word's first and last occurrences, grouped by class, each class shuffled with seed 0, taking one offset per class in turn;
	\item remaining internal offsets, shuffled, in class 0.
\end{enumerate}

\noindent\textbf{Observation.} Rows accepting $\kappa$ earn zero credit but enter the class mean; incomplete rows are discarded (\Cref{alg:observe}).

\begin{algorithm}[h]
	\caption{\textsc{Observe}: collect controlled evidence}
	\label{alg:observe}
	\small
	\begin{algorithmic}[1]
		\Require $S$, oracle $\test$, candidate words $W$, control $\kappa$, scheduled offsets, call limit $\mathit{lim}$
		\Ensure Complete observation rows $D$
		\Statex \textit{Before a query would exceed $\mathit{lim}$ calls, return $D$.}
		\State $D\gets$ empty list
		\For{each scheduled pair $(b,g)$ of class and offset}
			\State $C_g\gets\varnothing$
			\If{$\test(S\oplus_g\kappa)=\REJECT$}
				\State Query $\test(S\oplus_g u)$ for each $u\in W$ in order; $U\gets$ the accepted words
				\State $v_1\gets\REJECT$
				\For{each distinct length $\ell>1$ in $U$}
					\State $v_\ell\gets\test(S\oplus_g\kappa^\ell)$
				\EndFor
				\State $C_g\gets\{u\in U:v_{|u|}=\REJECT\}$
			\EndIf
			\State Append $(b,g,C_g)$ to $D$
		\EndFor
		\State \Return $D$
	\end{algorithmic}
\end{algorithm}

\noindent\textbf{Scoring.} Let $W=(u_0,u_1,\ldots)$ be the candidate order, $r$ its number of families, and $k_g=|\{\rho(u):u\in C_g\}|$. For nonempty $D$, let $D_b$ be the rows of class $b$, $\mathcal B$ the observed classes, and $\lambda_i=\log((i+1)(i+2))$. Define
\begin{equation}\label{eq:discovery-score}
	\begin{aligned}
		w_i(g) &=
		\begin{cases}
			\log\dfrac{r+1}{k_g}, & u_i\in C_g, \\
			0, & \text{otherwise},
		\end{cases} \\
		s_i &= \operatorname*{mean}_{b\in\mathcal B}\;\operatorname*{mean}_{g\in D_b} w_i(g), \\
		e_i &= |D|\,s_i-\mathbf{1}[|D|<N-1]\,\lambda_i.
	\end{aligned}
\end{equation}
The $+1$ keeps credit positive even when every family earns it; the rank penalty vanishes after all internal offsets have complete rows. For $r=3$, crediting $u_0$ alone at one of two class-0 rows gives weights $(\log4,0)$; crediting it alongside one other family at the sole class-1 row gives $\log2$. Then $s_0=\tfrac12(\tfrac12\log4+\log2)=\log2$: classes weigh equally despite unequal samples. If offsets remain unobserved, $e_0=3\log2-\log2=\log4$; later candidates pay a larger penalty.

\noindent\textbf{Confirmation.} \textsc{Sites} scans crediting rows in input order, retaining one offset per run and up to four distinct surroundings. With fewer than two surroundings, it adds further runs to reach two sites where possible. With two or three sites, \textsc{RefineSites} probes at most two occurrences with new surroundings, in input order, appending accepted sites up to four, without controls. \textsc{ReuseTests} takes the first five of these insertions at the sites of $T$, in order:
\begin{itemize}
	\item one site: 2, 4, 8, 16, and 32 copies there;
	\item two sites: two copies at both together, then two at each alone, then four at each alone;
	\item three or four sites: one copy at each site of every triple, in lexicographic order of indices in $T$, then two copies at each site alone, then four at each alone.
\end{itemize}
Two words need at most $2(2+5)=14$ confirmation calls, within the 16-call reserve; unspent observation calls are also available.

\FloatBarrier
\subsection{Tokenization}\label{app:tokenization}

\noindent\textbf{Contexts.} At offset $g$, let $\ell_k,r_k$ be its $k$ atoms to the left and right, $v^\pm=V[g\pm1]$, and $\rho_s$ the share of classified offsets within $s$ atoms, excluding $g$, that are boundaries. Bucket $\rho_s$ as none, at most a third, at most two thirds, fewer than all, and all. \textsc{Estimate} uses these 14 contexts, read down each column:
\begin{center}
	\small
	\begin{tabular}{@{}r@{\hspace{6pt}}l@{\hspace{24pt}}r@{\hspace{6pt}}l@{\hspace{24pt}}r@{\hspace{6pt}}l@{}}
		1 & $(v^-,v^+)$ & 6 & $(\ell_2,r_1)$ & 11 & $(\ell_3,r_3)$ \\
		2 & $\ell_1$ & 7 & $(\ell_1,r_1,\rho_{32})$ & 12 & $(\ell_1,r_1,\rho_{16})$ \\
		3 & $r_1$ & 8 & $(\ell_1,r_2)$ & 13 & $(\ell_1,r_1,\rho_8)$ \\
		4 & $(\ell_1,r_1)$ & 9 & $(\ell_2,r_2)$ & 14 & $(\ell_1,r_1,v^-,v^+)$ \\
		5 & $\ell_3$ & 10 & $r_2$ & & \\
	\end{tabular}
\end{center}
Out-of-range atoms, unknown verdicts, and empty density neighborhoods use markers. $V$ includes predictions and tentative pooled boundaries as well as settled verdicts.

\noindent\textbf{Estimate.} $M[j,h]=(t,b)$ records $t$ settled offsets whose $j$th context had key $h$ at settlement, $b$ of them boundaries. For the keys $h_j$ at $g$, write $(t_j,b_j)=M[j,h_j]$ and compute
\begin{equation}\label{eq:estimate}
	p_0=\frac12,\qquad
	p_j=\frac{b_j+2p_{j-1}}{t_j+2},\qquad
	\hat p=\sigma\Bigl(\sum_{j=1}^{14}\alpha_j\operatorname{logit}p_j\Bigr).
\end{equation}
Each rate is smoothed toward its predecessor with weight two; an unseen context leaves it unchanged. Initialize $\alpha=(0,\ldots,0,1)$, giving $\hat p=p_{14}$. On settling verdict $v\in\{0,1\}$, first update $V$ and recompute the context keys and estimates from the current $V$, $M$, and weights. Take the log-loss gradient step $\alpha_j\gets\alpha_j+0.01\,(v-\hat p)\operatorname{logit}p_j$, then update $(t,b)\gets(t+1,b+v)$ in each of the 14 entries. Clip probabilities to $[10^{-6},1-10^{-6}]$ before log-odds, and log-odds to $[-30,30]$ before $\sigma$.

\noindent\textbf{Information.} Write $H(q)=-q\log_2q-(1-q)\log_2(1-q)$. Under the estimates, predicting at $\hat p\ge0.98$ or $\hat p\le0.02$ skips queries worth at most $H(0.02)\approx0.14$ bits. A pool triggered by $\prod\hat p\le1/2$ has estimated acceptance probability in $[1/4,1/2]$, since the preceding product exceeds $1/2$ and each pooled estimate is at least $1/2$. Its estimated yield is therefore at least $H(1/4)\approx0.81$ bits.

\noindent\textbf{Pooling.} Under factorization (\Cref{sec:prediction}), a rejected pool contains a non-boundary. \textsc{TestPool} maintains such a set $T$, splitting by \emph{doubt}, $-\sum\log\hat p_h$. It queries the shortest proper prefix holding at least half the doubt, or all but the last offset if no proper prefix does. After an accepted prefix, the suffix still contains a non-boundary.

\begin{samepage}
For estimates $(0.9,0.8,0.8,0.8)$, the fourth offset triggers a query (product $0.4608$). The shortest prefix holding half the doubt comprises the first three offsets. Rejecting the pool and accepting that prefix classifies three boundaries and one non-boundary in two queries. \par
\end{samepage}

\begin{algorithm}[h]
	\caption{\textsc{TestPool}: test the pool}
	\label{alg:flush}
	\small
	\begin{algorithmic}[1]
		\Require Pool $R$ with estimates $\hat p_h$, verdict map $V$
		\While{$R$ is nonempty}
			\State $T\gets R$, and $R\gets$ empty
			\If{$\test$ accepts $S$ with $w$ inserted at every offset of $T$}
				\State $V[h]\gets\ACCEPT$ for each $h\in T$
			\Else
				\While{$|T|>1$}
					\State Split $T=T_1T_2$ by half the doubt, as above
					\If{$\test$ accepts $S$ with $w$ at every offset of $T_1$}
						\State $V[h]\gets\ACCEPT$ for each $h\in T_1$, and $T\gets T_2$
					\Else
						\State Prepend $T_2$ to $R$, and $T\gets T_1$
					\EndIf
				\EndWhile
				\State $V[h]\gets\REJECT$ for the offset $h$ left in $T$
			\EndIf
		\EndWhile
	\end{algorithmic}
\end{algorithm}

\noindent\textbf{Query bound.} Since each pooled estimate is below $0.98$ and $0.98^{35}<1/2$, pools hold at most 35 offsets. Each outer iteration settles at least one offset using one pool test and at most 34 split queries. Individual probes also settle an offset, giving at most $35(N-1)=O(N)$ queries for $N\ge1$.

\FloatBarrier
\subsection{Pair Learning}\label{app:pairs}

\noindent\textbf{Hypotheses.} Enumerate single pairs by occurrence-count classes $x\le y$. For families, split each non-blank token at every inner position and group prefixes, or suffixes, sharing at least two nonempty names.

\noindent\textbf{Candidate intervals.} Assign each position to its innermost strictly enclosing admitted interval, or to the outside region; an interval's own delimiters belong to its surrounding region. Symmetric pairing also starts at the second occurrence when a region has an odd count, allowing a stray occurrence, and drops intervals strictly nested in one with an earlier start.

\noindent\textbf{Balance prior.} For $u$ openings, $v$ closings, and $d$ unmatched closings, a uniform shuffle leaves more than $d$ closings unmatched exactly when its walk ($+1$ per opening, $-1$ per closing) reaches $-(d+1)$. Reflecting up to the first visit there gives a bijection to shuffles with $u+d+1$ openings and $v-d-1$ closings, of which there are $\binom{u+v}{v-d-1}$. Hence
\begin{equation}\label{eq:balance}
	\textsc{Balance}(u,v,d)=-\log\left(1-\binom{u+v}{v-d-1}\Big/\binom{u+v}{v}\right),
\end{equation}
with out-of-range binomial coefficients zero. Families sum \textsc{Balance} over names, using their counts and unmatched closings on the shared stack. This sum is heuristic: one name's stray delimiters can block another's matches.

\noindent\textbf{Parity prior.} For a symmetric hypothesis occurring in $r$ regions, $k$ of them odd, independent region parities with probability $1/2$ give
\begin{equation}\label{eq:parity}
	\textsc{Parity}(r,k)=-\log\left(2^{-r}\sum_{i=0}^{k}\binom{r}{i}\right).
\end{equation}
The cap $\log(1+|U|)$ of \Cref{sec:pairs} equals $\textsc{Balance}(|U|,|U|,0)$.

\noindent\textbf{Ranking bounds.} Scoring scans occurrences, so \textsc{Hypotheses} visits count-class pairs and families by decreasing upper bounds on their priors:
\begin{equation}\label{eq:bounds}
	\log\frac{y+1}{y-x+1} \quad\text{and}\quad \log\Bigl(1+\sum_{(\opn,\cls)\in H}\min\bigl(|S|_{\opn},|S|_{\cls}\bigr)\Bigr).
\end{equation}
For counts $x\le y$, the smallest feasible $d$ is $y-x$ with $x$ openings, or 0 with $y$ openings; either gives $-\log(1-x/(y+1))$. Since \textsc{Balance} decreases with $d$, this bounds both orientations. The family bound follows from the cap and $|U|\le\sum\min(|S|_{\opn},|S|_{\cls})$. A min-heap retains the $K$ best hypotheses, replacing only on strict improvement; stop when the next bound falls below its minimum. Final ties follow discovery order. Symmetric hypotheses are all scored.

\noindent\textbf{Sensitivity.} \textsc{Sensitive} duplicates each distinct delimiter of one representative pair at its occurrence nearest a multi-atom token, measuring distance in token positions. Ties, or absence of multi-atom tokens, select the earliest occurrence. Multi-atom tokens mark restrictive surroundings. Tests are individual and may use unrelated occurrences.

\noindent\textbf{Confirmation.} \textsc{Coupled} queries controls only after a paired edit is accepted, the second only if the first is rejected. In the empty-or-delete row, deletion is tried only if emptying is rejected. This costs at most three queries per row, four for empty-or-delete, and ten per interval. \textsc{Test}'s context is an interval's tokens plus one on either side. \textsc{Spread} samples fractions $0,1/2,1/4,3/4,\ldots$ of the list length, the van der Corput binary-digit-reversal sequence, rounding down and stepping cyclically past used indices.

\begin{algorithm}[h]
	\caption{\textsc{Test}: seek two coupling witnesses}
	\label{alg:pair-test}
	\small
	\begin{algorithmic}[1]
		\Require Compatible candidate intervals $A$, patience $\pi$
		\Ensure \textbf{true} after two coupling witnesses, \textbf{false} otherwise
		\State $L\gets A$ sorted by (length, position), retaining the first interval per context
		\State Reorder $L$ with $\Call{Spread}{|L|}$, and $\mathit{hits}\gets0$
		\For{$j=1,\ldots,\min(|L|,2\pi)$}
			\State \textbf{if} $\Call{Coupled}{L_j}$ \textbf{then} $\mathit{hits}\gets\mathit{hits}+1$
			\State \textbf{if} $\mathit{hits}=2$ \textbf{then return true}
			\State \textbf{if} $j=\pi$ and $\mathit{hits}=0$ \textbf{then return false}
		\EndFor
		\State \Return \textbf{false}
	\end{algorithmic}
\end{algorithm}

\noindent\textbf{Query bound.} Each of at most $2K$ hypotheses uses at most two sensitivity queries and $2\pi\cdot10$ coupling queries. The cap gives $\pi\le2+\ln(1+N)$, hence at most $2K\bigl(2+20(2+\ln(1+N))\bigr)=O(K\log N)$ queries.

\clearpage
\section{Supplementary Evaluation}\label{app:results}

\Cref{tab:stages} decomposes the costs in \Cref{tab:results}; \Cref{tab:budget,tab:other-pairs} detail the budget sweep and exceptional admissions.

\begin{table}[H]
	\caption{Per-input stage costs on the \worded{} returned-word inputs}
	\label{tab:stages}
	\tablestyle
	{\tablemedium\renewcommand{\arraystretch}{0.85}\setlength{\subtablesep}{3pt}\begin{tabularx}{\linewidth}{|L|N|N|N|N|N|N|N|N|N|N|N|N|N|}
	\hline
	\multirow{2}{*}{\textbf{Case}} & \multirow{2}{*}{\textbf{Input}} & \multicolumn{2}{c|}{\textbf{Discovery}} & \multicolumn{2}{c|}{\textbf{Tokenization}} & \multicolumn{2}{c|}{\textbf{Pair learning}} & \multicolumn{3}{c|}{\textbf{Laminar reduction}} & \multicolumn{3}{c|}{\textbf{Shrinking}} \\
	 & & \emph{Calls} & \emph{Time} & \emph{Calls} & \emph{Time} & \emph{Calls} & \emph{Time} & \emph{Output} & \emph{Calls} & \emph{Time} & \emph{Output} & \emph{Calls} & \emph{Time} \\
	\hline
	\subtablegap
	\multicolumn{14}{l}{\textit{asn1}} \\
	\hline
	\case{asn1-aah055-nested} & 580 & 287 & 17\,s & 125 & 8.4\,s & 1 & 0.1\,s & 116 & 26 & 1.6\,s & 112 & 184 & 6.3\,s \\
	\hline
	\subtablegap
	\multicolumn{14}{l}{\textit{c}} \\
	\hline
	\case{02} & 3,623 & 288 & 37\,s & 606 & 81\,s & 171 & 31\,s & 570 & 1,701 & 4\,m & 457 & 266 & 28\,s \\
	\case{10} & 3,071 & 290 & 33\,s & 514 & 66\,s & 194 & 32\,s & 369 & 902 & 90\,s & 281 & 226 & 25\,s \\
	\case{tree-ssa-structalias\_c\_4773\_96K} & 97,658 & 289 & 22\,s & 733 & 75\,s & 101 & 11\,s & 213 & 1,771 & 40\,s & 151 & 129 & 1.5\,s \\
	\case{tree-vect-transform\_c\_6496\_55K} & 55,923 & 288 & 12\,s & 1,072 & 56\,s & 117 & 6.2\,s & 160 & 4,090 & 3\,m & 114 & 91 & 1.7\,s \\
	\case{fold-const\_c\_8943\_23K} & 23,171 & 289 & 12\,s & 494 & 26\,s & 119 & 5.3\,s & 184 & 1,428 & 30\,s & 101 & 117 & 1.2\,s \\
	\case{recog\_c\_1990\_63K} & 63,509 & 289 & 13\,s & 518 & 35\,s & 82 & 5.9\,s & 214 & 1,504 & 45\,s & 138 & 140 & 1.5\,s \\
	\case{simplify-rtx\_c\_4922\_43K} & 43,720 & 289 & 8.2\,s & 453 & 19\,s & 113 & 4.6\,s & 214 & 2,034 & 58\,s & 135 & 140 & 1.6\,s \\
	\case{gimple\_c\_2343\_122K} & 124,593 & 288 & 24\,s & 988 & 2\,m & 83 & 9.2\,s & 208 & 8,795 & 11\,m & 142 & 148 & 4.7\,s \\
	\case{emit-rtl\_c\_1950\_75K} & 76,263 & 289 & 16\,s & 716 & 72\,s & 93 & 9.9\,s & 235 & 1,450 & 66\,s & 155 & 142 & 2.0\,s \\
	\case{ira-emit\_c\_371\_17K} & 16,424 & 289 & 8.4\,s & 455 & 17\,s & 118 & 4.4\,s & 350 & 2,097 & 45\,s & 199 & 151 & 2.5\,s \\
	\hline
	\subtablegap
	\multicolumn{14}{l}{\textit{crashjs}} \\
	\hline
	\case{7} & 1,012 & 289 & 68\,s & 332 & 61\,s & 163 & 28\,s & 230 & 830 & 2\,m & 159 & 168 & 28\,s \\
	\case{10} & 830 & 289 & 63\,s & 314 & 59\,s & 144 & 26\,s & 162 & 554 & 2\,m & 113 & 140 & 25\,s \\
	\case{commander-4} & 778 & 289 & 45\,s & 371 & 62\,s & 78 & 12\,s & 244 & 524 & 81\,s & 166 & 172 & 26\,s \\
	\case{commander-31} & 1,055 & 289 & 46\,s & 473 & 79\,s & 111 & 17\,s & 226 & 338 & 50\,s & 156 & 160 & 25\,s \\
	\case{javascript-algorithms-5} & 745 & 289 & 59\,s & 342 & 61\,s & 84 & 14\,s & 327 & 552 & 92\,s & 220 & 268 & 45\,s \\
	\case{javascript-algorithms-14} & 814 & 289 & 58\,s & 247 & 41\,s & 71 & 12\,s & 294 & 379 & 64\,s & 252 & 247 & 91\,s \\
	\case{moment-1} & 800 & 289 & 2\,m & 325 & 2\,m & 105 & 29\,s & 187 & 653 & 4\,m & 181 & 149 & 38\,s \\
	\case{moment-17} & 1,340 & 289 & 2\,m & 457 & 2\,m & 136 & 46\,s & 221 & 652 & 4\,m & 188 & 234 & 62\,s \\
	\case{express-1} & 644 & 289 & 64\,s & 341 & 67\,s & 69 & 14\,s & 102 & 346 & 62\,s & 98 & 137 & 24\,s \\
	\case{express-48} & 903 & 289 & 60\,s & 425 & 79\,s & 149 & 32\,s & 127 & 344 & 67\,s & 117 & 129 & 23\,s \\
	\hline
	\subtablegap
	\multicolumn{14}{l}{\textit{graphviz}} \\
	\hline
	\case{issue-1436} & 165,583 & 289 & 98\,s & 11,297 & 2.1\,h & 158 & 80\,s & 2,473 & 45,263 & 62\,m & 2,270 & 933 & 35\,s \\
	\case{issue-2723} & 2,141 & 289 & 16\,s & 120 & 6.5\,s & 119 & 5.9\,s & 90 & 468 & 19\,s & 90 & 15 & 0.3\,s \\
	\hline
	\subtablegap
	\multicolumn{14}{l}{\textit{lua}} \\
	\hline
	\case{luajit-913} & 1,248 & 289 & 15\,s & 444 & 19\,s & 151 & 4.3\,s & 130 & 380 & 12\,s & 72 & 61 & 1.7\,s \\
	\case{luajit-1116} & 889 & 288 & 13\,s & 96 & 4.1\,s & 98 & 0.9\,s & 391 & 1,841 & 39\,s & 377 & 62 & 1.1\,s \\
	\hline
	\subtablegap
	\multicolumn{14}{l}{\textit{pdf}} \\
	\hline
	\case{pdf-fuzzer-aah052} & 15,421 & 289 & 32\,s & 226 & 24\,s & 195 & 30\,s & 118 & 679 & 31\,s & 118 & 30 & 1.0\,s \\
	\case{pdf-fuzzer-jch207} & 15,505 & 289 & 11\,s & 12 & 1.4\,s & 255 & 13\,s & 7 & 116 & 3.6\,s & 7 & 0 & 0.0\,s \\
	\case{pdfimages-jch207-kw6} & 4,543 & 291 & 5.8\,s & 480 & 10\,s & 230 & 5.8\,s & 3 & 106 & 2.0\,s & 3 & 0 & 0.0\,s \\
	\case{pdftoppm-aah052} & 15,495 & 291 & 15\,s & 344 & 25\,s & 192 & 22\,s & 143 & 679 & 23\,s & 139 & 70 & 1.7\,s \\
	\case{pdftoppm-jch207} & 15,532 & 289 & 11\,s & 313 & 14\,s & 226 & 9.8\,s & 806 & 816 & 19\,s & 794 & 3,089 & 90\,s \\
	\hline
	\subtablegap
	\multicolumn{14}{l}{\textit{python}} \\
	\hline
	\case{ackermann} & 483 & 288 & 30\,s & 174 & 24\,s & 35 & 3.3\,s & 151 & 498 & 42\,s & 123 & 118 & 8.7\,s \\
	\case{alloc} & 1,125 & 288 & 23\,s & 350 & 32\,s & 144 & 11\,s & 400 & 1,268 & 97\,s & 304 & 323 & 23\,s \\
	\case{dict} & 265 & 288 & 21\,s & 129 & 11\,s & 89 & 6.5\,s & 136 & 610 & 44\,s & 135 & 116 & 8.0\,s \\
	\case{mroref} & 383 & 288 & 24\,s & 194 & 17\,s & 43 & 3.4\,s & 205 & 354 & 26\,s & 170 & 173 & 12\,s \\
	\case{recursion} & 212 & 288 & 23\,s & 109 & 9.4\,s & 63 & 5.2\,s & 62 & 290 & 22\,s & 59 & 65 & 4.7\,s \\
	\hline
	\subtablegap
	\multicolumn{14}{l}{\textit{smt}} \\
	\hline
	\case{btor2-bug-16966-2630} & 2,435 & 289 & 1.0\,s & 161 & 0.8\,s & 33 & 0.2\,s & 1,096 & 7,882 & 41\,s & 862 & 267 & 1.0\,s \\
	\case{btor2-bug-21435-401} & 3,546 & 289 & 1.1\,s & 190 & 1.2\,s & 40 & 0.3\,s & 1,314 & 8,688 & 47\,s & 1,058 & 357 & 1.4\,s \\
	\case{btor2-bug-16965-103} & 9,773 & 289 & 1.2\,s & 220 & 1.9\,s & 16 & 0.2\,s & 1,853 & 30,317 & 3\,m & 1,495 & 465 & 2.0\,s \\
	\case{btor2-bug-19025-698} & 63,795 & 291 & 4.1\,s & 265 & 9.6\,s & 15 & 5.9\,s & 3,689 & 183,766 & 55\,m & 2,772 & 987 & 6.2\,s \\
	\case{btor4-bug-29933-6674} & 95,337 & 289 & 3.3\,s & 445 & 9.5\,s & 142 & 28\,s & 1,608 & 29,211 & 7\,m & 1,364 & 438 & 1.7\,s \\
	\case{btor4-bug-1428-13899} & 162,928 & 289 & 5.8\,s & 270 & 13\,s & 11 & 64\,s & 844 & 143,766 & 66\,m & 712 & 259 & 1.0\,s \\
	\case{btor4-bug-29800-10218} & 224,151 & 289 & 8.0\,s & 454 & 21\,s & 118 & 3\,m & 894 & 96,398 & 39\,m & 708 & 324 & 1.2\,s \\
	\case{smtlib384a03} & 43,696 & 289 & 15\,s & 657 & 78\,s & 51 & 7.7\,s & 338 & 996 & 55\,s & 172 & 104 & 6.0\,s \\
	\case{smtlib46f14a} & 43,323 & 289 & 14\,s & 659 & 78\,s & 78 & 12\,s & 338 & 996 & 54\,s & 172 & 104 & 6.0\,s \\
	\case{smtlibf957ea} & 43,937 & 289 & 14\,s & 679 & 80\,s & 78 & 12\,s & 288 & 701 & 40\,s & 151 & 94 & 5.3\,s \\
	\hline
	\subtablegap
	\multicolumn{14}{l}{\textit{sql}} \\
	\hline
	\case{postgresql-18398} & 277 & 289 & 7.2\,s & 98 & 2.9\,s & 48 & 1.7\,s & 192 & 251 & 6.4\,s & 126 & 111 & 3.2\,s \\
	\case{mysql-112460} & 538 & 288 & 21\,s & 155 & 15\,s & 101 & 14\,s & 257 & 745 & 39\,s & 208 & 197 & 18\,s \\
	\case{sqlite-b7914e5795} & 503 & 289 & 4.0\,s & 150 & 2.1\,s & 85 & 1.2\,s & 247 & 1,625 & 23\,s & 233 & 151 & 2.1\,s \\
	\case{sqlflex-sqlite-79} & 128 & 289 & 3.8\,s & 74 & 1.0\,s & 25 & 0.4\,s & 71 & 118 & 1.6\,s & 64 & 73 & 1.0\,s \\
	\case{sqlflex-sqlite-165} & 163 & 289 & 3.8\,s & 74 & 1.0\,s & 20 & 0.3\,s & 39 & 81 & 1.1\,s & 38 & 40 & 0.6\,s \\
	\hline
\end{tabularx} 
}
	\tablenote{Sizes are in bytes; outputs exclude whitespace.
		Calls and times are per stage; \Cref{tab:results} sums all five.}
\end{table}

\begin{table}[H]
	\continuedcaption{tab:stages}{Per-input stage costs on the \worded{} returned-word inputs}
	\tablestyle
	{\tablemedium\renewcommand{\arraystretch}{0.85}\setlength{\subtablesep}{3pt}\begin{tabularx}{\linewidth}{|L|N|N|N|N|N|N|N|N|N|N|N|N|N|}
	\hline
	\multirow{2}{*}{\textbf{Case}} & \multirow{2}{*}{\textbf{Input}} & \multicolumn{2}{c|}{\textbf{Discovery}} & \multicolumn{2}{c|}{\textbf{Tokenization}} & \multicolumn{2}{c|}{\textbf{Pair learning}} & \multicolumn{3}{c|}{\textbf{Laminar reduction}} & \multicolumn{3}{c|}{\textbf{Shrinking}} \\
	 & & \emph{Calls} & \emph{Time} & \emph{Calls} & \emph{Time} & \emph{Calls} & \emph{Time} & \emph{Output} & \emph{Calls} & \emph{Time} & \emph{Output} & \emph{Calls} & \emph{Time} \\
	\hline
	\subtablegap
	\multicolumn{14}{l}{\textit{sql}} \\
	\hline
	\case{sqlflex-sqlite-256} & 104 & 289 & 3.9\,s & 59 & 0.8\,s & 20 & 0.3\,s & 69 & 113 & 1.6\,s & 69 & 73 & 1.1\,s \\
	\case{sqlflex-sqlite-622} & 53 & 289 & 3.8\,s & 35 & 0.5\,s & 11 & 0.2\,s & 37 & 62 & 0.9\,s & 36 & 35 & 0.5\,s \\
	\case{sqlflex-sqlite-686} & 179 & 289 & 3.8\,s & 95 & 1.3\,s & 14 & 0.2\,s & 58 & 179 & 2.4\,s & 58 & 63 & 0.9\,s \\
	\case{sqlflex-sqlite-808} & 51 & 289 & 3.8\,s & 39 & 0.6\,s & 9 & 0.2\,s & 37 & 58 & 0.8\,s & 36 & 35 & 0.5\,s \\
	\case{sqlflex-sqlite-909} & 144 & 289 & 3.9\,s & 83 & 1.2\,s & 24 & 0.4\,s & 84 & 221 & 3.0\,s & 83 & 74 & 1.0\,s \\
	\hline
	\subtablegap
	\multicolumn{14}{l}{\textit{xml}} \\
	\hline
	\case{1.2} & 3,974 & 289 & 22\,s & 505 & 30\,s & 77 & 11\,s & 66 & 371 & 19\,s & 66 & 80 & 4.1\,s \\
	\case{3.2} & 3,835 & 289 & 23\,s & 505 & 35\,s & 64 & 11\,s & 157 & 545 & 32\,s & 149 & 177 & 5.9\,s \\
	\case{4.1} & 3,933 & 289 & 23\,s & 533 & 33\,s & 64 & 11\,s & 150 & 1,292 & 58\,s & 140 & 130 & 5.9\,s \\
	\case{5.2} & 4,052 & 289 & 20\,s & 501 & 30\,s & 82 & 12\,s & 131 & 540 & 27\,s & 130 & 159 & 6.9\,s \\
	\case{071d221-1} & 101,759 & 289 & 42\,s & 3,452 & 6\,m & 29 & 16\,s & 15 & 106 & 14\,s & 15 & 11 & 4.1\,s \\
	\case{2d4ec80-1} & 108,560 & 289 & 19\,s & 3,198 & 99\,s & 32 & 10\,s & 63 & 663 & 39\,s & 63 & 39 & 3.6\,s \\
	\case{327c8af-1} & 103,527 & 289 & 27\,s & 3,282 & 3\,m & 40 & 15\,s & 60 & 513 & 36\,s & 60 & 44 & 4.4\,s \\
	\case{4c99b96-1} & 104,930 & 289 & 84\,s & 3,352 & 11\,m & 35 & 23\,s & 66 & 252 & 33\,s & 62 & 65 & 4.1\,s \\
	\case{8ede045-1} & 106,398 & 288 & 24\,s & 3,095 & 2\,m & 37 & 13\,s & 70 & 602 & 37\,s & 60 & 43 & 5.5\,s \\
	\case{f053486-1} & 109,614 & 289 & 74\,s & 4,727 & 12\,m & 24 & 15\,s & 33 & 157 & 18\,s & 33 & 33 & 4.3\,s \\
	\hline
	\subtablegap
	\multicolumn{14}{l}{\textit{yaml}} \\
	\hline
	\case{9J7A} & 16 & 85 & 0.2\,s & 13 & 0.0\,s & 2 & 0.0\,s & 11 & 30 & 0.1\,s & 11 & 8 & 0.0\,s \\
	\case{9FMG} & 36 & 288 & 0.5\,s & 17 & 0.0\,s & 10 & 0.0\,s & 15 & 69 & 0.1\,s & 15 & 12 & 0.0\,s \\
	\case{G4RS} & 177 & 290 & 0.6\,s & 60 & 0.1\,s & 9 & 0.0\,s & 153 & 35 & 0.1\,s & 153 & 288 & 0.6\,s \\
	\case{FBC9} & 148 & 288 & 0.5\,s & 82 & 0.2\,s & 9 & 0.0\,s & 130 & 65 & 0.1\,s & 130 & 258 & 0.5\,s \\
	\case{6JQW} & 44 & 288 & 0.5\,s & 37 & 0.1\,s & 0 & 0.0\,s & 21 & 24 & 0.0\,s & 21 & 35 & 0.1\,s \\
	\case{96L6} & 63 & 289 & 0.5\,s & 32 & 0.1\,s & 0 & 0.0\,s & 46 & 6 & 0.0\,s & 46 & 88 & 0.2\,s \\
	\case{3GZX} & 108 & 289 & 0.5\,s & 50 & 0.1\,s & 27 & 0.1\,s & 94 & 39 & 0.1\,s & 94 & 130 & 0.3\,s \\
	\case{ZH7C} & 16 & 98 & 0.2\,s & 11 & 0.0\,s & 0 & 0.0\,s & 10 & 21 & 0.0\,s & 10 & 6 & 0.0\,s \\
	\case{6FWR} & 20 & 151 & 0.3\,s & 17 & 0.0\,s & 0 & 0.0\,s & 10 & 9 & 0.0\,s & 9 & 25 & 0.1\,s \\
	\case{NAT4} & 86 & 288 & 0.5\,s & 23 & 0.1\,s & 61 & 0.1\,s & 35 & 193 & 0.4\,s & 35 & 17 & 0.0\,s \\
	\hline
\end{tabularx} 
}
	\tablenote{Sizes are in bytes; outputs exclude whitespace. Calls and times are per stage.}
\end{table}

\noindent\textbf{Discovery budget.} Each family contributes its smallest and largest inputs and any with a minority found/abstained outcome in the main run: \budgetInputs{} inputs, \budgetTextInputs{} text and 17 binary. Only the budget varies; seed 0 and the 16-call confirmation reserve are fixed. \Cref{tab:budget} gives the tested grid behind the thresholds in \Cref{sec:rqRecovery}. At 50 and 100 calls, C's \case{10} and PDF's \case{pdftoppm-jch207} abstain. At 600, only \case{asn1-aah055-nested} differs from the main run, returning four zero bytes instead of its 10-byte word: more budget need not preserve the choice.

\begin{table}[H]
	\begin{minipage}[t]{0.32\linewidth}
		\caption{Discovery budget sweep}
		\label{tab:budget}
	\end{minipage}\hfill
	\begin{minipage}[t]{0.65\linewidth}
		\caption{Admitted pairs other than format delimiters and conditional operators}
		\label{tab:other-pairs}
	\end{minipage}\par
	\begin{minipage}[t]{0.32\linewidth}
		\vspace{0pt}
		\tablestyle
		\begin{tabularx}{\linewidth}{|L|N|N|N|}
	\hline
	\textbf{Budget} & \textbf{Text} & \textbf{Binary} & \textbf{Same} \\
	\hline
	50 & 18/20 & 0/17 & 34/37 \\
	100 & 18/20 & 0/17 & 34/37 \\
	200 & 20/20 & 0/17 & 36/37 \\
	300 & 20/20 & 1/17 & 37/37 \\
	400 & 20/20 & 1/17 & 37/37 \\
	600 & 20/20 & 1/17 & 36/37 \\
	\hline
\end{tabularx}

		\tablenote{Text and Binary count inputs returning a word.
			Same counts identical words or abstentions to the main run.}
	\end{minipage}\hfill
	\begin{minipage}[t]{0.65\linewidth}
		\vspace{0pt}
		\tablestyle
		\begin{tabularx}{\linewidth}{|l|L|N|N|}
	\hline
	\textbf{Family} & \textbf{Input} & \textbf{Open} & \textbf{Close} \\
	\hline
	c & \case{recog\_c\_1990\_63K} & \texttt{*} & \texttt{*} \\
	c & \case{simplify-rtx\_c\_4922\_43K} & \texttt{-} & \texttt{-} \\
	python & \case{ackermann} & \texttt{=} & \texttt{ans} \\
	python & \case{mroref} & \texttt{:} & \texttt{\sq{}} \\
	smt & \case{btor2-bug-21435-401} & \texttt{ite} & \texttt{\#b1} \\
	sql & \case{sqlflex-sqlite-909} & \texttt{rt0} & \texttt{.} \\
	yaml & \case{NAT4} & \texttt{\textbackslash{}n} & \texttt{\textbackslash{}n} \\
	\hline
\end{tabularx}

		\tablenote{Token spellings are literal; \texttt{\textbackslash n} denotes a newline.}
	\end{minipage}
\end{table}

\noindent\textbf{Output size against input size.} \Cref{fig:sizesAutodidact,fig:sizesTools} plot output size against input size, for \drdd{} and \autodidact{} and for \autodidact{} and the format-aware tools. \drdd{}'s outputs grow almost in step with the input, to a median of 18,345 bytes on the \largeInputs{} inputs of at least 10\,KiB. On the same inputs, \autodidact{}'s outputs have a median of 141 bytes, close to C-Reduce's 157. HDD's median is 149 bytes on the \largeHddInputs{} of these inputs it covers.

\noindent\textbf{Oracle calls against input size.} \Cref{fig:callsAutodidact,fig:callsTools} plot oracle calls against input size for the same reducers. \drdd{}'s calls grow with the input, to a median of about 108,000 on the \largeInputs{} inputs of at least 10\,KiB. On the same inputs, \autodidact{}'s calls, inference included, have a median of about 3,900, below C-Reduce's 8,750. HDD's median is 945 on the \largeHddInputs{} of these inputs it covers.

\begin{figure}[H]
	\centering
	\definecolor{reducer-drdd}{HTML}{D7191C}
\definecolor{reducer-autodidact}{HTML}{0044CC}
\pgfplotslegendfromname{sizes-autodidact-legend}
\par\smallskip
\begin{tikzpicture}
\begin{axis}[
	width=\textwidth,
	height=0.36\textwidth,
	xmode=linear,
	ymode=linear,
	xmin=0,
	xmax=201164,
	ymin=0,
	ymax=64464,
	scaled x ticks=base 10:-3,
	scaled y ticks=base 10:-3,
	xtick scale label code/.code={},
	ytick scale label code/.code={},
	xticklabel style={/pgf/number format/fixed},
	yticklabel style={/pgf/number format/fixed},
	xlabel={input size (kB)},
	ylabel={output size (kB)},
	label style={font=\footnotesize},
	tick label style={font=\footnotesize},
	major grid style={line width=0.2pt, draw=black!12},
	axis line style={draw=black!45},
	tick style={draw=black!45},
	legend cell align=left,
	grid=major,
	legend to name=sizes-autodidact-legend,
	legend columns=11,
	legend style={font=\scriptsize, draw=none, nodes={text height=1.2ex, text depth=0.3ex}, /tikz/every even column/.append style={column sep=4pt}},
]
\addplot[no marks, line width=0.4pt, black!45, dashed, forget plot] coordinates {
	(0, 0)
	(64464, 64464)
};
\addplot[only marks, mark=pentagon*, mark size=1.5pt, color=reducer-drdd, forget plot, mark options={line width=0.7pt}] coordinates {
	(579, 164)
};
\addplot[only marks, mark=*, mark size=1.5pt, color=reducer-drdd, forget plot, mark options={line width=0.7pt}] coordinates {
	(2863, 824)
	(2466, 528)
	(79771, 45061)
	(44448, 13940)
	(19082, 9620)
	(51918, 31646)
	(35960, 17308)
	(99824, 31151)
	(62460, 35434)
	(13507, 8914)
};
\addplot[only marks, mark=square, mark size=1.5pt, color=reducer-drdd, forget plot, mark options={line width=0.7pt}] coordinates {
	(840, 333)
	(702, 332)
	(630, 313)
	(885, 297)
	(627, 497)
	(698, 331)
	(664, 434)
	(1110, 483)
	(528, 210)
	(760, 181)
};
\addplot[only marks, mark=diamond, mark size=1.5pt, color=reducer-drdd, forget plot, mark options={line width=0.7pt}] coordinates {
	(111516, 12949)
	(1706, 257)
};
\addplot[only marks, mark=triangle, mark size=1.5pt, color=reducer-drdd, forget plot, mark options={line width=0.7pt}] coordinates {
	(1045, 320)
	(391, 377)
};
\addplot[only marks, mark=diamond*, mark size=1.5pt, color=reducer-drdd, forget plot, mark options={line width=0.7pt}] coordinates {
	(14372, 118)
	(14481, 7)
	(3885, 3)
	(14457, 120)
	(14510, 799)
};
\addplot[only marks, mark=pentagon, mark size=1.5pt, color=reducer-drdd, forget plot, mark options={line width=0.7pt}] coordinates {
	(332, 216)
	(749, 466)
	(191, 148)
	(265, 211)
	(124, 113)
};
\addplot[only marks, mark=triangle*, mark size=1.5pt, color=reducer-drdd, forget plot, mark options={line width=0.7pt}] coordinates {
	(2049, 1991)
	(2984, 2887)
	(8278, 7591)
	(54669, 52001)
	(81240, 11027)
	(139161, 61394)
	(191584, 32617)
	(39139, 4723)
	(38813, 3943)
	(39326, 3556)
};
\addplot[only marks, mark=o, mark size=1.5pt, color=reducer-drdd, forget plot, mark options={line width=0.7pt}] coordinates {
	(232, 218)
	(408, 368)
	(407, 375)
	(112, 101)
	(147, 135)
	(93, 90)
	(47, 42)
	(156, 108)
	(45, 42)
	(124, 106)
};
\addplot[only marks, mark=square*, mark size=1.5pt, color=reducer-drdd, forget plot, mark options={line width=0.7pt}] coordinates {
	(3483, 1079)
	(3250, 1035)
	(3189, 811)
	(3403, 1004)
	(81948, 19966)
	(87051, 19980)
	(83663, 19576)
	(84240, 20000)
	(82087, 19841)
	(87945, 19381)
};
\addplot[only marks, mark=x, mark size=1.5pt, color=reducer-drdd, forget plot, mark options={line width=0.7pt}] coordinates {
	(11, 11)
	(15, 15)
	(153, 153)
	(130, 130)
	(30, 21)
	(46, 46)
	(94, 94)
	(10, 10)
	(10, 9)
	(35, 35)
};
\addplot[only marks, mark=pentagon*, mark size=1.5pt, color=reducer-autodidact, forget plot, mark options={line width=0.7pt}] coordinates {
	(579, 112)
};
\addplot[only marks, mark=*, mark size=1.5pt, color=reducer-autodidact, forget plot, mark options={line width=0.7pt}] coordinates {
	(2863, 457)
	(2466, 281)
	(79771, 151)
	(44448, 114)
	(19082, 101)
	(51918, 138)
	(35960, 135)
	(99824, 142)
	(62460, 155)
	(13507, 199)
};
\addplot[only marks, mark=square, mark size=1.5pt, color=reducer-autodidact, forget plot, mark options={line width=0.7pt}] coordinates {
	(840, 159)
	(702, 113)
	(630, 166)
	(885, 156)
	(627, 220)
	(698, 252)
	(664, 181)
	(1110, 188)
	(528, 98)
	(760, 117)
};
\addplot[only marks, mark=diamond, mark size=1.5pt, color=reducer-autodidact, forget plot, mark options={line width=0.7pt}] coordinates {
	(111516, 2270)
	(1706, 90)
};
\addplot[only marks, mark=triangle, mark size=1.5pt, color=reducer-autodidact, forget plot, mark options={line width=0.7pt}] coordinates {
	(1045, 72)
	(391, 377)
};
\addplot[only marks, mark=diamond*, mark size=1.5pt, color=reducer-autodidact, forget plot, mark options={line width=0.7pt}] coordinates {
	(14372, 118)
	(14481, 7)
	(3885, 3)
	(14457, 139)
	(14510, 794)
};
\addplot[only marks, mark=pentagon, mark size=1.5pt, color=reducer-autodidact, forget plot, mark options={line width=0.7pt}] coordinates {
	(332, 123)
	(749, 304)
	(191, 135)
	(265, 170)
	(124, 59)
};
\addplot[only marks, mark=triangle*, mark size=1.5pt, color=reducer-autodidact, forget plot, mark options={line width=0.7pt}] coordinates {
	(2049, 862)
	(2984, 1058)
	(8278, 1495)
	(54669, 2772)
	(81240, 1364)
	(139161, 712)
	(191584, 708)
	(39139, 172)
	(38813, 172)
	(39326, 151)
};
\addplot[only marks, mark=o, mark size=1.5pt, color=reducer-autodidact, forget plot, mark options={line width=0.7pt}] coordinates {
	(232, 126)
	(408, 208)
	(407, 233)
	(112, 64)
	(147, 38)
	(93, 69)
	(47, 36)
	(156, 58)
	(45, 36)
	(124, 83)
};
\addplot[only marks, mark=square*, mark size=1.5pt, color=reducer-autodidact, forget plot, mark options={line width=0.7pt}] coordinates {
	(3483, 66)
	(3250, 149)
	(3189, 140)
	(3403, 130)
	(81948, 15)
	(87051, 63)
	(83663, 60)
	(84240, 62)
	(82087, 60)
	(87945, 33)
};
\addplot[only marks, mark=x, mark size=1.5pt, color=reducer-autodidact, forget plot, mark options={line width=0.7pt}] coordinates {
	(11, 11)
	(15, 15)
	(153, 153)
	(130, 130)
	(30, 21)
	(46, 46)
	(94, 94)
	(10, 10)
	(10, 9)
	(35, 35)
};
\addlegendimage{only marks, mark=pentagon*, mark size=1.4pt, color=black!70, mark options={line width=0.45pt}}
\addlegendentry{asn1}
\addlegendimage{only marks, mark=*, mark size=1.4pt, color=black!70, mark options={line width=0.45pt}}
\addlegendentry{c}
\addlegendimage{only marks, mark=square, mark size=1.4pt, color=black!70, mark options={line width=0.45pt}}
\addlegendentry{crashjs}
\addlegendimage{only marks, mark=diamond, mark size=1.4pt, color=black!70, mark options={line width=0.45pt}}
\addlegendentry{graphviz}
\addlegendimage{only marks, mark=triangle, mark size=1.4pt, color=black!70, mark options={line width=0.45pt}}
\addlegendentry{lua}
\addlegendimage{only marks, mark=diamond*, mark size=1.4pt, color=black!70, mark options={line width=0.45pt}}
\addlegendentry{pdf}
\addlegendimage{only marks, mark=pentagon, mark size=1.4pt, color=black!70, mark options={line width=0.45pt}}
\addlegendentry{python}
\addlegendimage{only marks, mark=triangle*, mark size=1.4pt, color=black!70, mark options={line width=0.45pt}}
\addlegendentry{smt}
\addlegendimage{only marks, mark=o, mark size=1.4pt, color=black!70, mark options={line width=0.45pt}}
\addlegendentry{sql}
\addlegendimage{only marks, mark=square*, mark size=1.4pt, color=black!70, mark options={line width=0.45pt}}
\addlegendentry{xml}
\addlegendimage{only marks, mark=x, mark size=1.4pt, color=black!70, mark options={line width=0.45pt}}
\addlegendentry{yaml}
\node[anchor=north west, font=\scriptsize, fill=white, inner sep=2pt] at (rel axis cs:0.01,0.97) {\textcolor{reducer-autodidact}{\rule{1.3ex}{1.3ex}}\,\autodidact{}\quad \textcolor{reducer-drdd}{\rule{1.3ex}{1.3ex}}\,\drdd{}};
\end{axis}
\end{tikzpicture}
	\caption{Output size against input size for \drdd{} and \autodidact{}
		on the \worded{} returned-word inputs, in non-whitespace kB.
		\drdd{}'s outputs grow with the input, and \autodidact{}'s stay flat.
		Marks distinguish families,
		and the dashed diagonal marks no reduction.}
	\Description{A linear scatter plot of output size against input size
		for byte-level delta debugging and AutoDiDact,
		one point per input and reducer, marked by family.}
	\label{fig:sizesAutodidact}

	\bigskip
	\definecolor{reducer-creduce}{HTML}{C0007A}
\definecolor{reducer-hdd}{HTML}{E65C00}
\definecolor{reducer-autodidact}{HTML}{0044CC}
\pgfplotslegendfromname{sizes-tools-legend}
\par\smallskip
\begin{tikzpicture}
\begin{axis}[
	width=\textwidth,
	height=0.36\textwidth,
	xmode=linear,
	ymode=linear,
	xmin=0,
	xmax=201164,
	ymin=0,
	ymax=64464,
	scaled x ticks=base 10:-3,
	scaled y ticks=base 10:-3,
	xtick scale label code/.code={},
	ytick scale label code/.code={},
	xticklabel style={/pgf/number format/fixed},
	yticklabel style={/pgf/number format/fixed},
	xlabel={input size (kB)},
	ylabel={output size (kB)},
	label style={font=\footnotesize},
	tick label style={font=\footnotesize},
	major grid style={line width=0.2pt, draw=black!12},
	axis line style={draw=black!45},
	tick style={draw=black!45},
	legend cell align=left,
	grid=major,
	legend to name=sizes-tools-legend,
	legend columns=11,
	legend style={font=\scriptsize, draw=none, nodes={text height=1.2ex, text depth=0.3ex}, /tikz/every even column/.append style={column sep=4pt}},
]
\addplot[no marks, line width=0.4pt, black!45, dashed, forget plot] coordinates {
	(0, 0)
	(64464, 64464)
};
\addplot[only marks, mark=pentagon*, mark size=1.5pt, color=reducer-creduce, forget plot, mark options={line width=0.7pt}] coordinates {
	(579, 579)
};
\addplot[only marks, mark=*, mark size=1.5pt, color=reducer-creduce, forget plot, mark options={line width=0.7pt}] coordinates {
	(2863, 399)
	(2466, 224)
	(79771, 132)
	(44448, 98)
	(19082, 126)
	(51918, 65)
	(35960, 76)
	(99824, 91)
	(62460, 135)
	(13507, 229)
};
\addplot[only marks, mark=square, mark size=1.5pt, color=reducer-creduce, forget plot, mark options={line width=0.7pt}] coordinates {
	(840, 127)
	(702, 134)
	(630, 208)
	(885, 184)
	(627, 240)
	(698, 194)
	(664, 144)
	(1110, 207)
	(528, 107)
	(760, 212)
};
\addplot[only marks, mark=diamond, mark size=1.5pt, color=reducer-creduce, forget plot, mark options={line width=0.7pt}] coordinates {
	(111516, 51)
	(1706, 85)
};
\addplot[only marks, mark=triangle, mark size=1.5pt, color=reducer-creduce, forget plot, mark options={line width=0.7pt}] coordinates {
	(1045, 58)
	(391, 389)
};
\addplot[only marks, mark=diamond*, mark size=1.5pt, color=reducer-creduce, forget plot, mark options={line width=0.7pt}] coordinates {
	(14372, 113)
	(14481, 115)
	(3885, 3)
	(14457, 111)
	(14510, 3)
};
\addplot[only marks, mark=pentagon, mark size=1.5pt, color=reducer-creduce, forget plot, mark options={line width=0.7pt}] coordinates {
	(332, 78)
	(749, 220)
	(191, 92)
	(265, 182)
	(124, 62)
};
\addplot[only marks, mark=triangle*, mark size=1.5pt, color=reducer-creduce, forget plot, mark options={line width=0.7pt}] coordinates {
	(2049, 719)
	(2984, 1185)
	(8278, 957)
	(54669, 1925)
	(81240, 655)
	(139161, 407)
	(191584, 449)
	(39139, 157)
	(38813, 157)
	(39326, 157)
};
\addplot[only marks, mark=o, mark size=1.5pt, color=reducer-creduce, forget plot, mark options={line width=0.7pt}] coordinates {
	(232, 101)
	(408, 197)
	(407, 177)
	(112, 35)
	(147, 35)
	(93, 41)
	(47, 35)
	(156, 38)
	(45, 30)
	(124, 78)
};
\addplot[only marks, mark=square*, mark size=1.5pt, color=reducer-creduce, forget plot, mark options={line width=0.7pt}] coordinates {
	(3483, 78)
	(3250, 1416)
	(3189, 214)
	(3403, 141)
	(81948, 28960)
	(87051, 19437)
	(83663, 4426)
	(84240, 27401)
	(82087, 26684)
	(87945, 28116)
};
\addplot[only marks, mark=x, mark size=1.5pt, color=reducer-creduce, forget plot, mark options={line width=0.7pt}] coordinates {
	(11, 11)
	(15, 15)
	(153, 153)
	(130, 130)
	(30, 21)
	(46, 46)
	(94, 94)
	(10, 10)
	(10, 9)
	(35, 35)
};
\addplot[only marks, mark=*, mark size=1.5pt, color=reducer-hdd, forget plot, mark options={line width=0.7pt}] coordinates {
	(2863, 481)
	(2466, 277)
	(79771, 182)
	(44448, 111)
	(19082, 152)
	(51918, 107)
	(35960, 146)
	(99824, 194)
	(62460, 185)
	(13507, 315)
};
\addplot[only marks, mark=square, mark size=1.5pt, color=reducer-hdd, forget plot, mark options={line width=0.7pt}] coordinates {
	(840, 126)
	(702, 151)
	(630, 249)
	(885, 209)
	(627, 354)
	(698, 204)
	(664, 130)
	(1110, 222)
	(528, 122)
	(760, 220)
};
\addplot[only marks, mark=diamond, mark size=1.5pt, color=reducer-hdd, forget plot, mark options={line width=0.7pt}] coordinates {
	(111516, 1614)
	(1706, 89)
};
\addplot[only marks, mark=triangle, mark size=1.5pt, color=reducer-hdd, forget plot, mark options={line width=0.7pt}] coordinates {
	(1045, 114)
	(391, 357)
};
\addplot[only marks, mark=pentagon, mark size=1.5pt, color=reducer-hdd, forget plot, mark options={line width=0.7pt}] coordinates {
	(332, 142)
	(749, 426)
	(191, 150)
	(265, 225)
	(124, 74)
};
\addplot[only marks, mark=triangle*, mark size=1.5pt, color=reducer-hdd, forget plot, mark options={line width=0.7pt}] coordinates {
	(39139, 223)
	(38813, 223)
	(39326, 234)
};
\addplot[only marks, mark=o, mark size=1.5pt, color=reducer-hdd, forget plot, mark options={line width=0.7pt}] coordinates {
	(407, 210)
	(112, 59)
	(147, 41)
	(93, 45)
	(47, 39)
	(156, 63)
	(45, 39)
	(124, 71)
};
\addplot[only marks, mark=square*, mark size=1.5pt, color=reducer-hdd, forget plot, mark options={line width=0.7pt}] coordinates {
	(3483, 29)
	(3250, 118)
	(3189, 117)
	(3403, 104)
	(81948, 15)
	(87051, 47)
	(83663, 48)
	(84240, 106)
	(82087, 45)
	(87945, 15)
};
\addplot[only marks, mark=pentagon*, mark size=1.5pt, color=reducer-autodidact, forget plot, mark options={line width=0.7pt}] coordinates {
	(579, 112)
};
\addplot[only marks, mark=*, mark size=1.5pt, color=reducer-autodidact, forget plot, mark options={line width=0.7pt}] coordinates {
	(2863, 457)
	(2466, 281)
	(79771, 151)
	(44448, 114)
	(19082, 101)
	(51918, 138)
	(35960, 135)
	(99824, 142)
	(62460, 155)
	(13507, 199)
};
\addplot[only marks, mark=square, mark size=1.5pt, color=reducer-autodidact, forget plot, mark options={line width=0.7pt}] coordinates {
	(840, 159)
	(702, 113)
	(630, 166)
	(885, 156)
	(627, 220)
	(698, 252)
	(664, 181)
	(1110, 188)
	(528, 98)
	(760, 117)
};
\addplot[only marks, mark=diamond, mark size=1.5pt, color=reducer-autodidact, forget plot, mark options={line width=0.7pt}] coordinates {
	(111516, 2270)
	(1706, 90)
};
\addplot[only marks, mark=triangle, mark size=1.5pt, color=reducer-autodidact, forget plot, mark options={line width=0.7pt}] coordinates {
	(1045, 72)
	(391, 377)
};
\addplot[only marks, mark=diamond*, mark size=1.5pt, color=reducer-autodidact, forget plot, mark options={line width=0.7pt}] coordinates {
	(14372, 118)
	(14481, 7)
	(3885, 3)
	(14457, 139)
	(14510, 794)
};
\addplot[only marks, mark=pentagon, mark size=1.5pt, color=reducer-autodidact, forget plot, mark options={line width=0.7pt}] coordinates {
	(332, 123)
	(749, 304)
	(191, 135)
	(265, 170)
	(124, 59)
};
\addplot[only marks, mark=triangle*, mark size=1.5pt, color=reducer-autodidact, forget plot, mark options={line width=0.7pt}] coordinates {
	(2049, 862)
	(2984, 1058)
	(8278, 1495)
	(54669, 2772)
	(81240, 1364)
	(139161, 712)
	(191584, 708)
	(39139, 172)
	(38813, 172)
	(39326, 151)
};
\addplot[only marks, mark=o, mark size=1.5pt, color=reducer-autodidact, forget plot, mark options={line width=0.7pt}] coordinates {
	(232, 126)
	(408, 208)
	(407, 233)
	(112, 64)
	(147, 38)
	(93, 69)
	(47, 36)
	(156, 58)
	(45, 36)
	(124, 83)
};
\addplot[only marks, mark=square*, mark size=1.5pt, color=reducer-autodidact, forget plot, mark options={line width=0.7pt}] coordinates {
	(3483, 66)
	(3250, 149)
	(3189, 140)
	(3403, 130)
	(81948, 15)
	(87051, 63)
	(83663, 60)
	(84240, 62)
	(82087, 60)
	(87945, 33)
};
\addplot[only marks, mark=x, mark size=1.5pt, color=reducer-autodidact, forget plot, mark options={line width=0.7pt}] coordinates {
	(11, 11)
	(15, 15)
	(153, 153)
	(130, 130)
	(30, 21)
	(46, 46)
	(94, 94)
	(10, 10)
	(10, 9)
	(35, 35)
};
\addlegendimage{only marks, mark=pentagon*, mark size=1.4pt, color=black!70, mark options={line width=0.45pt}}
\addlegendentry{asn1}
\addlegendimage{only marks, mark=*, mark size=1.4pt, color=black!70, mark options={line width=0.45pt}}
\addlegendentry{c}
\addlegendimage{only marks, mark=square, mark size=1.4pt, color=black!70, mark options={line width=0.45pt}}
\addlegendentry{crashjs}
\addlegendimage{only marks, mark=diamond, mark size=1.4pt, color=black!70, mark options={line width=0.45pt}}
\addlegendentry{graphviz}
\addlegendimage{only marks, mark=triangle, mark size=1.4pt, color=black!70, mark options={line width=0.45pt}}
\addlegendentry{lua}
\addlegendimage{only marks, mark=diamond*, mark size=1.4pt, color=black!70, mark options={line width=0.45pt}}
\addlegendentry{pdf}
\addlegendimage{only marks, mark=pentagon, mark size=1.4pt, color=black!70, mark options={line width=0.45pt}}
\addlegendentry{python}
\addlegendimage{only marks, mark=triangle*, mark size=1.4pt, color=black!70, mark options={line width=0.45pt}}
\addlegendentry{smt}
\addlegendimage{only marks, mark=o, mark size=1.4pt, color=black!70, mark options={line width=0.45pt}}
\addlegendentry{sql}
\addlegendimage{only marks, mark=square*, mark size=1.4pt, color=black!70, mark options={line width=0.45pt}}
\addlegendentry{xml}
\addlegendimage{only marks, mark=x, mark size=1.4pt, color=black!70, mark options={line width=0.45pt}}
\addlegendentry{yaml}
\node[anchor=north west, font=\scriptsize, fill=white, inner sep=2pt] at (rel axis cs:0.01,0.97) {\textcolor{reducer-autodidact}{\rule{1.3ex}{1.3ex}}\,\autodidact{}\quad \textcolor{reducer-hdd}{\rule{1.3ex}{1.3ex}}\,HDD\quad \textcolor{reducer-creduce}{\rule{1.3ex}{1.3ex}}\,C-Reduce};
\end{axis}
\end{tikzpicture}
	\caption{Output size against input size for \autodidact{}, HDD, and C-Reduce,
		on the axes of \Cref{fig:sizesAutodidact}.
		All three stay flat,
		except C-Reduce on XML.}
	\Description{A linear scatter plot of output size against input size
		for AutoDiDact, HDD, and C-Reduce,
		one point per input and reducer, marked by family.}
	\label{fig:sizesTools}
\end{figure}

\begin{figure}[H]
	\centering
	\definecolor{reducer-drdd}{HTML}{D7191C}
\definecolor{reducer-autodidact}{HTML}{0044CC}
\pgfplotslegendfromname{calls-autodidact-legend}
\par\smallskip
\begin{tikzpicture}
\begin{axis}[
	width=\textwidth,
	height=0.36\textwidth,
	xmode=linear,
	ymode=linear,
	xmin=0,
	xmax=201164,
	ymin=0,
	ymax=515703,
	scaled x ticks=base 10:-3,
	scaled y ticks=base 10:-3,
	xtick scale label code/.code={},
	ytick scale label code/.code={},
	xticklabel style={/pgf/number format/fixed},
	yticklabel style={/pgf/number format/fixed},
	xlabel={input size (kB)},
	ylabel={oracle calls (thousands)},
	label style={font=\footnotesize},
	tick label style={font=\footnotesize},
	major grid style={line width=0.2pt, draw=black!12},
	axis line style={draw=black!45},
	tick style={draw=black!45},
	legend cell align=left,
	grid=major,
	legend to name=calls-autodidact-legend,
	legend columns=11,
	legend style={font=\scriptsize, draw=none, nodes={text height=1.2ex, text depth=0.3ex}, /tikz/every even column/.append style={column sep=4pt}},
]
\addplot[only marks, mark=pentagon*, mark size=1.5pt, color=reducer-drdd, forget plot, mark options={line width=0.7pt}] coordinates {
	(579, 455)
};
\addplot[only marks, mark=*, mark size=1.5pt, color=reducer-drdd, forget plot, mark options={line width=0.7pt}] coordinates {
	(2863, 2679)
	(2466, 1859)
	(79771, 232803)
	(44448, 146979)
	(19082, 42493)
	(51918, 156700)
	(35960, 66093)
	(99824, 284208)
	(62460, 188661)
	(13507, 38223)
};
\addplot[only marks, mark=square, mark size=1.5pt, color=reducer-drdd, forget plot, mark options={line width=0.7pt}] coordinates {
	(840, 1966)
	(702, 2246)
	(630, 1062)
	(885, 1185)
	(627, 1637)
	(698, 1045)
	(664, 1694)
	(1110, 2000)
	(528, 1185)
	(760, 904)
};
\addplot[only marks, mark=diamond, mark size=1.5pt, color=reducer-drdd, forget plot, mark options={line width=0.7pt}] coordinates {
	(111516, 137314)
	(1706, 1461)
};
\addplot[only marks, mark=triangle, mark size=1.5pt, color=reducer-drdd, forget plot, mark options={line width=0.7pt}] coordinates {
	(1045, 1934)
	(391, 1991)
};
\addplot[only marks, mark=diamond*, mark size=1.5pt, color=reducer-drdd, forget plot, mark options={line width=0.7pt}] coordinates {
	(14372, 688)
	(14481, 111)
	(3885, 63)
	(14457, 701)
	(14510, 3929)
};
\addplot[only marks, mark=pentagon, mark size=1.5pt, color=reducer-drdd, forget plot, mark options={line width=0.7pt}] coordinates {
	(332, 1073)
	(749, 1856)
	(191, 765)
	(265, 935)
	(124, 484)
};
\addplot[only marks, mark=triangle*, mark size=1.5pt, color=reducer-drdd, forget plot, mark options={line width=0.7pt}] coordinates {
	(2049, 7914)
	(2984, 11445)
	(8278, 38428)
	(54669, 287101)
	(81240, 78179)
	(139161, 444404)
	(191584, 158747)
	(39139, 17826)
	(38813, 14513)
	(39326, 11455)
};
\addplot[only marks, mark=o, mark size=1.5pt, color=reducer-drdd, forget plot, mark options={line width=0.7pt}] coordinates {
	(232, 751)
	(408, 1333)
	(407, 1164)
	(112, 310)
	(147, 387)
	(93, 218)
	(47, 119)
	(156, 417)
	(45, 117)
	(124, 419)
};
\addplot[only marks, mark=square*, mark size=1.5pt, color=reducer-drdd, forget plot, mark options={line width=0.7pt}] coordinates {
	(3483, 4110)
	(3250, 3896)
	(3189, 3491)
	(3403, 3919)
	(81948, 108336)
	(87051, 108244)
	(83663, 106657)
	(84240, 108471)
	(82087, 108616)
	(87945, 107771)
};
\addplot[only marks, mark=x, mark size=1.5pt, color=reducer-drdd, forget plot, mark options={line width=0.7pt}] coordinates {
	(11, 41)
	(15, 87)
	(153, 407)
	(130, 421)
	(30, 49)
	(46, 92)
	(94, 272)
	(10, 44)
	(10, 41)
	(35, 189)
};
\addplot[only marks, mark=pentagon*, mark size=1.5pt, color=reducer-autodidact, forget plot, mark options={line width=0.7pt}] coordinates {
	(579, 623)
};
\addplot[only marks, mark=*, mark size=1.5pt, color=reducer-autodidact, forget plot, mark options={line width=0.7pt}] coordinates {
	(2863, 3032)
	(2466, 2126)
	(79771, 3023)
	(44448, 5658)
	(19082, 2447)
	(51918, 2533)
	(35960, 3029)
	(99824, 10302)
	(62460, 2690)
	(13507, 3110)
};
\addplot[only marks, mark=square, mark size=1.5pt, color=reducer-autodidact, forget plot, mark options={line width=0.7pt}] coordinates {
	(840, 1782)
	(702, 1441)
	(630, 1434)
	(885, 1371)
	(627, 1535)
	(698, 1233)
	(664, 1521)
	(1110, 1768)
	(528, 1182)
	(760, 1336)
};
\addplot[only marks, mark=diamond, mark size=1.5pt, color=reducer-autodidact, forget plot, mark options={line width=0.7pt}] coordinates {
	(111516, 57940)
	(1706, 1011)
};
\addplot[only marks, mark=triangle, mark size=1.5pt, color=reducer-autodidact, forget plot, mark options={line width=0.7pt}] coordinates {
	(1045, 1325)
	(391, 2385)
};
\addplot[only marks, mark=diamond*, mark size=1.5pt, color=reducer-autodidact, forget plot, mark options={line width=0.7pt}] coordinates {
	(14372, 1419)
	(14481, 672)
	(3885, 1107)
	(14457, 1576)
	(14510, 4733)
};
\addplot[only marks, mark=pentagon, mark size=1.5pt, color=reducer-autodidact, forget plot, mark options={line width=0.7pt}] coordinates {
	(332, 1113)
	(749, 2373)
	(191, 1232)
	(265, 1052)
	(124, 815)
};
\addplot[only marks, mark=triangle*, mark size=1.5pt, color=reducer-autodidact, forget plot, mark options={line width=0.7pt}] coordinates {
	(2049, 8632)
	(2984, 9564)
	(8278, 31307)
	(54669, 185324)
	(81240, 30525)
	(139161, 144595)
	(191584, 97583)
	(39139, 2097)
	(38813, 2126)
	(39326, 1841)
};
\addplot[only marks, mark=o, mark size=1.5pt, color=reducer-autodidact, forget plot, mark options={line width=0.7pt}] coordinates {
	(232, 797)
	(408, 1486)
	(407, 2300)
	(112, 579)
	(147, 504)
	(93, 554)
	(47, 432)
	(156, 640)
	(45, 430)
	(124, 691)
};
\addplot[only marks, mark=square*, mark size=1.5pt, color=reducer-autodidact, forget plot, mark options={line width=0.7pt}] coordinates {
	(3483, 1322)
	(3250, 1580)
	(3189, 2308)
	(3403, 1571)
	(81948, 3887)
	(87051, 4221)
	(83663, 4168)
	(84240, 3993)
	(82087, 4065)
	(87945, 5230)
};
\addplot[only marks, mark=x, mark size=1.5pt, color=reducer-autodidact, forget plot, mark options={line width=0.7pt}] coordinates {
	(11, 138)
	(15, 396)
	(153, 682)
	(130, 702)
	(30, 384)
	(46, 415)
	(94, 535)
	(10, 136)
	(10, 202)
	(35, 582)
};
\addlegendimage{only marks, mark=pentagon*, mark size=1.4pt, color=black!70, mark options={line width=0.45pt}}
\addlegendentry{asn1}
\addlegendimage{only marks, mark=*, mark size=1.4pt, color=black!70, mark options={line width=0.45pt}}
\addlegendentry{c}
\addlegendimage{only marks, mark=square, mark size=1.4pt, color=black!70, mark options={line width=0.45pt}}
\addlegendentry{crashjs}
\addlegendimage{only marks, mark=diamond, mark size=1.4pt, color=black!70, mark options={line width=0.45pt}}
\addlegendentry{graphviz}
\addlegendimage{only marks, mark=triangle, mark size=1.4pt, color=black!70, mark options={line width=0.45pt}}
\addlegendentry{lua}
\addlegendimage{only marks, mark=diamond*, mark size=1.4pt, color=black!70, mark options={line width=0.45pt}}
\addlegendentry{pdf}
\addlegendimage{only marks, mark=pentagon, mark size=1.4pt, color=black!70, mark options={line width=0.45pt}}
\addlegendentry{python}
\addlegendimage{only marks, mark=triangle*, mark size=1.4pt, color=black!70, mark options={line width=0.45pt}}
\addlegendentry{smt}
\addlegendimage{only marks, mark=o, mark size=1.4pt, color=black!70, mark options={line width=0.45pt}}
\addlegendentry{sql}
\addlegendimage{only marks, mark=square*, mark size=1.4pt, color=black!70, mark options={line width=0.45pt}}
\addlegendentry{xml}
\addlegendimage{only marks, mark=x, mark size=1.4pt, color=black!70, mark options={line width=0.45pt}}
\addlegendentry{yaml}
\node[anchor=north east, font=\scriptsize, fill=white, inner sep=2pt] at (rel axis cs:0.99,0.97) {\textcolor{reducer-autodidact}{\rule{1.3ex}{1.3ex}}\,\autodidact{}\quad \textcolor{reducer-drdd}{\rule{1.3ex}{1.3ex}}\,\drdd{}};
\end{axis}
\end{tikzpicture}
	\caption{Oracle calls against input size for \drdd{} and \autodidact{}
		on the \worded{} returned-word inputs.
		\autodidact{}'s calls include its inference.
		\drdd{}'s calls grow with the input,
		and \autodidact{}'s stay low
		except on five SMT inputs and one Graphviz input.}
	\Description{A linear scatter plot of oracle calls against input size
		for byte-level delta debugging and AutoDiDact,
		one point per input and reducer, marked by family.}
	\label{fig:callsAutodidact}

	\bigskip
	\definecolor{reducer-creduce}{HTML}{C0007A}
\definecolor{reducer-hdd}{HTML}{E65C00}
\definecolor{reducer-autodidact}{HTML}{0044CC}
\pgfplotslegendfromname{calls-tools-legend}
\par\smallskip
\begin{tikzpicture}
\begin{axis}[
	width=\textwidth,
	height=0.36\textwidth,
	xmode=linear,
	ymode=linear,
	xmin=0,
	xmax=201164,
	ymin=0,
	ymax=515703,
	scaled x ticks=base 10:-3,
	scaled y ticks=base 10:-3,
	xtick scale label code/.code={},
	ytick scale label code/.code={},
	xticklabel style={/pgf/number format/fixed},
	yticklabel style={/pgf/number format/fixed},
	xlabel={input size (kB)},
	ylabel={oracle calls (thousands)},
	label style={font=\footnotesize},
	tick label style={font=\footnotesize},
	major grid style={line width=0.2pt, draw=black!12},
	axis line style={draw=black!45},
	tick style={draw=black!45},
	legend cell align=left,
	grid=major,
	legend to name=calls-tools-legend,
	legend columns=11,
	legend style={font=\scriptsize, draw=none, nodes={text height=1.2ex, text depth=0.3ex}, /tikz/every even column/.append style={column sep=4pt}},
]
\addplot[only marks, mark=pentagon*, mark size=1.5pt, color=reducer-creduce, forget plot, mark options={line width=0.7pt}] coordinates {
	(579, 3280)
};
\addplot[only marks, mark=*, mark size=1.5pt, color=reducer-creduce, forget plot, mark options={line width=0.7pt}] coordinates {
	(2863, 13856)
	(2466, 4319)
	(79771, 8174)
	(44448, 5993)
	(19082, 5791)
	(51918, 2795)
	(35960, 4586)
	(99824, 12537)
	(62460, 6507)
	(13507, 15364)
};
\addplot[only marks, mark=square, mark size=1.5pt, color=reducer-creduce, forget plot, mark options={line width=0.7pt}] coordinates {
	(840, 2016)
	(702, 1524)
	(630, 5163)
	(885, 3473)
	(627, 2690)
	(698, 1897)
	(664, 1943)
	(1110, 2267)
	(528, 1888)
	(760, 2872)
};
\addplot[only marks, mark=diamond, mark size=1.5pt, color=reducer-creduce, forget plot, mark options={line width=0.7pt}] coordinates {
	(111516, 6781)
	(1706, 2048)
};
\addplot[only marks, mark=triangle, mark size=1.5pt, color=reducer-creduce, forget plot, mark options={line width=0.7pt}] coordinates {
	(1045, 1789)
	(391, 10087)
};
\addplot[only marks, mark=diamond*, mark size=1.5pt, color=reducer-creduce, forget plot, mark options={line width=0.7pt}] coordinates {
	(14372, 3053)
	(14481, 246)
	(3885, 541)
	(14457, 2702)
	(14510, 2794)
};
\addplot[only marks, mark=pentagon, mark size=1.5pt, color=reducer-creduce, forget plot, mark options={line width=0.7pt}] coordinates {
	(332, 2784)
	(749, 7729)
	(191, 3476)
	(265, 5328)
	(124, 1314)
};
\addplot[only marks, mark=triangle*, mark size=1.5pt, color=reducer-creduce, forget plot, mark options={line width=0.7pt}] coordinates {
	(2049, 37289)
	(2984, 62063)
	(8278, 94137)
	(54669, 491145)
	(81240, 92026)
	(139161, 255725)
	(191584, 170158)
	(39139, 8742)
	(38813, 8742)
	(39326, 8763)
};
\addplot[only marks, mark=o, mark size=1.5pt, color=reducer-creduce, forget plot, mark options={line width=0.7pt}] coordinates {
	(232, 1866)
	(408, 5291)
	(407, 5689)
	(112, 786)
	(147, 824)
	(93, 791)
	(47, 422)
	(156, 483)
	(45, 420)
	(124, 1534)
};
\addplot[only marks, mark=square*, mark size=1.5pt, color=reducer-creduce, forget plot, mark options={line width=0.7pt}] coordinates {
	(3483, 7633)
	(3250, 8976)
	(3189, 12071)
	(3403, 10737)
	(81948, 142284)
	(87051, 125681)
	(83663, 137325)
	(84240, 131567)
	(82087, 209083)
	(87945, 135287)
};
\addplot[only marks, mark=x, mark size=1.5pt, color=reducer-creduce, forget plot, mark options={line width=0.7pt}] coordinates {
	(11, 42)
	(15, 216)
	(153, 423)
	(130, 1029)
	(30, 185)
	(46, 232)
	(94, 382)
	(10, 114)
	(10, 68)
	(35, 506)
};
\addplot[only marks, mark=*, mark size=1.5pt, color=reducer-hdd, forget plot, mark options={line width=0.7pt}] coordinates {
	(2863, 5219)
	(2466, 1515)
	(79771, 1902)
	(44448, 7046)
	(19082, 1178)
	(51918, 513)
	(35960, 1116)
	(99824, 14342)
	(62460, 2134)
	(13507, 3467)
};
\addplot[only marks, mark=square, mark size=1.5pt, color=reducer-hdd, forget plot, mark options={line width=0.7pt}] coordinates {
	(840, 252)
	(702, 363)
	(630, 757)
	(885, 394)
	(627, 602)
	(698, 313)
	(664, 405)
	(1110, 707)
	(528, 186)
	(760, 281)
};
\addplot[only marks, mark=diamond, mark size=1.5pt, color=reducer-hdd, forget plot, mark options={line width=0.7pt}] coordinates {
	(111516, 5512)
	(1706, 722)
};
\addplot[only marks, mark=triangle, mark size=1.5pt, color=reducer-hdd, forget plot, mark options={line width=0.7pt}] coordinates {
	(1045, 308)
	(391, 3529)
};
\addplot[only marks, mark=pentagon, mark size=1.5pt, color=reducer-hdd, forget plot, mark options={line width=0.7pt}] coordinates {
	(332, 1117)
	(749, 2124)
	(191, 1087)
	(265, 1014)
	(124, 461)
};
\addplot[only marks, mark=triangle*, mark size=1.5pt, color=reducer-hdd, forget plot, mark options={line width=0.7pt}] coordinates {
	(39139, 945)
	(38813, 945)
	(39326, 834)
};
\addplot[only marks, mark=o, mark size=1.5pt, color=reducer-hdd, forget plot, mark options={line width=0.7pt}] coordinates {
	(407, 846)
	(112, 110)
	(147, 61)
	(93, 126)
	(47, 58)
	(156, 112)
	(45, 60)
	(124, 143)
};
\addplot[only marks, mark=square*, mark size=1.5pt, color=reducer-hdd, forget plot, mark options={line width=0.7pt}] coordinates {
	(3483, 96)
	(3250, 360)
	(3189, 595)
	(3403, 184)
	(81948, 28)
	(87051, 80)
	(83663, 237)
	(84240, 260)
	(82087, 232)
	(87945, 20)
};
\addplot[only marks, mark=pentagon*, mark size=1.5pt, color=reducer-autodidact, forget plot, mark options={line width=0.7pt}] coordinates {
	(579, 623)
};
\addplot[only marks, mark=*, mark size=1.5pt, color=reducer-autodidact, forget plot, mark options={line width=0.7pt}] coordinates {
	(2863, 3032)
	(2466, 2126)
	(79771, 3023)
	(44448, 5658)
	(19082, 2447)
	(51918, 2533)
	(35960, 3029)
	(99824, 10302)
	(62460, 2690)
	(13507, 3110)
};
\addplot[only marks, mark=square, mark size=1.5pt, color=reducer-autodidact, forget plot, mark options={line width=0.7pt}] coordinates {
	(840, 1782)
	(702, 1441)
	(630, 1434)
	(885, 1371)
	(627, 1535)
	(698, 1233)
	(664, 1521)
	(1110, 1768)
	(528, 1182)
	(760, 1336)
};
\addplot[only marks, mark=diamond, mark size=1.5pt, color=reducer-autodidact, forget plot, mark options={line width=0.7pt}] coordinates {
	(111516, 57940)
	(1706, 1011)
};
\addplot[only marks, mark=triangle, mark size=1.5pt, color=reducer-autodidact, forget plot, mark options={line width=0.7pt}] coordinates {
	(1045, 1325)
	(391, 2385)
};
\addplot[only marks, mark=diamond*, mark size=1.5pt, color=reducer-autodidact, forget plot, mark options={line width=0.7pt}] coordinates {
	(14372, 1419)
	(14481, 672)
	(3885, 1107)
	(14457, 1576)
	(14510, 4733)
};
\addplot[only marks, mark=pentagon, mark size=1.5pt, color=reducer-autodidact, forget plot, mark options={line width=0.7pt}] coordinates {
	(332, 1113)
	(749, 2373)
	(191, 1232)
	(265, 1052)
	(124, 815)
};
\addplot[only marks, mark=triangle*, mark size=1.5pt, color=reducer-autodidact, forget plot, mark options={line width=0.7pt}] coordinates {
	(2049, 8632)
	(2984, 9564)
	(8278, 31307)
	(54669, 185324)
	(81240, 30525)
	(139161, 144595)
	(191584, 97583)
	(39139, 2097)
	(38813, 2126)
	(39326, 1841)
};
\addplot[only marks, mark=o, mark size=1.5pt, color=reducer-autodidact, forget plot, mark options={line width=0.7pt}] coordinates {
	(232, 797)
	(408, 1486)
	(407, 2300)
	(112, 579)
	(147, 504)
	(93, 554)
	(47, 432)
	(156, 640)
	(45, 430)
	(124, 691)
};
\addplot[only marks, mark=square*, mark size=1.5pt, color=reducer-autodidact, forget plot, mark options={line width=0.7pt}] coordinates {
	(3483, 1322)
	(3250, 1580)
	(3189, 2308)
	(3403, 1571)
	(81948, 3887)
	(87051, 4221)
	(83663, 4168)
	(84240, 3993)
	(82087, 4065)
	(87945, 5230)
};
\addplot[only marks, mark=x, mark size=1.5pt, color=reducer-autodidact, forget plot, mark options={line width=0.7pt}] coordinates {
	(11, 138)
	(15, 396)
	(153, 682)
	(130, 702)
	(30, 384)
	(46, 415)
	(94, 535)
	(10, 136)
	(10, 202)
	(35, 582)
};
\addlegendimage{only marks, mark=pentagon*, mark size=1.4pt, color=black!70, mark options={line width=0.45pt}}
\addlegendentry{asn1}
\addlegendimage{only marks, mark=*, mark size=1.4pt, color=black!70, mark options={line width=0.45pt}}
\addlegendentry{c}
\addlegendimage{only marks, mark=square, mark size=1.4pt, color=black!70, mark options={line width=0.45pt}}
\addlegendentry{crashjs}
\addlegendimage{only marks, mark=diamond, mark size=1.4pt, color=black!70, mark options={line width=0.45pt}}
\addlegendentry{graphviz}
\addlegendimage{only marks, mark=triangle, mark size=1.4pt, color=black!70, mark options={line width=0.45pt}}
\addlegendentry{lua}
\addlegendimage{only marks, mark=diamond*, mark size=1.4pt, color=black!70, mark options={line width=0.45pt}}
\addlegendentry{pdf}
\addlegendimage{only marks, mark=pentagon, mark size=1.4pt, color=black!70, mark options={line width=0.45pt}}
\addlegendentry{python}
\addlegendimage{only marks, mark=triangle*, mark size=1.4pt, color=black!70, mark options={line width=0.45pt}}
\addlegendentry{smt}
\addlegendimage{only marks, mark=o, mark size=1.4pt, color=black!70, mark options={line width=0.45pt}}
\addlegendentry{sql}
\addlegendimage{only marks, mark=square*, mark size=1.4pt, color=black!70, mark options={line width=0.45pt}}
\addlegendentry{xml}
\addlegendimage{only marks, mark=x, mark size=1.4pt, color=black!70, mark options={line width=0.45pt}}
\addlegendentry{yaml}
\node[anchor=north east, font=\scriptsize, fill=white, inner sep=2pt] at (rel axis cs:0.99,0.97) {\textcolor{reducer-autodidact}{\rule{1.3ex}{1.3ex}}\,\autodidact{}\quad \textcolor{reducer-hdd}{\rule{1.3ex}{1.3ex}}\,HDD\quad \textcolor{reducer-creduce}{\rule{1.3ex}{1.3ex}}\,C-Reduce};
\end{axis}
\end{tikzpicture}
	\caption{Oracle calls against input size for \autodidact{}, HDD, and C-Reduce,
		on the axes of \Cref{fig:callsAutodidact}.
		HDD makes fewer calls than \autodidact{} on 44 of its 50 inputs,
		and \autodidact{} fewer than C-Reduce on 59 of 75.}
	\Description{A linear scatter plot of oracle calls against input size
		for AutoDiDact, HDD, and C-Reduce,
		one point per input and reducer, marked by family.}
	\label{fig:callsTools}
\end{figure}

\end{document}